\documentclass[11pt,a4paper]{article}
\usepackage[margin=25mm]{geometry}
\usepackage{amsmath,amssymb,amsthm}
\usepackage{booktabs,longtable,array}
\usepackage{enumitem}
\usepackage[hidelinks]{hyperref}

\newtheorem{theorem}{Theorem}
\newtheorem{proposition}[theorem]{Proposition}
\newtheorem{lemma}[theorem]{Lemma}
\newtheorem{corollary}[theorem]{Corollary}
\newtheorem{approximation}[theorem]{Approximation}
\theoremstyle{definition}
\newtheorem{definition}{Definition}
\newtheorem{assumption}{Assumption}

\theoremstyle{remark}
\newtheorem{remark}{Remark}

\newcommand{\R}{\mathbb{R}}
\newcommand{\Prob}{\mathrm{Pr}}
\newcommand{\SD}{\mathrm{SD}}
\DeclareMathOperator*{\argmax}{arg\,max}

\title{Why flattening the $P$-value histogram works: a theory of the standard-deviation optimization in tensor-decomposition-based unsupervised feature extraction}
\author{Y-h. Taguchi$^{1,*}$ and Turki Turki$^2$\\[4pt]
\small $^{1}$Bioinformatics Center, Institute for Chemical Research, Kyoto University, Gokasho, Uji, Kyoto 611-0011, Japan\\
\small Permanent address: Department of Physics, Chuo University, Tokyo, Japan\\
\small $^2$ Department of Computer Science, King Abdulaziz University, Jeddah 21589, Saudi Arabia \\
\small $^{*}$Corresponding author. Email: \texttt{tag@granular.com}}
\date{}

\begin{document}
\maketitle

\begin{abstract}
\noindent\textbf{Background:} Tensor-decomposition-based unsupervised feature extraction (TD-based UFE) selects genes by attributing $P$-values to the components of a gene singular vector under a null normal distribution. The standard deviation $\sigma$ of this distribution is unknown and is chosen so that the histogram of the $P$-values of the genes that are not selected becomes as flat as possible. This heuristic has worked well in practice but has had no theoretical justification.

\noindent\textbf{Results:} We prove, for a model of signal, mean expression and noise, that the higher-order singular value decomposition has a component that reproduces the experimental design, that the components of its gene singular vector belonging to background genes are approximately normal with standard deviation $\sigma_0=\sqrt d/s$, where $d$ is the noise variance and $s$ the singular value, and that signal genes separate from the background even when they form a positive fraction of all genes. The accuracy of these approximations is governed by dimensionless quantities that can be estimated from the data. We derive the limit of the objective function of the flattening and prove, by a computer-assisted argument based on interval arithmetic, that for 100 histogram bins, Benjamini--Hochberg level $0.01$ and every fraction of background genes between $0.1$ and $0.99$, the limit has a unique minimizer within $1.1\%$ of $\sigma_0$. We formalize the practice of taking the valley to the right of the peak of the objective as a maximum-drop rule and prove that the resulting estimator is consistent up to this bounded bias. In simulations the sampling fluctuation of the estimate was $0.5$--$3.6\%$ for $2\times10^{3}$ to $10^{5}$ genes. Two effects specific to uncentred data, contamination of background genes and cancellation of signals by the mean expression, are identified.

\noindent\textbf{Conclusions:} Flattening the $P$-value histogram estimates the noise level of the background genes in the gene singular vector, with a small and bounded systematic bias, provided that the signals are strong enough to be detected individually and that the noise variances of background genes are nearly equal. We give diagnostics and two predictions that can be checked with existing analyses.
\end{abstract}


\section{Background}
TD-based UFE \cite{Taguchi2024} proceeds in four steps. (1) A tensor of omics measurements, for example genes $\times$ samples $\times$ conditions, is decomposed by the higher-order singular value decomposition (HOSVD) \cite{DeLathauwer2000,Kolda2009}. (2) Among the singular vectors attached to the samples and conditions, one looks for those whose patterns reflect the experimental design, e.g.\ a vector that takes different values for patients and controls. (3) Through the core tensor of the HOSVD, the gene singular vector associated with these sample vectors is identified. (4) Each component of this gene singular vector is regarded as a score of the corresponding gene, a $P$-value is attributed to it under the assumption that the scores of genes without signal follow a normal distribution with mean zero, and genes with small corrected $P$-values are selected.

Step (4) requires the standard deviation of that normal distribution, which is not known. In \cite{Taguchi2022} it was proposed to choose the standard deviation $\sigma$ so that the histogram of the $P$-values of the genes that are \emph{not} selected is as flat as possible. The idea is that genes without signal should have uniformly distributed $P$-values if, and only if, the correct $\sigma$ is used. The procedure has been used in many applications, but why it works, under what conditions, and how accurately it recovers the true standard deviation have remained open questions.

The question is related to the estimation of an empirical null distribution \cite{Efron2004,JinCai2007} and of the proportion of null hypotheses \cite{Storey2002}. Two features, however, are specific to TD-based UFE. First, the quantity being tested is a component of a singular vector, which is normalized to unit length and whose components are therefore dependent. Second, the null standard deviation is estimated from the same data, and with the same selection rule, that are used to select the genes.

This paper answers these questions. Its contributions are as follows.
\begin{enumerate}
\item We prove that, when the signal is strong, the HOSVD has a component $\ell_1^{*}$ whose sample-side singular vector reproduces the pattern of the design, that the background components of its gene singular vector are approximately normal with $\sigma_0=\sqrt{d}/s_{\ell_1^{*}}$ ($d$ is the noise variance and $s_{\ell_1^{*}}$ the singular value of the component), and that signal genes with strong effective signals separate from the background (Theorem~\ref{thm:main}); we also give conditions under which TD-based UFE selects this component (Proposition~\ref{prop:select}). Because $N$ and $M$ are fixed numbers in any real data set, we express the accuracy of these approximations through quantities computable from the data (Table~\ref{tab:diag}).
\item We find a condition specific to uncentred data: the singular vector of interest can be contaminated by the mean expression, which shifts background genes in proportion to their mean expression (Approximation~\ref{prop:contam}).
\item We derive the limit of the objective function and prove that its minimizer is unique and lies within $1.1\%$ of the true value for all fractions $\pi_0\in[0.1,0.99]$ of background genes, in the limit of large data sets with perfectly separated signals and for $K=100$ histogram bins and BH level $\alpha=0.01$ (Theorem~\ref{thm:cap}). Part of the proof is computer-assisted.
\item We show that, given normal background scores and separated signal genes, the estimator converges to $\tau^{*}\sigma_0$, where $\tau^{*}\in[0.989,1]$ is the minimizer of the limiting objective (Theorem~\ref{thm:consist}), and that dividing by $\tau^{*}$ gives a consistent estimator of $\sigma_0$ (Corollary~\ref{cor:biascorr}). We formalize the practice of selecting the valley to the right of the peak as a maximum-drop rule and prove the same convergence for it (Theorem~\ref{thm:drop}).
\item We derive two predictions and a set of diagnostics that can be checked with existing analyses (Sections~\ref{sec:diag} and \ref{sec:pred}).
\end{enumerate}

Section~\ref{sec:outline} states what has to be shown and which partial results are needed for it. The Methods section describes TD-based UFE, the standard-deviation optimization and the statistical model, and the Results section establishes the partial results one by one.

\subsection{What is to be shown}\label{sec:outline}
This section states the goal of the paper, the chain of results needed to reach it, and where each result is established.

\paragraph{Model.} Let $X\in\R^{N\times M}$ be the gene-mode unfolding of the data tensor (Section~\ref{sec:method}): $M=JQ$, and the column index $r$ enumerates the pairs $(j,q)$ of sample and condition indices. We model it as
\begin{equation}\label{eq:model}
X=\frac{1}{\sqrt{M}}\,a^{0}\mathbf{1}_M^{\top}+a\,b^{\top}+Z .
\end{equation}
Here $a\in\R^{N}$ is the gene pattern of the signal, non-zero only for signal genes, and $b\in\R^{M}$ is a unit vector describing the pattern of the signal over samples and conditions. Because the columns of $X$ are indexed by pairs $(j,q)$, $b$ is the unfolding of a $J\times Q$ array, $b_r=b_{(j,q)}$. For example, if the design is described by a pattern $b^{(2)}\in\R^{J}$ in mode~2 (over samples) and a pattern $b^{(3)}\in\R^{Q}$ in mode~3 (over conditions), then $b_{(j,q)}=b^{(2)}_jb^{(3)}_q$. $a^{0}\in\R^{N}$ is the mean expression and $Z$ is noise. We write $\mathbf{1}_J\in\R^{J}$, $\mathbf{1}_Q\in\R^{Q}$ and $\mathbf{1}_M\in\R^{M}$ for vectors of ones, and
\begin{equation}\label{eq:evec}
e_J=\frac{\mathbf{1}_J}{\sqrt J},\qquad e_Q=\frac{\mathbf{1}_Q}{\sqrt Q},\qquad e_M=\frac{\mathbf{1}_M}{\sqrt M}=e_J\otimes e_Q
\end{equation}
for the constant vectors of unit length in mode~2, in mode~3 and over the columns of $X$; in particular the mean term of \eqref{eq:model} is $a^{0}e_M^{\top}$. All terms are explained in Section~\ref{sec:growM}.

\paragraph{Goal.} TD-based UFE works for data of the form \eqref{eq:model} if the following three statements hold.
\begin{enumerate}[label=\textbf{(G\arabic*)}]
\item \emph{The HOSVD captures the signal in one component, and TD-based UFE finds it.} There is an index $\ell_1^{*}$ such that the right singular vector $v_{\ell_1^{*}}$ paired with the gene singular vector $u_{\ell_1^{*}\cdot}$ reproduces the pattern $b$, i.e.\ $\langle b,v_{\ell_1^{*}}\rangle\approx1$; the gene singular vector then carries the gene pattern $a$. That TD-based UFE finds this $\ell_1^{*}$ follows from the way the HOSVD is computed: the gene singular vectors are the left singular vectors of $X$, and the paired right singular vector is the combination of sample-side singular vectors weighted by the core tensor, Eq.~\eqref{eq:vhat}. This does not follow from $v_{\ell_1^{*}}\approx b$ alone, because the sample-side singular vectors come from other unfoldings of the tensor. Under additional conditions on the sample modes (Proposition~\ref{prop:select}), namely that $b$ is a product $b_{(j,q)}=b^{(2)}_jb^{(3)}_q$, that each of $b^{(2)}$ and $b^{(3)}$ is orthogonal or equal to the constant vector of its mode, and that the signal is uncorrelated with the mean expression over genes, there are sample-side singular vectors $u_{\ell_2^{*}j}\approx b^{(2)}_j$ and $u_{\ell_3^{*}q}\approx b^{(3)}_q$, and the largest $|G(\ell_1\ell_2^{*}\ell_3^{*})|$ over $\ell_1$ is attained at $\ell_1=\ell_1^{*}$. This is how the component is selected.
\item \emph{The scores have the assumed distribution.} For this component $\ell_1^{*}$, the scores $u_{\ell_1^{*} i}$ of background genes are approximately $N(0,\sigma_0^2)$ with a common $\sigma_0$, whereas those of signal genes are much larger. This justifies the $P$-values \eqref{eq:pval}. A common $\sigma_0$ presupposes that the noise variances of background genes are nearly equal; this is made an explicit assumption (Assumption~\ref{as:het}).
\item \emph{$\sigma_0$ can be estimated.} The $P$-values require $\sigma_0$, the standard deviation of the scores of the background genes $i\in\mathcal{I}_0$ (the set of background genes, Section~\ref{sec:method}) \emph{only}. But $\mathcal{I}_0$ is what is to be found, and the standard deviation of all scores overestimates $\sigma_0$, because the signal genes take up part of the unit length of $u_{\ell_1^{*}\cdot}$ (Corollary~\ref{cor:local}). Therefore $\sigma$ must be estimated without knowing $\mathcal{I}_0$. This is the role of the standard-deviation optimization, which for this reason comes \emph{before} the final selection by BH. We must show that flattening the histogram recovers $\sigma_0$.
\end{enumerate}
The procedure is accordingly: (1) compute the HOSVD; (2) select $\hat\ell_2$ and $\hat\ell_3$ from the design and $\hat\ell_1$ through the core tensor; (3) estimate $\hat\sigma$ by flattening the $P$-value histogram, where for each candidate $\sigma$ the genes selected by BH at that $\sigma$ are left out; (4) compute the $P$-values with $\hat\sigma$ and select genes by BH.

\paragraph{Several signals.} Model \eqref{eq:model} contains a single signal. If there are several, $X=\frac{1}{\sqrt{M}}a^{0}\mathbf{1}_M^{\top}+\sum_{m}a^{(m)}b^{(m)\top}+Z$ with nearly orthogonal patterns $b^{(m)}$ and well-separated strengths, each signal is expected to be captured by its own component $\ell_1^{(m)}$, to which the analysis of this paper applies separately. We do not prove this here.

\paragraph{Logical structure.} The argument is divided into two parts that can be checked independently. Sections~\ref{sec:limit}--\ref{sec:drop} show that flattening recovers $\sigma_0$ (up to a bias of at most $1.1\%$) under two explicit conditions on the scores: the background scores are normal (condition (N)) and the signal genes are separated from the background (condition (S)), stated precisely at the beginning of Section~\ref{sec:limit}. These results do not depend on where (N) and (S) come from. Section~\ref{sec:growM} derives (N) and (S) from the model \eqref{eq:model} (Theorem~\ref{thm:main}): the normal law of the background scores arises because each score is a weighted sum of the noise over the $M$ columns of $X$ (its projection on a deterministic vector is exactly normal for Gaussian noise and approximately normal otherwise by the central limit theorem, and the score differs from it by a term whose mean square is small), and separation arises from strong effective signals, Eq.~\eqref{eq:mueff}.

\paragraph{Partial results needed, and where they are shown.}
\begin{enumerate}
\item The HOSVD algorithm, in particular the fact that the gene singular vectors are the left singular vectors of $X$ and the relation \eqref{eq:vhat} between the paired right singular vectors and the core tensor; the definitions of the $P$-values and of the objective of the optimization (Section~\ref{sec:method}).
\item Basic facts: flattening recovers $\sigma_0$ when there is no signal; it fails when signal and background overlap; for Gaussian noise, the normal law of the background follows from rotational invariance (Section~\ref{sec:prelim}).
\item For a given data set: (G1) (Theorem~\ref{thm:main} and Proposition~\ref{prop:select}), (G2) in the form of conditions (N) and (S), and the reason for (G3) (Theorem~\ref{thm:main}, Corollary~\ref{cor:local}); the quantities that control the accuracy of these approximations, and how to check them (Section~\ref{sec:growM}).
\item (G3) in the limit of large data sets: the objective of the optimization has a unique minimizer, within $1.1\%$ of $\sigma_0$ (Section~\ref{sec:limit}).
\item (G3) for real data: the objective computed from data converges to its limit; effects of finite $N$ and finite signal strength (Section~\ref{sec:sep}); a rule for selecting the minimizer that does not require knowing $\sigma_0$ (Section~\ref{sec:drop}).
\item Predictions that can be checked with existing analyses (Section~\ref{sec:pred}).
\end{enumerate}

\paragraph{Empirical observations explained.} In addition to justifying the procedure, the analysis explains three observations that have been known empirically from applications of TD-based UFE: why the standard deviation of all components of the gene singular vector cannot be used as $\sigma$ (Corollary~\ref{cor:local}); why the objective of the optimization decreases again at small $\sigma$ (Lemma~\ref{lem:unimodal}); and why the valley to the right of the peak of the objective is the correct choice (Theorem~\ref{thm:drop}).

\section{Methods}\label{sec:method}
This section describes TD-based UFE and the standard-deviation optimization, the statistical model analysed in the Results, and how the numerical results were obtained.
\subsection{HOSVD and notation}
Let $x_{ijq}\in\R$, $i=1,\dots,N$, $j=1,\dots,J$, $q=1,\dots,Q$, be a data tensor in which $i$ indexes genes and $j$ and $q$ index, for example, samples and conditions. Tensors of higher order are treated in the same way. The HOSVD of $x_{ijq}$ is
\begin{equation}\label{eq:hosvd}
x_{ijq}=\sum_{\ell_1=1}^{N}\sum_{\ell_2=1}^{J}\sum_{\ell_3=1}^{Q}G(\ell_1\ell_2\ell_3)\,u_{\ell_1 i}\,u_{\ell_2 j}\,u_{\ell_3 q},
\end{equation}
where $G\in\R^{N\times J\times Q}$ is the core tensor and $u_{\ell_1 i}\in\R^{N\times N}$, $u_{\ell_2 j}\in\R^{J\times J}$ and $u_{\ell_3 q}\in\R^{Q\times Q}$ are orthogonal matrices:
\begin{equation}
\sum_{i=1}^{N}u_{\ell_1 i}u_{\ell_1' i}=\delta_{\ell_1\ell_1'},\quad
\sum_{j=1}^{J}u_{\ell_2 j}u_{\ell_2' j}=\delta_{\ell_2\ell_2'},\quad
\sum_{q=1}^{Q}u_{\ell_3 q}u_{\ell_3' q}=\delta_{\ell_3\ell_3'} .
\end{equation}
In particular each gene singular vector $(u_{\ell_1 1},\dots,u_{\ell_1 N})$ has unit length; this normalization will matter later.

The tensor has three \emph{modes}: mode~1, indexed by $i$ (genes); mode~2, indexed by $j$ (e.g.\ samples); and mode~3, indexed by $q$ (e.g.\ conditions). Modes~2 and 3 are called the \emph{sample-side modes}. For each mode the tensor can be rearranged into a matrix, its \emph{mode-$n$ unfolding}, whose rows are indexed by that mode and whose columns by the pairs of the other two indices: with $M=JQ$,
\begin{equation}\label{eq:unfold}
\begin{aligned}
X^{(1)}&\in\R^{N\times JQ}, & X^{(1)}_{i,(j,q)}&=x_{ijq},\\
X^{(2)}&\in\R^{J\times NQ}, & X^{(2)}_{j,(i,q)}&=x_{ijq},\\
X^{(3)}&\in\R^{Q\times NJ}, & X^{(3)}_{q,(i,j)}&=x_{ijq}.
\end{aligned}
\end{equation}
The analysis is easiest in terms of the mode-1 unfolding, which we also call the \emph{gene-mode unfolding} and denote by $X=X^{(1)}\in\R^{N\times M}$; its entries are $X_{ir}=x_{ijq}$, where $r=1,\dots,M$ enumerates the pairs $(j,q)$, and each row collects all measurements of one gene. The HOSVD is computed as follows \cite{DeLathauwer2000}. The singular vectors $u_{\ell_1 i}$, $u_{\ell_2 j}$ and $u_{\ell_3 q}$ are the left singular vectors of $X^{(1)}$, $X^{(2)}$ and $X^{(3)}$, respectively. The core tensor is then
\begin{equation}
G(\ell_1\ell_2\ell_3)=\sum_{i=1}^{N}\sum_{j=1}^{J}\sum_{q=1}^{Q}x_{ijq}\,u_{\ell_1 i}\,u_{\ell_2 j}\,u_{\ell_3 q}.
\end{equation}
In particular, \emph{the gene singular vectors of the HOSVD are exactly the left singular vectors of $X$}, so everything about $u_{\ell_1 i}$ can be studied through the ordinary singular value decomposition of $X$. Substituting \eqref{eq:hosvd} into $X_{ir}=x_{ijq}$ gives this decomposition explicitly, $X=\sum_{\ell_1}s_{\ell_1}u_{\ell_1\cdot}v_{\ell_1}^{\top}$, with singular values
\begin{equation}\label{eq:sv}
s_{\ell_1}=\left(\sum_{\ell_2=1}^{J}\sum_{\ell_3=1}^{Q}G(\ell_1\ell_2\ell_3)^{2}\right)^{1/2}
\end{equation}
and right singular vectors
\begin{equation}\label{eq:vhat}
(v_{\ell_1})_{(j,q)}=\frac{1}{s_{\ell_1}}\sum_{\ell_2=1}^{J}\sum_{\ell_3=1}^{Q}G(\ell_1\ell_2\ell_3)\,u_{\ell_2 j}\,u_{\ell_3 q}.
\end{equation}
These vectors are orthonormal because the slices $G(\ell_1\cdot\cdot)$ for different $\ell_1$ are mutually orthogonal (all-orthogonality of the core tensor \cite{DeLathauwer2000}) and the sample-side singular vectors are orthonormal. The right singular vector paired with $u_{\ell_1\cdot}$ is thus the combination of the products $u_{\ell_2 j}u_{\ell_3 q}$ weighted by the core tensor.
In TD-based UFE one first selects indices $\hat\ell_2$ and $\hat\ell_3$ such that the sample-side singular vectors $u_{\hat\ell_2 j}$ and $u_{\hat\ell_3 q}$ reflect the design, and then selects the index $\hat\ell_1$ maximizing $|G(\ell_1\hat\ell_2\hat\ell_3)|$ over $\ell_1$. By \eqref{eq:vhat}, $G(\ell_1\ell_2\ell_3)/s_{\ell_1}$ is the weight of the product $u_{\ell_2 j}u_{\ell_3 q}$ in $v_{\ell_1}$, so, for given $\ell_2$ and $\ell_3$, this selects the component for which $s_{\ell_1}|\langle v_{\ell_1},u_{\ell_2\cdot}\otimes u_{\ell_3\cdot}\rangle|$ is largest. When the chosen sample-side vectors reproduce the design, this is the component whose right singular vector reproduces the design (Proposition~\ref{prop:select}). We call $\hat\ell_1$, $\hat\ell_2$, $\hat\ell_3$ the indices selected in this way.

\paragraph{Indices.} Three kinds of indices must be distinguished. Indices without decoration, $\ell_1$, $\ell_2$, $\ell_3$, are running indices, as in \eqref{eq:hosvd}, \eqref{eq:sv} and \eqref{eq:vhat}. The component that actually carries the signal is denoted by $\ell_1^{*}$; its existence is shown in Theorem~\ref{thm:main}. The sample-side components that reproduce the patterns of the design in modes~2 and 3 are denoted by $\ell_2^{*}$ and $\ell_3^{*}$. The indices of the three factors of the mean term $a^{0}\otimes e_J\otimes e_Q$ in modes~1, 2 and 3 are denoted by $\ell_1^{\mathrm{mean}}$, $\ell_2^{\mathrm{mean}}$ and $\ell_3^{\mathrm{mean}}$; for uncentred data they are usually all $1$. As with $\ell_1^{*},\ell_2^{*},\ell_3^{*}$, the common label indicates factors of the same term, not equal values. When the signal and the mean term share a factor, for example $b^{(3)}=e_Q$ for a mode of replicates, the corresponding indices coincide, $\ell_3^{\mathrm{mean}}=\ell_3^{*}$. The indices selected by TD-based UFE are $\hat\ell_1$, $\hat\ell_2$, $\hat\ell_3$; Proposition~\ref{prop:select} shows that $\hat\ell_1=\ell_1^{*}$ with probability tending to one when $\hat\ell_2=\ell_2^{*}$ and $\hat\ell_3=\ell_3^{*}$. Throughout the paper, $s_{\ell_1^{*}}$ is the singular value of the signal component, $v_{\ell_1^{*}}$ the paired right singular vector, and $u_{\ell_1^{*}i}$ is called the \emph{score} of gene $i$.

\subsection{\texorpdfstring{$P$}{P}-values and selection}
For a candidate standard deviation $\sigma>0$, the $P$-value of gene $i$ is
\begin{equation}\label{eq:pval}
P_i(\sigma)=\Prob\left[\chi^{2}_{1}>\frac{u_{\ell_1^{*} i}^{2}}{\sigma^{2}}\right]=2\Phi\left(-\frac{|u_{\ell_1^{*} i}|}{\sigma}\right),
\end{equation}
where $\chi^2_1$ is a chi-squared variable with one degree of freedom, $\Phi$ is the standard normal distribution function and $\phi$ its density. This is the two-sided $P$-value of the score under the null hypothesis $u_{\ell_1^{*} i}\sim N(0,\sigma^2)$. The $P$-values are corrected by the Benjamini--Hochberg (BH) procedure \cite{BH1995}, and genes whose corrected $P$-values are below $\alpha$ are selected. We write $\mathcal{S}(\sigma)$ for the set of selected genes; in practice $\alpha=0.01$.

\begin{remark}\label{rem:bhmono}
The selected set changes with $\sigma$ in a simple way. Since $P_i(\sigma)$ increases with $\sigma$ and the ordering of $|u_{\ell_1^{*} i}|$ does not depend on $\sigma$, the ranks of the $P$-values do not change with $\sigma$. Order the genes so that $|u_{\ell_1^{*}(1)}|\ge|u_{\ell_1^{*}(2)}|\ge\cdots$. Gene $(k)$ satisfies the BH condition $P_{(k)}(\sigma)\le\alpha k/N$ if and only if
\begin{equation}
\sigma\le\sigma_k\equiv\frac{|u_{\ell_1^{*}(k)}|}{\Phi^{-1}\left(1-\alpha k/(2N)\right)} .
\end{equation}
Hence the number of selected genes is a non-increasing step function of $\sigma$ that changes only at the finitely many values $\sigma_k$. Increasing $\sigma$ can only remove genes from $\mathcal{S}(\sigma)$, never add new ones.
\end{remark}

\subsection{The standard-deviation optimization}
Divide $[0,1]$ into $K$ bins of equal width $1/K$. For each $\sigma$, let $h_k(\sigma)$, $k=1,\dots,K$, be the number of genes \emph{not} selected ($i\notin\mathcal{S}(\sigma)$) whose $P$-value $P_i(\sigma)$ falls into the $k$th bin; bin $k=1$ contains the smallest $P$-values. (The original description bins $1-P$ instead of $P$; this only reverses the order of the bins and changes nothing below.) The objective function is the standard deviation of these counts over the bins,
\begin{equation}\label{eq:obj}
L_N(\sigma)=\frac{1}{N}\,\SD_k\left[h_k(\sigma)\right],
\end{equation}
divided by $N$ so that it remains of order one as $N$ grows. Here $\SD_k$ is the standard deviation over the $K$ bins with divisor $K$, $\SD_k(x_k)=\left(\frac1K\sum_{k=1}^{K}(x_k-\bar x)^{2}\right)^{1/2}$ with $\bar x=\frac1K\sum_kx_k$; the sample standard deviation with divisor $K-1$ differs only by the constant factor $\sqrt{K/(K-1)}$, which changes neither the shape of $L_N$ nor its minimizer. A flat histogram has equal counts in all bins and $L_N=0$; the more uneven the histogram, the larger $L_N$. For fixed total count, minimizing $L_N$ is equivalent to minimizing Pearson's goodness-of-fit statistic against the uniform distribution. In practice $K=100$. The estimate $\hat\sigma$ is taken at a minimum of $L_N$; the precise rule is given in Section~\ref{sec:drop}.

\subsection{Statistical model}
The genes are divided into \emph{background} genes, which carry no signal and should not be selected, and \emph{signal} genes. We write $\mathcal{I}_0$ for the set of background genes, $\mathcal{I}_1=\{1,\dots,N\}\setminus\mathcal{I}_0$ for the set of signal genes, $n_0=|\mathcal{I}_0|$, and
\begin{equation}
\pi_0=\frac{n_0}{N}
\end{equation}
for the \emph{fraction of background genes}; $1-\pi_0$ is the fraction of signal genes. (In the asymptotic statements of Section~\ref{sec:seq}, $n_0/N\to\pi_0$ along a sequence of data sets.) The goal of the optimization is to recover the standard deviation $\sigma_0$ of the scores $u_{\ell_1^{*} i}$ of background genes. For a given data set $\sigma_0$ is a fixed number. Its size, however, depends on the data set: because the singular vector has unit length, its $N$ components are of order $N^{-1/2}$, and so is $\sigma_0$. For this reason we express all results in terms of the ratio
\begin{equation}
\tau=\frac{\sigma}{\sigma_0}
\end{equation}
of a candidate $\sigma$ to the true value, which does not depend on this normalization. The correct choice is $\tau=1$.

Table~\ref{tab:notation} summarizes the notation used in the paper. A subscript $0$ always refers to background genes ($\mathcal{I}_0$, $n_0$, $\pi_0$, $\sigma_0$). The mean expression is written $a^{0}$ with a superscript, and quantities derived from it are labelled `mean', as in $\bar\mu_{\mathrm{mean}}$ of Section~\ref{sec:eps}. The noise variance per entry of $X$ is $d$.

\paragraph{Convention on limits.} Every real data set has fixed $N$ and $M$, and for a single data set the statements of this paper are approximations whose accuracy is controlled by the quantities of Section~\ref{sec:eps}. Asymptotic statements are made along a sequence of data sets indexed by $\nu=1,2,\dots$ with $N_\nu\to\infty$, defined precisely in Section~\ref{sec:seq}. Throughout the paper, ``$\to$'', ``converges in probability'', ``with probability tending to one'' and $O_P(\cdot)$ refer to $\nu\to\infty$ along such a sequence, unless another limit is stated explicitly (as in Proposition~\ref{prop:sphere}, where $n_0\to\infty$). Quantities that depend on the data set, such as indices of selected components, may depend on $\nu$.

\begin{table}[!htbp]
\centering
\caption{Notation}\label{tab:notation}
\small
\begin{tabular}{p{0.2\textwidth}p{0.72\textwidth}}
\toprule
Symbol & Meaning\\
\midrule
$N$, $J$, $Q$ & numbers of genes, and of values of the other two tensor indices\\
$M=JQ$ & number of columns of the gene-mode unfolding $X\in\R^{N\times M}$\\
$\|\cdot\|$ & length of a vector; for a matrix, its largest singular value (operator norm)\\
$X^{(1)},X^{(2)},X^{(3)}$ & mode-1 ($=X$), mode-2 and mode-3 unfoldings of the tensor, Eq.~\eqref{eq:unfold}\\
$\mathbf{1}_J,\mathbf{1}_Q,\mathbf{1}_M$; $e_J,e_Q,e_M$ & vectors of ones; constant vectors of unit length in mode~2, mode~3 and over the columns of $X$, $e_M=e_J\otimes e_Q$, Eq.~\eqref{eq:evec}\\
$\ell_1,\ell_2,\ell_3$; $\ell_1^{*},\ell_2^{*},\ell_3^{*}$; $\hat\ell_1,\hat\ell_2,\hat\ell_3$ & running indices; components carrying the signal; indices selected by TD-based UFE (Section~\ref{sec:method})\\
$\ell_1^{\mathrm{mean}}$, $\ell_2^{\mathrm{mean}}$, $\ell_3^{\mathrm{mean}}$ & indices of the factors of the mean term $a^{0}\otimes e_J\otimes e_Q$ in modes~1, 2, 3 (usually all $1$); $s_{\ell_1^{\mathrm{mean}}}$ is the singular value of the mean component in the data $X$ (the noiseless value is $\bar s_{\ell_1^{\mathrm{mean}}}$)\\
$u_{\ell_1^{*} i}$, $v_{\ell_1^{*}}$, $s_{\ell_1^{*}}$ & signal component $\ell_1^{*}$ of $X$: left singular vector (score of gene $i$), paired right singular vector, and singular value; size of $s_{\ell_1^{*}}$ in Eq.~\eqref{eq:s2}\\
$\bar v_{\ell_1^{*}}$, $\bar s_{\ell_1^{*}}$, $\bar s_{\ell_1^{\mathrm{mean}}}$, $\theta$ & noiseless part $a^{0}e_M^{\top}+ab^{\top}$ (Eq.~\eqref{eq:vbar}): $\bar v_{\ell_1^{*}}$, right singular vector of the signal component; $\bar s_{\ell_1^{*}}$, its singular value; $\bar s_{\ell_1^{\mathrm{mean}}}$, singular value of the mean component; $\theta$, mixing angle. A bar denotes the noiseless value\\
$P_i(\sigma)$, $\alpha$, $\mathcal{S}(\sigma)$ & $P$-value of gene $i$, BH level, set of selected genes\\
$K$, $h_k(\sigma)$, $L_N(\sigma)$ & number of bins, count of non-selected genes in bin $k$, objective function\\
$\mathcal{I}_0$, $\mathcal{I}_1$, $n_0$, $\pi_0$ & background genes, signal genes, number and fraction of background genes\\
$\sigma_0$, $\tau=\sigma/\sigma_0$ & true standard deviation of background scores (components of the unit vector $u_{\ell_1^{*}\cdot}$), scaled candidate; $\sigma_0\approx\sqrt{d}/s_{\ell_1^{*}}$\\
$a^{0}$, $a$, $b$, $Z$, $d$ & mean expression (superscript $0$: mean, not background), signal, contrast vector, noise matrix, mean noise variance per entry of $X$ over background genes (Section~\ref{sec:growM}); $d$ is not the same as $\sigma_0$\\
$d_i$, $\eta_i$ & noise variance of gene $i$, $d_i=d(1+\eta_i)$\\
$\mu_i$ & standardized signal strength: length of the signal in row $i$ of $X$ in units of the noise standard deviation, Eq.~\eqref{eq:mu}; $\mu_i\approx u_{\ell_1^{*} i}/\sigma_0$\\
$\mu_i^{\mathrm{eff}}$ & effective signal: the part of the standardized score of gene $i$ retained from the signal and the mean term, Eq.~\eqref{eq:mueff}\\
$\delta_i$ & effect size per entry: root mean square over the $M$ entries of the signal per entry, in noise units, $\delta_i=\mu_i/\sqrt{M}$ (Remark~\ref{rem:delta})\\
$\bar\mu$, $\bar\delta$ & root mean square of $\mu_i$ over signal genes, $\bar\delta=\bar\mu/\sqrt{M}$, Eq.~\eqref{eq:mubar}\\
$b^{(2)}$, $b^{(3)}$, $\rho$ & patterns in mode~2 and mode~3 with $b=b^{(2)}\otimes b^{(3)}$; correlation of $a$ and $a^{0}$ over genes, Eq.~\eqref{eq:rho}\\
$\varepsilon_{\mathrm{sig}}$, $\varepsilon_{\mathrm{loo}}$, $\varepsilon_{\mathrm{spk}}$, $\varepsilon_{\mathrm{con}}$, $\varepsilon_{\mathrm{mix}}$, $\varepsilon_{\mathrm{ng}}$, $\varepsilon_{\mathrm{side}}$, $\varepsilon_{\mathrm{gap}}$, $\varepsilon_{\mathrm{het}}$ & quantities controlling the approximations (Section~\ref{sec:eps}, Table~\ref{tab:epsdep})\\
$\bar\mu_{\mathrm{mean}}$, $R_i$, $R$, $\kappa_3$, $\beta_{\max}$ & standardized size of the mean expression, relative mean expression of gene $i$ and its maximum (dynamic range), third moment of the noise, spread of the design pattern (Section~\ref{sec:eps})\\
$\nu$ & index of a sequence of data sets used for asymptotic statements\\
$z(t)$, $F_\tau$, $f_\tau$ & $z(t)=\Phi^{-1}(1-t/2)$; distribution function and density of null $P$-values\\
$t^{*}$, $E$ & limit of the BH threshold, fraction of background genes excluded by BH\\
$\beta_k$, $m_k$, $L(\tau)$ & bin probability of null $P$-values, limiting bin mass, limiting objective\\
$\tau_{\max}$, $\tau^{*}$ & local maximum and minimizer of $L$\\
$D(\tau)$ & a quantity with the sign of $L'(\tau)$, Eq.~\eqref{eq:D}\\
\bottomrule
\end{tabular}
\end{table}

\subsection{Numerical computations}\label{sec:numerics}
The computer-assisted part of the proof of Theorem~\ref{thm:cap} uses rigorous interval arithmetic (python-flint \cite{Johansson2017}); all other numerical results (the limiting objective, Tables~\ref{tab:bias}, \ref{tab:finite} and \ref{tab:leak}, the simulations of Tables~\ref{tab:sim} and \ref{tab:finiteN}, the example of Remark~\ref{rem:cancel} and the check of Proposition~\ref{prop:lambda}) use ordinary floating-point arithmetic in Python with NumPy and SciPy. Every number reported in this paper can be reproduced with the scripts in Additional file~2, whose README lists which script produces which result.

\subsection{Use of large language models}
Large language models (Claude, Anthropic; ChatGPT, OpenAI) were used to assist in drafting and revising the text, in deriving and checking proofs, and in writing the verification code. All derivations, proofs and code were checked by the authors, who take full responsibility for the content.

\section{Results}\label{sec:results}
\subsection{Preliminaries}\label{sec:prelim}
This section collects three basic facts. Flattening recovers the correct $\sigma$ when every gene is background (Lemma~\ref{lem:pit}); it can fail when signal and background overlap (Proposition~\ref{prop:nonid}); and, for Gaussian noise, the normal law of the background scores follows from a symmetry of the noise (Proposition~\ref{prop:sphere}).

\subsubsection{Uniformity of null \texorpdfstring{$P$}{P}-values and identifiability}
Let $z(t)=\Phi^{-1}(1-t/2)$ for $t\in(0,1]$ and $z(0)=\infty$; $z(t)$ is the absolute $z$-value whose two-sided $P$-value is $t$, and it decreases from $\infty$ to $0$ as $t$ increases from $0$ to $1$.
\begin{lemma}\label{lem:pit}
Suppose $u_{\ell_1^{*} i}\sim N(0,\sigma_0^{2})$ and let $\tau=\sigma/\sigma_0$. The distribution function of $P_i(\sigma)$ is
\begin{equation}\label{eq:Ftau}
F_\tau(t)=2\Phi\left(-\tau z(t)\right),\qquad 0\le t\le1,
\end{equation}
and its density is
\begin{equation}\label{eq:ftau}
f_\tau(t)=\tau\exp\left(-\frac{(\tau^{2}-1)z(t)^{2}}{2}\right).
\end{equation}
For every $t\in(0,1)$, $F_\tau(t)>t$ if $\tau<1$, $F_\tau(t)=t$ if $\tau=1$, and $F_\tau(t)<t$ if $\tau>1$. In particular the $P$-values are uniformly distributed if and only if $\sigma=\sigma_0$.
\end{lemma}
\begin{proof}
$P_i\le t$ if and only if $|u_{\ell_1^{*} i}|/\sigma\ge z(t)$, i.e.\ $|u_{\ell_1^{*} i}|/\sigma_0\ge\tau z(t)$, which has probability $2\Phi(-\tau z(t))$. Since $2\Phi(-z(t))=t$ and $z(t)>0$ for $t<1$, the inequalities follow from the monotonicity of $\Phi$. Differentiating with $\mathrm{d}z/\mathrm{d}t=-1/(2\phi(z))$ gives $f_\tau=\tau\phi(\tau z)/\phi(z)$, which is \eqref{eq:ftau}.
\end{proof}
The lemma has a simple interpretation. If $\sigma$ is too small ($\tau<1$), every score looks too large relative to $\sigma$, the $P$-values are too small, and the histogram has an excess near $P=0$. If $\sigma$ is too large ($\tau>1$), the histogram has an excess near $P=1$. Only the correct $\sigma$ gives a flat histogram. This justifies flattening when all genes are background.

When signal genes are present, the scores follow a mixture $\pi_0\,N(0,\sigma_0^2)+(1-\pi_0)f_1$, where $f_1$ is the distribution of the scores of signal genes. Without restrictions on $f_1$, flattening cannot work, as the following example shows.

\begin{proposition}[non-identifiability]\label{prop:nonid}
Let $\varphi_v$ be the density of $N(0,v)$, $\pi_0=1/2$ and $f_1(x)=2\varphi_2(x)-\varphi_1(x)$. Then $f_1$ is a probability density and $\tfrac12\varphi_1+\tfrac12 f_1=\varphi_2$.
\end{proposition}
\begin{proof}
$2\varphi_2(x)/\varphi_1(x)=\sqrt{2}\,e^{x^{2}/4}\ge1$, so $f_1\ge0$, and $\int f_1=1$. The identity follows directly.
\end{proof}
In this example the background variance is $1$, but the scores of all genes together are exactly $N(0,2)$. Nothing in the data distinguishes this situation from one with no signal and background variance $2$, so any procedure based on the distribution of the scores, flattening included, returns $\sqrt{2}$. The problem is that the signal scores overlap the background. Some separation between signal and background is therefore indispensable; Section~\ref{sec:sep} shows that strong standardized signals provide it.

\subsubsection{Normality of the background from rotational invariance}
Why should background scores be normal at all? The following proposition shows that normality results from a symmetry of the noise.

\begin{proposition}\label{prop:sphere}
Let $X=A+Z$, where $A\in\R^{N\times M}$ is the signal with $A_{i\cdot}=0$ for $i\in\mathcal{I}_0$ and $Z\in\R^{N\times M}$ is noise. Suppose that the background rows $Z_{\mathcal{I}_0\cdot}$ have a distribution invariant under rotations of the background genes, i.e.\ $QZ_{\mathcal{I}_0\cdot}\overset{d}{=}Z_{\mathcal{I}_0\cdot}$ for every orthogonal $Q\in\R^{n_0\times n_0}$; independent Gaussian entries are an example. Let $u^{(0)}_{\ell_1}\in\R^{n_0}$ be the background part of a left singular vector $u_{\ell_1\cdot}$ with a simple singular value, and let $\varrho=\|u^{(0)}_{\ell_1}\|>0$ be its length. Then $u^{(0)}_{\ell_1}/\varrho$ is uniformly distributed on the unit sphere $S^{n_0-1}$; for each $i\in\mathcal{I}_0$,
\begin{equation}
\frac{u_{\ell_1 i}^{2}}{\varrho^{2}}\sim\mathrm{Beta}\left(\frac12,\frac{n_0-1}{2}\right);
\end{equation}
and $\sqrt{n_0}\,u_{\ell_1 i}/\varrho\to N(0,1)$ in distribution as $n_0\to\infty$. The empirical distribution of these components also converges to $N(0,1)$.
\end{proposition}
\begin{proof}
Rotating the background rows of $Z$ rotates $u^{(0)}_{\ell_1}$ in the same way and leaves the distribution of $X$ unchanged. Up to the choice of sign, the distribution of $u^{(0)}_{\ell_1}/\varrho$ is therefore invariant under all rotations, and the only such distribution on the sphere is the uniform one. A uniform vector on the sphere is distributed as $g/\|g\|$ with $g\sim N(0,I_{n_0})$, which gives the Beta law; since $\|g\|^2/n_0\to1$ by the law of large numbers, each component multiplied by $\sqrt{n_0}$ tends to $N(0,1)$ (the Poincar\'e--Borel lemma; see \cite{DiaconisFreedman1987}).
\end{proof}
The normal distribution thus arises because the background part of the singular vector points in a random direction in a space of high dimension $n_0$. This argument, however, covers essentially only Gaussian noise: if the entries of $Z$ are independent, invariance of their joint distribution under all rotations holds only when they are Gaussian (Maxwell's theorem). For non-Gaussian noise the normal law has a different origin, the summation of the noise over the $M$ columns of $X$; this is shown in Theorem~\ref{thm:main}(ii).

\begin{remark}\label{rem:hetero}
Without rotational invariance, normality may fail. If half of the background genes have noise variance $1$ and the other half $4$, the background scores behave as the mixture $\tfrac12N(0,v)+\tfrac12N(0,4v)$ for some $v$, which is not normal, and no single $\sigma$ flattens the histogram. Theorem~\ref{thm:main}(ii) therefore requires the noise variances of background genes to be nearly equal, $\varepsilon_{\mathrm{het}}\to0$ (Section~\ref{sec:eps}, Assumption~\ref{as:het}); unequal variances are allowed as long as their relative spread is small.
\end{remark}

\begin{remark}\label{rem:weak}
Normality of the background alone is not enough. If the standardized signal strengths stay bounded, the scores of signal genes are of the same order as those of background genes. The two populations then overlap, and the situation of Proposition~\ref{prop:nonid} arises: the histogram cannot tell which part of the overlap is background. What is needed in addition is a large standardized signal, made precise in the next section.
\end{remark}

\subsection{Normality and separation for a given data set}\label{sec:growM}
In any real data set the numbers $N$ and $M$ are fixed. The results of this section are therefore approximations, and the relevant question is how accurate they are for the data at hand. We answer it by identifying dimensionless quantities that control the errors (Section~\ref{sec:eps}) and that can be computed from an existing analysis (Section~\ref{sec:diag}). Statements involving limits are made along a sequence of data sets (Section~\ref{sec:seq}); such limits are a device to express that these quantities are small, not an assumption about how a real data set is generated.

\subsubsection{Model}
We now examine in detail the model \eqref{eq:model} of Section~\ref{sec:outline},
\[
X=\frac{1}{\sqrt{M}}\,a^{0}\mathbf{1}_M^{\top}+a\,b^{\top}+Z .
\]
The three terms have the following meaning.
\begin{itemize}
\item $a^{0}\in\R^{N}$ is the mean expression of each gene, and $\mathbf{1}_M\in\R^{M}$ is the vector of ones; the first term gives gene $i$ the value $a^{0}_i/\sqrt{M}$ in every column. It is present because genes are not centred in TD-based UFE.
\item $b\in\R^{M}$ is a contrast vector of unit length, orthogonal to $\mathbf{1}_M$, that describes the pattern of the experimental design. It is the unfolding of a $J\times Q$ array of values over samples and conditions, $b_r=b_{(j,q)}$, for instance $b_{(j,q)}=b^{(2)}_jb^{(3)}_q$ (Section~\ref{sec:outline}); for a contrast that divides the $M$ columns $(j,q)$ into two sets of equal size, $b_r=\pm1/\sqrt{M}$. $a\in\R^{N}$ is the signal, with $a_i=0$ for background genes $i\in\mathcal{I}_0$. The second term adds $a_ib_r$ to entry $(i,r)$.
\item $Z\in\R^{N\times M}$ is noise of the form $Z_{ir}=\sqrt{d_i}\,W_{ir}$, where the $W_{ir}$ are independent and identically distributed with mean zero and variance $1$, and $d_i$ is the noise variance of gene $i$. We write $d$ for the mean of $d_i$ over background genes and $d_i=d(1+\eta_i)$, so that $\sum_{i\in\mathcal{I}_0}\eta_i=0$. The variances may differ between genes; how much they may differ among background genes is measured by the quantity $\varepsilon_{\mathrm{het}}$ of Section~\ref{sec:eps}.
\end{itemize}
We write $Z_i\in\R^{1\times M}$ for the $i$th row of $Z$; $v_{\ell_1^{*}}\in\R^{M}$ is the right singular vector of $X$ paired with $u_{\ell_1^{*} i}$, Eq.~\eqref{eq:vhat}. The case of several signals is discussed in Section~\ref{sec:outline}.

We also use the correlation over genes between the signal and the mean expression,
\begin{equation}\label{eq:rho}
\rho=\frac{|\langle a,a^{0}\rangle|}{\|a\|\,\|a^{0}\|}
\end{equation}
The strength of the signal of gene $i$ is measured by
\begin{equation}\label{eq:mu}
\mu_i=\frac{a_i\|b\|}{\sqrt{d}}=\frac{a_i}{\sqrt{d}},\qquad |\mu_i|=\frac{\left(\sum_{r=1}^{M}(a_ib_r)^{2}\right)^{1/2}}{\sqrt{d}} .
\end{equation}
Its absolute value is the length of the signal in row $i$ of $X$, in units of the noise standard deviation. It does not depend on how the signal is split into the factors $a$ and $b$: replacing $b$ by $cb$ and $a$ by $a/c$ leaves $\mu_i$ unchanged, and the convention $\|b\|=1$ only allows us to write it as $a_i/\sqrt{d}$. We call $\mu_i$ the \emph{standardized signal strength} of gene $i$. In terms of the singular vector, $\mu_i\approx u_{\ell_1^{*} i}/\sigma_0$ when the mean term is negligible (more precisely, $u_{\ell_1^{*} i}/\sigma_0$ approximates the effective signal $\mu_i^{\mathrm{eff}}$ of Section~\ref{sec:eps}, which equals $\mu_i$ up to small corrections when $\rho\ll1$): the score of a signal gene is then $\mu_i$ times the standard deviation of the background scores. $\mu_i$ is thus not the component of the unit vector $u_{\ell_1^{*}\cdot}$ itself but its ratio to the background fluctuation. For a given data set $\mu_i$ is a fixed number, which can be estimated by $u_{\ell_1^{*} i}/\hat\sigma$, and all conditions below are stated in terms of $\mu_i$. As shown below, what separates a signal gene from the background is $\mu_i^{\mathrm{eff}}$, which equals $\mu_i$ in the usual case $\rho\ll1$.

\paragraph{Size of the singular value.} The singular value $s_{\ell_1^{*}}$, which appears throughout, can be expressed through the quantities of the model. Let
\begin{equation}\label{eq:mubar}
\bar\mu=\left(\frac{1}{|\mathcal{I}_1|}\sum_{i\in\mathcal{I}_1}\mu_i^{2}\right)^{1/2},\qquad \bar\delta=\frac{\bar\mu}{\sqrt{M}},
\end{equation}
be the root mean square of the standardized signal strengths of the signal genes and the corresponding effect size per entry. Since $b$ has unit length, the singular value of the signal term $ab^{\top}$ alone is $\|a\|$, with
\begin{equation}
\|a\|^{2}=d\sum_{i\in\mathcal{I}_1}\mu_i^{2}=d(1-\pi_0)N\bar\mu^{2}.
\end{equation}
The noise adds to this. When the signal is much stronger than the noise, the squared singular value of a rank-one signal plus independent noise is approximately $\|a\|^{2}+d(N+M)$ \cite{BenaychNadakuditi2012}. Hence
\begin{equation}\label{eq:s2}
s_{\ell_1^{*}}^{2}\approx dN\left[(1-\pi_0)\bar\mu^{2}+1+\frac{M}{N}\right].
\end{equation}
This expression assumes $\rho\ll1$; in general $\bar\mu$ is to be replaced by the root mean square of $\mu_i^{\mathrm{eff}}$ over signal genes (Section~\ref{sec:eps}), i.e.\ by the part of the signal that survives orthogonalization to the mean component. Thus $s_{\ell_1^{*}}$ is of the order of $\sqrt{dN}$ multiplied by $\sqrt{1+(1-\pi_0)\bar\mu^{2}}$, a factor that measures the strength of the signal; its leading term is $s_{\ell_1^{*}}\approx\sqrt{d(1-\pi_0)N}\,\bar\mu$. Combined with $\sigma_0=\sqrt{d}/s_{\ell_1^{*}}$ (Eq.~\eqref{eq:sigma0} below), this gives
\begin{equation}\label{eq:Nsigma2}
N\sigma_0^{2}\approx\frac{1}{1+(1-\pi_0)\bar\mu^{2}+M/N}.
\end{equation}
Like \eqref{eq:s2}, this assumes $\rho\ll1$, so that the loss of signal energy by orthogonalization to the mean component, a factor $1-\rho^{2}$, is negligible. The same result follows from the unit length of the singular vector. A background score is about $g\sigma_0$ and a signal score about $(\mu_i+g)\sigma_0$ with $g\sim N(0,1)$ (Theorem~\ref{thm:main}), so the squared length is about $n_0\sigma_0^{2}+\sum_{i\in\mathcal{I}_1}(\mu_i^{2}+1)\sigma_0^{2}=N\sigma_0^{2}\left[1+(1-\pi_0)\bar\mu^{2}\right]$. Setting it to $1$ reproduces \eqref{eq:Nsigma2} apart from the small term $M/N$. Strong signals take up most of the unit length and leave little for the background, which is why $\sigma_0$ is smaller than $N^{-1/2}$. These relations also clarify two points. First, $d$ and $\sigma_0$ are different quantities: $d$ is the variance of the noise in each entry of $X$, measured in units of expression, whereas $\sigma_0$ is the standard deviation of the background components of the unit vector $u_{\ell_1^{*}\cdot}$, which has no units and is at most of order $N^{-1/2}$; $\sigma_0\approx\sqrt{d}/s_{\ell_1^{*}}$ converts $\sqrt{d}$ to the scale of the singular vector. Second, $\mu_i$ can grow while the shape of $u_{\ell_1^{*}\cdot}$ stays the same: for a fixed gene pattern $a$, the components of the signal genes in $u_{\ell_1^{*}\cdot}$ stay roughly the same, but the background fluctuation $\sigma_0$ shrinks as the signal grows, by \eqref{eq:Nsigma2}, and $\mu_i\approx u_{\ell_1^{*} i}/\sigma_0$ grows.

\begin{remark}[dependence of $\mu_i$ on the number of samples]\label{rem:delta}
How $\mu_i$ changes when samples are added is not determined by the model; it depends on the samples that are added. In any design, $d\mu_i^{2}=\sum_r(a_ib_r)^{2}$, so that $\delta_i=\mu_i/\sqrt{M}$ is the root mean square over the $M$ entries $(j,q)$ of the signal per entry, $\delta_i\sqrt d=\left((1/M)\sum_r(a_ib_r)^{2}\right)^{1/2}$, in units of the noise standard deviation. A simple example is $b_r=\pm1/\sqrt{M}$, in which the $M$ columns $(j,q)$ fall into two sets of equal size according to the sign of $b_r$, for instance the columns of treated and of control samples when each sample is measured under all $Q$ conditions. The signal is then $\pm\delta_i\sqrt d$ in every entry, and the two sets of columns differ by $2\delta_i$ noise standard deviations. If replicates of the same kind are added, the per-entry effect size $\delta_i=\mu_i/\sqrt{M}$ stays the same and $\mu_i$ grows like $\sqrt{M}$, just as a $t$-statistic grows with the square root of the number of samples. If the added samples or conditions carry no signal for gene $i$, the length of the signal in row $i$, and hence $\mu_i$, stays the same. Normalizing the data does not change $\mu_i$, because it rescales signal and noise by the same factor.

The standard deviation of the background scores behaves accordingly. By \eqref{eq:Nsigma2},
\begin{equation}\label{eq:sigma0M}
\sigma_0\approx\frac{1}{\sqrt{N\left[1+(1-\pi_0)\bar\mu^{2}+M/N\right]}},
\end{equation}
which depends on $M$ only through $M/N$ and through $\bar\mu$. If the added samples carry no signal, $\bar\mu$ is fixed and $\sigma_0$ hardly changes (it decreases slightly through $M/N$). If replicates of the same kind are added, $\bar\mu=\bar\delta\sqrt{M}$ and, for a strong signal, $\sigma_0\approx[(1-\pi_0)\bar\delta^{2}NM]^{-1/2}$ decreases like $M^{-1/2}$: the signal takes up more of the unit length of $u_{\ell_1^{*}\cdot}$ and leaves less for the background. This is why $\mu_i\approx u_{\ell_1^{*} i}/\sigma_0$ grows like $\sqrt{M}$ even though the shape of $u_{\ell_1^{*}\cdot}$ does not change. The leave-one-out correction (Remark~\ref{rem:loo}) multiplies $\sigma_0$ by $1+\varepsilon_{\mathrm{loo}}$, with $\varepsilon_{\mathrm{loo}}\le M/N$ (Section~\ref{sec:eps}).
\end{remark}

\subsubsection{Quantities that control the approximations}\label{sec:eps}

\paragraph{Noiseless singular vectors.} The noiseless part $a^{0}e_M^{\top}+ab^{\top}$ of \eqref{eq:model} has rank two. Let $\bar s_{\ell_1^{*}}$ be its singular value whose right singular vector is closer to $b$ (the signal component), $\bar s_{\ell_1^{\mathrm{mean}}}$ the other non-zero singular value (the mean component), and $\bar v_{\ell_1^{*}}$ the right singular vector for $\bar s_{\ell_1^{*}}$, with its sign fixed by $\langle b,\bar v_{\ell_1^{*}}\rangle>0$. Its right singular vectors lie in the plane spanned by $e_M$ and $b$, so that
\begin{equation}\label{eq:vbar}
\bar v_{\ell_1^{*}}=\cos\theta\,b-\sin\theta\,e_M,\qquad \tan2\theta=\frac{2\langle a,a^{0}\rangle}{\|a^{0}\|^{2}-\|a\|^{2}},
\end{equation}
where $\theta$ is the \emph{mixing angle} by which they are rotated away from $b$ and $e_M$ (proof of Theorem~\ref{thm:main}); $\theta=0$ if and only if $\langle a,a^{0}\rangle=0$.

\paragraph{Effective signal.} By \eqref{eq:model}, the score of gene $i$ decomposes exactly as
\begin{equation}\label{eq:scoredecomp}
s_{\ell_1^{*}}u_{\ell_1^{*} i}=(Xv_{\ell_1^{*}})_i=a_i\langle b,v_{\ell_1^{*}}\rangle+a^{0}_i\langle e_M,v_{\ell_1^{*}}\rangle+Z_iv_{\ell_1^{*}} .
\end{equation}
We call
\begin{equation}\label{eq:mueff}
\mu_i^{\mathrm{eff}}=\frac{a_i\langle b,v_{\ell_1^{*}}\rangle+a^{0}_i\langle e_M,v_{\ell_1^{*}}\rangle}{\sqrt{d}}
\end{equation}
the \emph{effective signal} of gene $i$: the part of its standardized score that the selected component actually retains from the signal and from the mean term. To leading order $v_{\ell_1^{*}}\approx\bar v_{\ell_1^{*}}=\cos\theta\,b-\sin\theta\,e_M$, so that $\mu_i^{\mathrm{eff}}\approx\mu_i\cos\theta-(a^{0}_i/\sqrt{d})\sin\theta$. It is $\mu_i^{\mathrm{eff}}$, not $\mu_i$, that separates a signal gene from the background. $\mu_i^{\mathrm{eff}}$ depends on the estimated $v_{\ell_1^{*}}$, so Assumption~\ref{as:weak} is a condition on a random quantity, not on the model parameters alone. Replacing $\mu_i^{\mathrm{eff}}$ by its deterministic counterpart $(a_i\cos\theta-a^{0}_i\sin\theta)/\sqrt d$ would require the difference, which is at most $(|a_i|+|a^{0}_i|)\,\|v_{\ell_1^{*}}-\bar v_{\ell_1^{*}}\|/\sqrt d$ with $\bar v_{\ell_1^{*}}$ as defined above, to be negligible; this need not hold for genes with large $a^{0}_i$, so we keep the assumption on $\mu_i^{\mathrm{eff}}$. The mean term in \eqref{eq:mueff} can shift background genes and can reduce or cancel signals; these effects, and why they are usually negligible, are discussed together in Section~\ref{sec:contam}. In the rest of the paper we write expressions in terms of $\mu_i$ where $\rho\ll1$ is assumed, and keep $\mu_i^{\mathrm{eff}}$ in the statements of the theorems.

We define nine dimensionless quantities. For $c>0$,
\begin{align}
\varepsilon_{\mathrm{sig}}(c)&=\frac{|\{i\in\mathcal{I}_1:|\mu_i^{\mathrm{eff}}|\le c\}|}{N}, &
\varepsilon_{\mathrm{loo}}&=\frac{Md}{s_{\ell_1^{*}}^{2}},\label{eq:eps1}\\
\varepsilon_{\mathrm{spk}}&=\frac{d(\sqrt N+\sqrt M)^{2}}{s_{\ell_1^{*}}^{2}}, &
\varepsilon_{\mathrm{con}}&=\max_{i\in\mathcal{I}_0}|\mu_i^{\mathrm{eff}}|=\frac{|\langle e_M,v_{\ell_1^{*}}\rangle|\,\max_{i\in\mathcal{I}_0}|a^{0}_i|}{\sqrt{d}} ,\label{eq:eps2}\\
\varepsilon_{\mathrm{mix}}&=\frac{|\langle a,a^{0}\rangle|}{\left|\|a^{0}\|^{2}-\|a\|^{2}\right|}, &
\varepsilon_{\mathrm{ng}}&=\kappa_3\max_{r}|\bar v_{\ell_1^{*}r}|,\quad \kappa_3=\mathbb{E}|W_{ir}|^{3} ,\label{eq:eps3}\\
\varepsilon_{\mathrm{side}}&=\frac{d\sqrt{NM}}{\|a\|^{2}} ,\label{eq:eps4}\\
\varepsilon_{\mathrm{gap}}&=\frac{d(\sqrt N+\sqrt M)^{2}}{(\bar s_{\ell_1^{\mathrm{mean}}}-\bar s_{\ell_1^{*}})^{2}} ,\label{eq:epsgap}\\
\varepsilon_{\mathrm{het}}&=\left(\frac{1}{n_0}\sum_{i\in\mathcal{I}_0}\eta_i^{2}\right)^{1/2} .\label{eq:epshet}
\end{align}
Their meanings are as follows.
\begin{itemize}
\item $\varepsilon_{\mathrm{sig}}(c)$ is the fraction of genes that are signal genes with weak effective signal, $|\mu_i^{\mathrm{eff}}|\le c$. Weak signal genes overlap the background.
\item $\varepsilon_{\mathrm{loo}}$ is the \emph{leave-one-out error}. The singular vector is computed from all rows of $X$, including the row of gene $i$ itself; the noise of gene $i$ therefore pulls the singular vector slightly towards gene $i$ and inflates its own score. $\varepsilon_{\mathrm{loo}}$ is the relative size of this inflation (Remark~\ref{rem:loo}).
\item $\varepsilon_{\mathrm{spk}}$ compares the largest singular value that pure noise would produce, about $\sqrt{d}(\sqrt N+\sqrt M)$ \cite{BenaychNadakuditi2012}, with the singular value $s_{\ell_1^{*}}$ of interest. When it is small, the singular vector is dominated by the signal (the ``spike'') and not by noise. By \eqref{eq:s2}, $\varepsilon_{\mathrm{spk}}\approx(1+\sqrt{M/N})^{2}/[1+(1-\pi_0)\bar\mu^{2}+M/N]$.
\item $\varepsilon_{\mathrm{con}}$ is the largest effective signal of a background gene, i.e.\ the largest shift of a background score caused by the mean expression, in units of the noise standard deviation. When $\|a^{0}\|\gg\|a\|$, $|\langle e_M,v_{\ell_1^{*}}\rangle|\approx|\langle a,a^{0}\rangle|/\|a^{0}\|^{2}$, in agreement with Approximation~\ref{prop:contam}.
\item $\varepsilon_{\mathrm{mix}}$ is the angle by which the mean expression rotates the right singular vector away from $b$ (Section~\ref{sec:contam}). Orthogonality of $b$ to $\mathbf{1}_M$ over the columns of $X$ is not enough to prevent this rotation; what matters is the overlap $\langle a,a^{0}\rangle$ of the two patterns over genes.
\item $\varepsilon_{\mathrm{ng}}$ measures how far a background score can be from normal when the noise is not Gaussian. A background score is a weighted sum $\sum_rZ_{ir}\bar v_{\ell_1^{*}r}$ of $M$ independent noise entries; by the Berry--Esseen inequality its distribution differs from the normal one by at most $C_0\kappa_3\sum_r|\bar v_{\ell_1^{*}r}|^{3}\le C_0\varepsilon_{\mathrm{ng}}$ with $C_0\le0.56$ \cite{Shevtsova2010}. When the pattern $b$ is spread over all columns, $\max_r|\bar v_{\ell_1^{*}r}|\approx M^{-1/2}$ and $\varepsilon_{\mathrm{ng}}\approx\kappa_3/\sqrt{M}$. Here $\bar v_{\ell_1^{*}}$ is the right singular vector of the noiseless part (Theorem~\ref{thm:main}), which $v_{\ell_1^{*}}$ approximates. For Gaussian noise the projections $Z_i\bar v_{\ell_1^{*}}$ are exactly normal and $\varepsilon_{\mathrm{ng}}$ plays no role.
\item $\varepsilon_{\mathrm{side}}$ compares the fluctuations of the noise in the Gram matrices of the sample-side unfoldings $X^{(2)}$ and $X^{(3)}$, which are of order $d\sqrt{NM}$, with the energy $\|a\|^{2}$ of the signal. When it is small, the sample-side singular vectors reproduce the patterns of the design (Proposition~\ref{prop:select}).
\item $\varepsilon_{\mathrm{gap}}$ compares the noise with the gap between the singular values of the signal component and of the mean component. When it is small, the noise cannot mix the two components (Theorem~\ref{thm:main}(i)).
\item $\varepsilon_{\mathrm{het}}$ is the relative spread of the noise variances of background genes, the root mean square of $(d_i-d)/d$. When it is not small, the background scores follow a mixture of normal laws with different variances rather than a single normal law (Remark~\ref{rem:hetero}).
\end{itemize}

The quantity $\varepsilon_{\mathrm{loo}}$ can also be written with $\sigma_0$. By \eqref{eq:sigma0} below, $\sigma_0\approx\sqrt{d}/s_{\ell_1^{*}}$, so $\varepsilon_{\mathrm{loo}}\approx M\sigma_0^{2}$. Since $\sigma_0$ itself shrinks as $N^{-1/2}$ because of the normalization of the singular vector (Section~\ref{sec:method}), it is clearer to separate this factor:
\begin{equation}\label{eq:loofactor}
\varepsilon_{\mathrm{loo}}\approx\frac{M}{N}\cdot N\sigma_0^{2}.
\end{equation}
The second factor $N\sigma_0^{2}$ does not shrink with $N$. By \eqref{eq:Nsigma2} it is about $1/[1+(1-\pi_0)\bar\mu^{2}]$, so
\begin{equation}\label{eq:loobound}
\varepsilon_{\mathrm{loo}}\approx\frac{M}{N}\cdot\frac{1}{1+(1-\pi_0)\bar\mu^{2}}\le\frac{M}{N},
\end{equation}
and the size of $\varepsilon_{\mathrm{loo}}$ is determined by the ratio $M/N$, further reduced by the strength of the signal. In particular, $\varepsilon_{\mathrm{loo}}$ decreases as $1/N$ if $M$ is fixed and stays constant if $M$ grows in proportion to $N$. Unlike the decrease of $\sigma_0$ itself, this decrease is not an artefact of the normalization: the factor $N\sigma_0^{2}$ carries no $N$-dependence of that kind, and what remains, $M/N$, measures how strongly a single gene influences the singular vector (Remark~\ref{rem:loo}). It is small whenever the number of genes greatly exceeds the number of columns of $X$, as is usual. For example, with $N=2\times10^{4}$, $M=100$, $\pi_0=0.7$ and $\bar\mu=8$, we have $(1-\pi_0)\bar\mu^{2}\approx19$ and $\varepsilon_{\mathrm{loo}}\approx2.5\times10^{-4}$; even without any signal it would be at most $M/N=0.005$.

\paragraph{Dependence on $N$, $M$ and the strength of the signal.} By \eqref{eq:s2}, the quantities above can be expressed through $N$, $M$, the standardized signal strength $\bar\mu$ and a few further quantities that characterize the data without referring to $N$ or $M$: the fraction $\pi_0$; the correlation $\rho$ between $a$ and $a^{0}$ over genes, Eq.~\eqref{eq:rho}; the standardized size $\bar\mu_{\mathrm{mean}}=\mathrm{rms}(a^{0})/\sqrt{d}$ of the mean expression, with $\mathrm{rms}(a^{0})=\|a^{0}\|/\sqrt{N}$; the relative mean expression $R_i=a^{0}_i/\mathrm{rms}(a^{0})$ of gene $i$ and its maximum, the dynamic range $R=\max_i|R_i|$; the moment $\kappa_3$ of the noise; and the spread $\beta_{\max}=\sqrt{M}\max_r|b_r|$ of the design pattern ($\beta_{\max}=1$ when $b_r=\pm1/\sqrt{M}$). Table~\ref{tab:epsdep} gives the resulting expressions. They show which conditions concern the strength of the signal, which the shape $M/N$ of the data matrix, and which the number $M$ of columns: $M$ appears explicitly only through $M/N$ and in $\varepsilon_{\mathrm{ng}}$, which concerns the central limit theorem over the $M$ columns. For a single data set, $\bar\mu$ and the other quantities are fixed numbers, and so are the entries of the table, which measure the errors of the approximations. The expressions assume a strong signal, $(1-\pi_0)\bar\mu^{2}\gg1$, and, for $\varepsilon_{\mathrm{mix}}$ and $\varepsilon_{\mathrm{con}}$, $\|a^{0}\|\gg\|a\|$. $\rho$ and $R$ may themselves depend on $N$: if up- and down-regulation balance, $\rho$ is of order $N^{-1/2}$, and $R$ grows slowly with $N$ at a rate set by the tail of the distribution of the mean expression. For example, with $N=2\times10^{4}$, $M=100$, $\pi_0=0.7$ and $\bar\mu=8$, $(1-\pi_0)\bar\mu^{2}\approx19$, $\varepsilon_{\mathrm{loo}}\approx2.5\times10^{-4}$, $\varepsilon_{\mathrm{spk}}\approx0.06$ and $\varepsilon_{\mathrm{side}}\approx0.004$.

\begin{table}[!htbp]
\centering
\caption{Approximation quantities expressed through $N$, $M$ and the signal strength}\label{tab:epsdep}
\small
\begin{tabular}{p{0.14\textwidth}p{0.4\textwidth}p{0.36\textwidth}}
\toprule
Quantity & Expression & Small when\\
\midrule
$\varepsilon_{\mathrm{sig}}(c)$ & fraction of genes that are signal genes with $|\mu_i^{\mathrm{eff}}|\le c$ & nearly all signal genes have $|\mu_i|\gg c$ and $|a^{0}_i\sin\theta|/\sqrt{d}\ll|\mu_i|$\\
$\varepsilon_{\mathrm{loo}}$ & $\dfrac{M/N}{1+(1-\pi_0)\bar\mu^{2}+M/N}$ & $M\ll N$ (it never exceeds $M/N$)\\
$\varepsilon_{\mathrm{spk}}$ & $\dfrac{(1+\sqrt{M/N})^{2}}{1+(1-\pi_0)\bar\mu^{2}+M/N}$ & $(1-\pi_0)\bar\mu^{2}\gg1$\\
$\varepsilon_{\mathrm{side}}$ & $\dfrac{\sqrt{M/N}}{(1-\pi_0)\bar\mu^{2}}$ & $M\ll N$ or $(1-\pi_0)\bar\mu^{2}\gg1$\\
$\varepsilon_{\mathrm{gap}}$ & $\dfrac{(1+\sqrt{M/N})^{2}}{\left(\bar\mu_{\mathrm{mean}}-\sqrt{1-\pi_0}\,\bar\mu\right)^{2}}$ & $|\bar\mu_{\mathrm{mean}}-\sqrt{1-\pi_0}\,\bar\mu|\gg1+\sqrt{M/N}$; usually $\bar\mu_{\mathrm{mean}}\gg\bar\mu$ for uncentred data\\
$\varepsilon_{\mathrm{mix}}$ & $\rho\sqrt{1-\pi_0}\,\bar\mu/\bar\mu_{\mathrm{mean}}$ & $\rho\bar\mu\ll\bar\mu_{\mathrm{mean}}$\\
$\varepsilon_{\mathrm{con}}$ & $\rho R\sqrt{1-\pi_0}\,\bar\mu$ & $\rho R\bar\mu\ll1$\\
$\varepsilon_{\mathrm{ng}}$ & $\kappa_3\beta_{\max}/\sqrt{M}$ & $M\gg\kappa_3^{2}$ (not needed for Gaussian noise)\\
$\rho$ & Eq.~\eqref{eq:rho} & signal uncorrelated with mean expression\\
$\varepsilon_{\mathrm{het}}$ & root mean square of $(d_i-d)/d$ over background genes; independent of $N$ and $M$ & noise variances nearly equal, e.g.\ after a variance-stabilizing transformation such as the logarithm\\
\bottomrule
\end{tabular}
\par\smallskip{\raggedright\footnotesize Quantities of Section~\ref{sec:eps}. The expressions assume $\rho\ll1$, so that $\mu_i^{\mathrm{eff}}\approx\mu_i$.\par}
\end{table}

\subsubsection{Assumptions and main theorem}\label{sec:assume}
\paragraph{Sequences of data sets.}\label{sec:seq} To state approximations as theorems, we consider a sequence of data sets indexed by $\nu=1,2,\dots$, each with its own $N_\nu$, $M_\nu$, $a^{0}_\nu$, $a_\nu$, $b_\nu$, $Z_\nu$, $d_{\nu}$ and $\mathcal{I}_{0,\nu}$, such that
\begin{equation}\label{eq:seq}
N_\nu\to\infty\qquad(\nu\to\infty),
\end{equation}
while $M_\nu$ may stay fixed or grow. The quantities \eqref{eq:eps1}--\eqref{eq:eps2} are computed for each $\nu$. As announced in Section~\ref{sec:method}, ``$\to$'', ``converges in probability'' and $O_P(\cdot)$ refer to $\nu\to\infty$, and the subscript $\nu$ is suppressed. Note that $\sigma_0=\sigma_{0,\nu}$ changes with $\nu$. Two kinds of decrease with $N$ must be distinguished. First, $\sigma_0$ tends to zero like $N_\nu^{-1/2}$. This is an artefact of normalizing the singular vector to unit length and has no statistical meaning; quantities from which this factor is removed, such as $\tau=\sigma/\sigma_0$ and $\sqrt{N}\sigma_0$, converge to non-zero values. Second, $\varepsilon_{\mathrm{loo}}\approx(M/N)(N\sigma_0^{2})$ decreases like $M/N$. This decrease is not caused by the normalization, since the factor $N\sigma_0^{2}$ is bounded; it expresses a real effect, namely that the influence of a single gene on the singular vector $v_{\ell_1^{*}}$ becomes smaller as the number of genes grows. $\varepsilon_{\mathrm{loo}}$ is therefore a small parameter, and Assumption~\ref{as:loo} below requires it to tend to zero.

The growth of $N$ is needed in addition to the conditions below: however small these quantities are, the empirical distribution of $n_0$ background scores fluctuates by $O_P(n_0^{-1/2})$, and this fluctuation disappears only as $n_0\to\infty$.

\begin{assumption}[strong signals]\label{as:weak}
$\varepsilon_{\mathrm{sig}}(c)\to0$ for every $c>0$.
\end{assumption}
This means that nearly all signal genes have a large effective signal $\mu_i^{\mathrm{eff}}$. A large $\mu_i$ alone is not sufficient: the mean term in \eqref{eq:mueff} must not cancel it (Remark~\ref{rem:cancel}). In practice $|\mu_i^{\mathrm{eff}}|\gtrsim6$ suffices (Table~\ref{tab:finite}). How $\mu_i$ depends on the number of samples is discussed in Remark~\ref{rem:delta}.
\begin{assumption}[spike separation]\label{as:spike}
The singular value $s_{\ell_1^{*}}$ is simple and separated from the other singular values, and $\varepsilon_{\mathrm{spk}}\to0$.
\end{assumption}
Assumption~\ref{as:spike} states that the signal genes jointly carry much more energy, $\|a\|^{2}=d(1-\pi_0)N\bar\mu^{2}$, than the noise can produce in a single direction, $d(\sqrt N+\sqrt M)^{2}$. By Table~\ref{tab:epsdep} it holds when $(1-\pi_0)\bar\mu^{2}\gg1$, i.e.\ when signal genes form a non-vanishing fraction of all genes and their signals are strong; Assumption~\ref{as:spike} therefore essentially follows from Assumption~\ref{as:weak}. For $\pi_0=0.7$, $\bar\mu=8$ and $M\ll N$, $\varepsilon_{\mathrm{spk}}\approx0.06$.
\begin{assumption}[no mean-component contamination]\label{as:contam}
$\varepsilon_{\mathrm{con}}\to0$.
\end{assumption}
\begin{assumption}[small leave-one-out error]\label{as:loo}
$\varepsilon_{\mathrm{loo}}\to0$.
\end{assumption}
\begin{assumption}[no mixing with the mean component]\label{as:mix}
$\varepsilon_{\mathrm{mix}}\to0$.
\end{assumption}
\begin{assumption}[normality of the noise sums]\label{as:ng}
The entries of $Z$ have finite third moments, and either they are Gaussian or $\varepsilon_{\mathrm{ng}}\to0$.
\end{assumption}
\begin{assumption}[gap to the other component]\label{as:gap}
$\varepsilon_{\mathrm{gap}}\to0$, with $\varepsilon_{\mathrm{gap}}$ of \eqref{eq:epsgap}.
\end{assumption}
\begin{assumption}[size of the noise]\label{as:norm}
$\|Z\|=O_P\left(\sqrt d(\sqrt N+\sqrt M)\right)$, where $\|Z\|$ is the largest singular value of $Z$ (its operator norm).
\end{assumption}
\begin{assumption}[nearly equal noise variances]\label{as:het}
$\varepsilon_{\mathrm{het}}\to0$, and the mean of $d_i/d$ over all genes is bounded.
\end{assumption}
Assumption~\ref{as:gap} requires the noise to be small compared with the gap between the signal component and the mean component; no ordering of the two is assumed. Assumption~\ref{as:mix} controls the mixing of these two components before noise is added; it does not prevent the noise from mixing them when their singular values are close. Assumption~\ref{as:het} allows the noise variances to differ between genes but requires the differences among background genes to be small; it is the assumption most likely to fail in real data, for example for untransformed counts, whose variance grows with the mean. Assumption~\ref{as:norm} holds, for example, when $\max_id_i/d$ is bounded and the $W_{ir}$ are Gaussian, or have a finite fourth moment and $M/N$ converges \cite{YinBaiKrishnaiah1988}, since $\|Z\|\le(\max_id_i)^{1/2}\|W\|$; a finite third moment, as in Assumption~\ref{as:ng}, is not sufficient for it in general.
Assumption~\ref{as:mix} excludes signal patterns that overlap the mean expression over genes. Assumption~\ref{as:contam} alone does not do so: if $a^{0}=a$, no background gene is shifted ($\varepsilon_{\mathrm{con}}=0$), but the signal part becomes $a(e_M+b)^{\top}$ and the right singular vector tends to $(e_M+b)/\sqrt2$, so that $\langle b,v_{\ell_1^{*}}\rangle\to1/\sqrt2$; in this case $\varepsilon_{\mathrm{mix}}=\infty$. Assumption~\ref{as:ng} requires, for non-Gaussian noise, that $M$ grows and that the pattern $b$ is not concentrated on a few columns. It is needed: with $M=2$, $b=(1,-1)/\sqrt2$ and noise entries equal to $\pm1$ with probability $1/2$, the background scores tend to $(Z_{i1}-Z_{i2})/\sqrt2$, which takes the values $0,\pm\sqrt2$ and is not normal.
By \eqref{eq:loobound}, Assumption~\ref{as:loo} holds whenever $M/N\to0$.

\begin{theorem}\label{thm:main}
Consider a sequence of data sets satisfying \eqref{eq:seq} with $M/N$ bounded and $n_0/N\to\pi_0\in(0,1)$, and let $\bar v_{\ell_1^{*}}$ be the right singular vector of the noiseless part $a^{0}e_M^{\top}+ab^{\top}$ for its signal component $\bar s_{\ell_1^{*}}$ (Section~\ref{sec:eps}), with its sign fixed by $\langle b,\bar v_{\ell_1^{*}}\rangle>0$. As $\nu\to\infty$:
\begin{enumerate}[label=(\roman*)]
\item \emph{(the HOSVD captures the signal)} under Assumptions~\ref{as:spike}, \ref{as:mix}, \ref{as:gap} and \ref{as:norm}, there is a component $\ell_1^{*}$, associated with the signal, whose right singular vector $v_{\ell_1^{*}}$ satisfies $\langle b,v_{\ell_1^{*}}\rangle\to1$; for a given data set, $1-\langle b,v_{\ell_1^{*}}\rangle$ is of the order of $\varepsilon_{\mathrm{spk}}+\varepsilon_{\mathrm{gap}}+\varepsilon_{\mathrm{mix}}^{2}$. Thus this component reproduces the pattern $b$ over samples and conditions. Whether TD-based UFE selects it is a separate question, answered in Proposition~\ref{prop:select};
\item \emph{(background scores are normal)} the empirical distribution of the standardized background scores $s_{\ell_1^{*}}u_{\ell_1^{*} i}/\sqrt{d}$, $i\in\mathcal{I}_0$, converges in probability to $N(0,1)$ (that is, the fraction of background genes whose standardized score is at most $x$ converges in probability to $\Phi(x)$, uniformly in $x$). Consequently, to leading order, the standard deviation of the normal law followed by the background scores is
\begin{equation}\label{eq:sigma0}
\sigma_0=\frac{\sqrt{d}}{s_{\ell_1^{*}}};
\end{equation}
\item \emph{(signal genes separate)} under the assumptions of (ii) and Assumption~\ref{as:weak}, for every $c>0$ the fraction of genes that are signal genes with $s_{\ell_1^{*}}|u_{\ell_1^{*} i}|/\sqrt{d}\le c$ tends to zero in probability: signal genes, whose effective signals $\mu_i^{\mathrm{eff}}$ are large, separate from the background, whose scores remain of order one.
\end{enumerate}
Part (ii) holds under the assumptions of (i) and Assumptions~\ref{as:contam}, \ref{as:ng} and \ref{as:het}. For a given data set, the proof gives the following bound. Let $G$ be the empirical distribution function of the standardized background scores, $F_{\omega}$ that of the standardized noise projections $\omega_i=W_i\bar v_{\ell_1^{*}}$ over background genes, and
\[
\mathcal{E}=\frac{1}{n_0}\sum_{i\in\mathcal{I}_0}\left[\xi_i+\left(\sqrt{1+\eta_i}-1\right)\omega_i\right]^{2}=O_P\left(\frac{(1+\sqrt{M/N})^{2}(\varepsilon_{\mathrm{spk}}+\varepsilon_{\mathrm{gap}})}{\pi_0}+\varepsilon_{\mathrm{het}}^{2}\right)
\]
the mean square of the error terms of \eqref{eq:threeterm} and \eqref{eq:zetahet}. Then, for every $\lambda>0$,
\begin{equation}\label{eq:devbound}
\sup_x|G(x)-\Phi(x)|\le\sup_x|F_{\omega}(x)-\Phi(x)|+\frac{\varepsilon_{\mathrm{con}}+\lambda}{\sqrt{2\pi}}+\frac{\mathcal{E}}{\lambda^{2}},\qquad \sup_x|F_{\omega}(x)-\Phi(x)|\le C_0\varepsilon_{\mathrm{ng}}+O_P(n_0^{-1/2}).
\end{equation}
Choosing $\lambda=\mathcal{E}^{1/3}$ gives an error of order $\varepsilon_{\mathrm{ng}}+\varepsilon_{\mathrm{con}}+\mathcal{E}^{1/3}+n_0^{-1/2}$; in this bound the mean-square error $\mathcal{E}$ appears through its cube root.
\end{theorem}

\begin{remark}[scope of the proof]\label{rem:proofstatus}
Theorem~\ref{thm:main} gives the background standard deviation to leading order, $\sigma_0=\sqrt d/s_{\ell_1^{*}}$. The leave-one-out refinement $\sigma_0\approx(1+\varepsilon_{\mathrm{loo}})\sqrt d/s_{\ell_1^{*}}$ of Remark~\ref{rem:loo} is a heuristic first-order correction, not covered by the proof, and Assumption~\ref{as:loo} is used only for it. Because the theorem does not specify how fast the individual error quantities decay, it gives no rate of convergence in terms of $N$ alone.
\end{remark}

In words: (i) the HOSVD has a component that reproduces the pattern of the design (the first half of goal (G1), the selection being treated in Proposition~\ref{prop:select}); (ii) the scores of background genes in this component are approximately normal with a standard deviation determined by the noise level and the singular value; and (iii) the scores of signal genes are larger, of the order of their effective signal $\mu_i^{\mathrm{eff}}$ (goal (G2)). The proof of (ii) and (iii) compares $v_{\ell_1^{*}}$ with the deterministic vector $\bar v_{\ell_1^{*}}$, so that the noise projections $Z_i\bar v_{\ell_1^{*}}$ are independent across genes; no leave-one-out argument is needed.

\begin{corollary}[why $\sigma$ must be estimated]\label{cor:local}
The standard deviation of all $N$ components of the unit vector $u_{\ell_1^{*}\cdot}$ is $\sigma_{\mathrm{all}}\approx N^{-1/2}$, whereas by \eqref{eq:Nsigma2}
\begin{equation}
\frac{\sigma_0}{\sigma_{\mathrm{all}}}\approx\sqrt{N}\sigma_0\approx\frac{1}{\sqrt{1+(1-\pi_0)\bar\mu^{2}}}<1 .
\end{equation}
Using all components therefore overestimates the background standard deviation, by a factor that grows with the strength of the signal: the signal genes take up part of the unit length of the vector, leaving less for the background. For $\pi_0=0.7$ and $\bar\mu=8$ the overestimate is about a factor of $4.5$.
\end{corollary}

\begin{proof}[Proof of part (i) of Theorem~\ref{thm:main}]
\emph{Part (i).} First consider the noiseless part $a^{0}e_M^{\top}+ab^{\top}$, with $e_M$ as in \eqref{eq:evec}. Its right singular vectors lie in the plane spanned by $e_M$ and $b$; since $e_M\perp b$, they are $e_M$ and $b$ themselves only if, in addition, $a^{0}\perp a$ over genes. In general they are rotated by an angle $\theta$ with
\begin{equation}
\tan2\theta=\frac{2\langle a,a^{0}\rangle}{\|a^{0}\|^{2}-\|a\|^{2}},
\end{equation}
so that $|\theta|\le\varepsilon_{\mathrm{mix}}$ and $1-\cos\theta=O(\varepsilon_{\mathrm{mix}}^{2})$; Assumption~\ref{as:mix} makes the rotation vanish. Adding the noise, let $\bar v_{\ell_1^{*}}$ be the right singular vector of the noiseless part for $\bar s_{\ell_1^{*}}$, and let $\delta=\min(\bar s_{\ell_1^{*}},|\bar s_{\ell_1^{\mathrm{mean}}}-\bar s_{\ell_1^{*}}|)$ be the gap of $\bar s_{\ell_1^{*}}$ to the other singular values of the noiseless part ($\bar s_{\ell_1^{\mathrm{mean}}}$ and $0$). By Weyl's inequality for singular values, $|s_{\ell_1^{*}}-\bar s_{\ell_1^{*}}|\le\|Z\|$. By Assumption~\ref{as:norm}, $\|Z\|=O_P(\sqrt d(\sqrt N+\sqrt M))$, so Assumption~\ref{as:spike} gives $\|Z\|/s_{\ell_1^{*}}=O_P(\varepsilon_{\mathrm{spk}}^{1/2})\to0$; hence $\bar s_{\ell_1^{*}}\ge s_{\ell_1^{*}}-\|Z\|$ and $\|Z\|/\bar s_{\ell_1^{*}}\to0$ as well. Together with Assumption~\ref{as:gap}, $\|Z\|/\delta\to0$. Wedin's theorem for singular vectors \cite{Wedin1972} then gives $\sin\angle(v_{\ell_1^{*}},\bar v_{\ell_1^{*}})\le C\|Z\|/\delta$ with an absolute constant $C$, so that $\sin^{2}\angle(v_{\ell_1^{*}},\bar v_{\ell_1^{*}})=O_P(\varepsilon_{\mathrm{spk}}+\varepsilon_{\mathrm{gap}})$. Together with $\langle\bar v_{\ell_1^{*}},b\rangle=\cos\theta$, this gives $1-\langle b,v_{\ell_1^{*}}\rangle=O_P(\varepsilon_{\mathrm{spk}}+\varepsilon_{\mathrm{gap}}+\varepsilon_{\mathrm{mix}}^{2})$, and hence $\langle b,v_{\ell_1^{*}}\rangle\to1$.
\end{proof}

\begin{proof}[Proof of parts (ii) and (iii) of Theorem~\ref{thm:main}]
The sign of $\bar v_{\ell_1^{*}}$ is fixed by $\langle b,\bar v_{\ell_1^{*}}\rangle>0$. Singular vectors are defined only up to sign, so we choose the sign of $v_{\ell_1^{*}}$, together with that of the paired left singular vector $u_{\ell_1^{*}\cdot}$, so that $\langle v_{\ell_1^{*}},\bar v_{\ell_1^{*}}\rangle\ge0$. By \eqref{eq:scoredecomp}, for every gene
\begin{equation}\label{eq:threeterm}
\frac{s_{\ell_1^{*}}u_{\ell_1^{*} i}}{\sqrt d}=\mu_i^{\mathrm{eff}}+\zeta_i+\xi_i,\qquad \zeta_i=\frac{Z_i\bar v_{\ell_1^{*}}}{\sqrt d},\qquad \xi_i=\frac{Z_i(v_{\ell_1^{*}}-\bar v_{\ell_1^{*}})}{\sqrt d}.
\end{equation}
\emph{The term $\zeta_i$.} Write $Z_i=\sqrt{d_i}\,W_i$ and $\omega_i=W_i\bar v_{\ell_1^{*}}$, so that
\begin{equation}\label{eq:zetahet}
\zeta_i=\sqrt{1+\eta_i}\,\omega_i=\omega_i+\left(\sqrt{1+\eta_i}-1\right)\omega_i .
\end{equation}
The vector $\bar v_{\ell_1^{*}}$ is deterministic and the rows $W_i$ are independent and identically distributed, so the $\omega_i$ are independent and identically distributed with mean $0$ and variance $\|\bar v_{\ell_1^{*}}\|^{2}=1$. For Gaussian noise $\omega_i\sim N(0,1)$ exactly; otherwise, by the Berry--Esseen inequality, the distribution function of $\omega_i$ differs from $\Phi$ by at most $C_0\kappa_3\sum_r|\bar v_{\ell_1^{*}r}|^{3}\le C_0\varepsilon_{\mathrm{ng}}\to0$ (Assumption~\ref{as:ng}). Since they are independent, the empirical distribution of $\omega_i$ over the $n_0$ background genes is within $O_P(n_0^{-1/2})$ of their common distribution function, uniformly (Dvoretzky--Kiefer--Wolfowitz inequality), and hence converges uniformly to $\Phi$. Since $|\sqrt{1+\eta}-1|\le|\eta|$, the second term of \eqref{eq:zetahet} has expected mean square over background genes at most $\frac{1}{n_0}\sum_{i\in\mathcal{I}_0}\eta_i^{2}=\varepsilon_{\mathrm{het}}^{2}$, so its mean square is $O_P(\varepsilon_{\mathrm{het}}^{2})\to0$ (Assumption~\ref{as:het}).

\emph{The term $\xi_i$.} By the proof of (i), $\|v_{\ell_1^{*}}-\bar v_{\ell_1^{*}}\|^{2}\le2\sin^{2}\angle(v_{\ell_1^{*}},\bar v_{\ell_1^{*}})=O_P(\varepsilon_{\mathrm{spk}}+\varepsilon_{\mathrm{gap}})$, and by Assumption~\ref{as:norm},
\begin{equation}
\frac1N\sum_{i=1}^{N}\xi_i^{2}=\frac{\|Z(v_{\ell_1^{*}}-\bar v_{\ell_1^{*}})\|^{2}}{Nd}\le\frac{\|Z\|^{2}\|v_{\ell_1^{*}}-\bar v_{\ell_1^{*}}\|^{2}}{Nd}=O_P\left(\left(1+\sqrt{M/N}\right)^{2}(\varepsilon_{\mathrm{spk}}+\varepsilon_{\mathrm{gap}})\right)\to0 .
\end{equation}
The terms $\xi_i$ need not be small for each gene, but their mean square is.

\emph{A perturbation lemma.} If the empirical distribution of $X_1,\dots,X_n$ converges uniformly to $\Phi$ in probability and $\frac1n\sum_iY_i^{2}\to0$ in probability, then so does the empirical distribution of $X_i+Y_i$. Quantitatively, if $|Y_i|\le\epsilon+|W_i|$ with $\frac1n\sum_iW_i^{2}=\mathcal{E}$, then $\sup_x|G_n(x)-\Phi(x)|\le\sup_x|F_n(x)-\Phi(x)|+(\epsilon+\lambda)/\sqrt{2\pi}+\mathcal{E}/\lambda^{2}$ for every $\lambda>0$, which gives \eqref{eq:devbound}. Indeed, for $\lambda>0$ the fraction of indices with $|Y_i|>\lambda$ is at most $\frac{1}{n\lambda^{2}}\sum_iY_i^{2}\to0$; therefore the empirical distribution function $G_n$ of $X_i+Y_i$ satisfies $F_n(x-\lambda)-o_P(1)\le G_n(x)\le F_n(x+\lambda)+o_P(1)$ uniformly in $x$, where $F_n$ is that of $X_i$, and since $\Phi$ is uniformly continuous and $\lambda$ is arbitrary, $\sup_x|G_n(x)-\Phi(x)|\to0$ in probability.

\emph{Part (ii).} For background genes, $|\mu_i^{\mathrm{eff}}|\le\varepsilon_{\mathrm{con}}\to0$ uniformly (Assumption~\ref{as:contam}), the mean square of $\xi_i$ over background genes is at most $\frac{N}{n_0}\cdot\frac1N\sum_i\xi_i^{2}\to0$, and that of $(\sqrt{1+\eta_i}-1)\omega_i$ tends to zero. Applying the lemma to $X_i=\omega_i$ with $Y_i=\mu_i^{\mathrm{eff}}+(\sqrt{1+\eta_i}-1)\omega_i+\xi_i$ gives the first statement of (ii). For each background gene $i\in\mathcal{I}_0$, $u_{\ell_1^{*} i}=(\sqrt d/s_{\ell_1^{*}})\cdot\left[s_{\ell_1^{*}}u_{\ell_1^{*} i}/\sqrt d\right]$, so the empirical distribution of the background components $u_{\ell_1^{*} i}$, $i\in\mathcal{I}_0$, approaches $N(0,d/s_{\ell_1^{*}}^{2})$; this gives \eqref{eq:sigma0}. The components of signal genes do not enter this distribution; they affect it only through the common normalization $s_{\ell_1^{*}}$. Here $\sigma_0$ means the scale of this limiting normal law; convergence of the sample standard deviation of the background scores is not claimed, since (ii) concerns distribution functions.

\emph{Part (iii).} Fix $c>0$ and $c'>0$. A signal gene with $|s_{\ell_1^{*}}u_{\ell_1^{*} i}|/\sqrt d\le c$ satisfies $|\mu_i^{\mathrm{eff}}|\le c+|\zeta_i|+|\xi_i|$, so it has $|\mu_i^{\mathrm{eff}}|\le c+2c'$, or $|\zeta_i|>c'$, or $|\xi_i|>c'$. As fractions of all $N$ genes, the first set has size $\varepsilon_{\mathrm{sig}}(c+2c')\to0$ (Assumption~\ref{as:weak}). The second is at most $\frac{1}{N{c'}^{2}}\sum_i\zeta_i^{2}$; the expectation of $\frac1N\sum_i\zeta_i^{2}$ is the mean of $d_i/d$ over all genes, which is bounded by Assumption~\ref{as:het}, so by Markov's inequality the second fraction is $O_P({c'}^{-2})$, uniformly along the sequence. The third is at most $\frac{1}{N{c'}^{2}}\sum_i\xi_i^{2}\to0$. Hence, for every $c'$, the fraction in question is at most $O_P({c'}^{-2})+o_P(1)$, and letting $c'\to\infty$ gives (iii).
\end{proof}

\begin{remark}[meaning of the leave-one-out error]\label{rem:loo}
The right singular vector $v_{\ell_1^{*}}$ is computed from all rows of $X$, including row $i$. It is therefore slightly tilted towards $Z_i$, so that $Z_iv_{\ell_1^{*}}$ is not exactly distributed like $Z_iv^{(-i)}_{\ell_1^{*}}$, which has mean zero and variance $d_i$ and is exactly $N(0,d_i)$ for Gaussian noise. The difference is the influence of gene $i$ on its own score. Adding row $i$ changes $X^{\top}X$ by $X_{i\cdot}^{\top}X_{i\cdot}$, with $X_{i\cdot}=a^{0}_ie_M^{\top}+Z_i$ for a background gene. The part along $e_M$ changes $v_{\ell_1^{*}}$ only through the large gap to the mean component and is negligible (heuristically); the remaining part is $Z_i^{\top}Z_i$. To first order, and up to corrections of order $\varepsilon_{\mathrm{spk}}$ (i.e.\ when $s_{\ell_1^{*}}^{2}$ dominates the other squared singular values), the right singular vector moves by
\begin{equation}
v_{\ell_1^{*}}-v^{(-i)}_{\ell_1^{*}}\approx\frac{(Z_iv_{\ell_1^{*}})\,P_\perp Z_i^{\top}}{s_{\ell_1^{*}}^{2}},
\end{equation}
where $P_\perp$ is the projection onto the orthogonal complement of $v_{\ell_1^{*}}$. Multiplying by $Z_i$ and using $\|P_\perp Z_i^{\top}\|^{2}=(M-1)d_i\left(1+O_P(M^{-1/2})\right)\approx Md_i\approx Md$ when $\varepsilon_{\mathrm{het}}$ is small, we obtain
\begin{equation}\label{eq:selfinfl}
Z_iv_{\ell_1^{*}}\approx Z_iv^{(-i)}_{\ell_1^{*}}+\frac{Md}{s_{\ell_1^{*}}^{2}}\,Z_iv_{\ell_1^{*}},\qquad\text{that is,}\qquad
Z_iv_{\ell_1^{*}}\approx\frac{Z_iv^{(-i)}_{\ell_1^{*}}}{1-\varepsilon_{\mathrm{loo}}}\approx(1+\varepsilon_{\mathrm{loo}})\,Z_iv^{(-i)}_{\ell_1^{*}} .
\end{equation}
Thus $\varepsilon_{\mathrm{loo}}$ is the fraction by which each background gene inflates its own score. Because $\|Z_i\|^{2}\approx Md_i$ and the $d_i$ of background genes are nearly equal, the inflation factor is nearly the same for all of them. It does not destroy normality; it rescales the background standard deviation, which to this order is
\begin{equation}\label{eq:sigma0refined}
\sigma_0\approx(1+\varepsilon_{\mathrm{loo}})\,\frac{\sqrt{d}}{s_{\ell_1^{*}}} .
\end{equation}
By \eqref{eq:loobound} the correction is at most $M/N$ and usually much smaller ($0.025\%$ in the example of Section~\ref{sec:eps}); \eqref{eq:sigma0} is the leading-order form.
\end{remark}

\begin{remark}\label{rem:rate}
Theorem~\ref{thm:main}(iii) does not require $|\mu_i|$ to grow at any particular rate, such as $\sqrt{2\log N}$. When signal genes form a positive fraction of all genes, the BH threshold stays at a constant of order $\alpha(1-\pi_0)$ (Section~\ref{sec:limit}), so a signal gene is selected as soon as its $P$-value tends to zero.
\end{remark}

\subsubsection{Selection of \texorpdfstring{$\ell_1$}{l1} through the sample modes}\label{sec:select}
Theorem~\ref{thm:main} shows that the singular value decomposition of $X$ has a component $\ell_1^{*}$ with $v_{\ell_1^{*}}\approx b$. TD-based UFE, however, does not look at $v_{\ell_1^{*}}$ directly. It selects indices $\hat\ell_2$ and $\hat\ell_3$ whose sample-side singular vectors reflect the design, and then the index $\hat\ell_1$ maximizing $|G(\ell_1\hat\ell_2\hat\ell_3)|$ over $\ell_1$. The sample-side singular vectors are obtained from other unfoldings of the tensor, so $v_{\ell_1^{*}}\approx b$ alone does not imply that they reproduce the design; additional conditions are needed.

We use the constant vectors $e_J$, $e_Q$ and $e_M=e_J\otimes e_Q$ of \eqref{eq:evec} and the correlation $\rho$ of \eqref{eq:rho}. The left singular vectors of the mode-2 unfolding $X^{(2)}\in\R^{J\times NQ}$ of \eqref{eq:unfold} are the $u_{\ell_2 j}$. In the model, the mean term is $a^{0}\otimes e_J\otimes e_Q$ and the signal term is $a\otimes b^{(2)}\otimes b^{(3)}$ when $b=b^{(2)}\otimes b^{(3)}$, so that, apart from noise,
\begin{equation}\label{eq:gram2}
X^{(2)}X^{(2)\top}\approx\|a^{0}\|^{2}e_Je_J^{\top}+\|a\|^{2}b^{(2)}b^{(2)\top}+\langle a,a^{0}\rangle\langle e_Q,b^{(3)}\rangle\left(e_Jb^{(2)\top}+b^{(2)} e_J^{\top}\right),
\end{equation}
and similarly for the mode-3 unfolding $X^{(3)}$. The conditions below make $b^{(2)}$ an eigenvector of this matrix.

\begin{proposition}[selection of $\ell_1$]\label{prop:select}
Consider a sequence of data sets satisfying the assumptions of Theorem~\ref{thm:main}, and suppose in addition:
\begin{enumerate}[label=\textbf{(B\arabic*)}]
\item \emph{(product pattern)} $b_{(j,q)}=b^{(2)}_jb^{(3)}_q$ with unit vectors $b^{(2)}\in\R^{J}$ and $b^{(3)}\in\R^{Q}$;
\item \emph{(constant or orthogonal to constant in each mode)} $b^{(2)}\perp e_J$ or $b^{(2)}=\pm e_J$, and $b^{(3)}\perp e_Q$ or $b^{(3)}=\pm e_Q$;
\item \emph{(uncorrelated with the mean expression)} $\rho\to0$;
\item \emph{(spike separation in the sample modes)} $\varepsilon_{\mathrm{side}}\to0$ (Eq.~\eqref{eq:eps4}; by Table~\ref{tab:epsdep} this holds when $M\ll N$ or $(1-\pi_0)\bar\mu^{2}\gg1$), and the eigenvalues of the sample-mode Gram matrices associated with $b^{(2)}$ and $b^{(3)}$ are simple and separated from the others.
\end{enumerate}
Then the following hold.
\begin{enumerate}[label=(\alph*)]
\item \emph{(for a given data set)} Let $\ell_2^{*}$ and $\ell_3^{*}$ be the indices maximizing $|\langle u_{\ell_2\cdot},b^{(2)}\rangle|$ over $\ell_2$ and $|\langle u_{\ell_3\cdot},b^{(3)}\rangle|$ over $\ell_3$. The angles between $u_{\ell_2^{*}\cdot}$ and $b^{(2)}$ and between $u_{\ell_3^{*}\cdot}$ and $b^{(3)}$ are, to leading order, at most of the order of $\varepsilon_{\mathrm{mix}}+\varepsilon_{\mathrm{side}}$. For these indices, the ratio of $|G(\ell_1\ell_2^{*}\ell_3^{*})|$ for any component $\ell_1$ other than the component $\ell_1^{*}$ of Theorem~\ref{thm:main} to $|G(\ell_1^{*}\ell_2^{*}\ell_3^{*})|$ is, to leading order, at most of the order of $\rho+\varepsilon_{\mathrm{side}}+\sqrt{\varepsilon_{\mathrm{spk}}}$.
\item \emph{(along the sequence)} As $\nu\to\infty$, with $\ell_2^{*}=\ell_2^{*}(\nu)$ and $\ell_3^{*}=\ell_3^{*}(\nu)$ chosen as in (a), $|\langle u_{\ell_2^{*}\cdot},b^{(2)}\rangle|\to1$ and $|\langle u_{\ell_3^{*}\cdot},b^{(3)}\rangle|\to1$ in probability, and with probability tending to one the index $\hat\ell_1$ maximizing $|G(\ell_1\ell_2^{*}\ell_3^{*})|$ over $\ell_1$ equals $\ell_1^{*}$. Hence, if TD-based UFE selects $\hat\ell_2=\ell_2^{*}$ and $\hat\ell_3=\ell_3^{*}$, it selects $\hat\ell_1=\ell_1^{*}$.
\end{enumerate}
In a single data set, no sample-side singular vector coincides exactly with $b^{(2)}$ or $b^{(3)}$; (a) states how close the best one is.
\end{proposition}
\begin{proof}[Proof (outline)]
\emph{Sample-side vectors.} Since $b\perp\mathbf{1}_M$, $\langle e_J,b^{(2)}\rangle\langle e_Q,b^{(3)}\rangle=0$, so by (B2) at least one of $b^{(2)}$, $b^{(3)}$ is orthogonal to the constant vector of its mode. Consider mode~2. If $b^{(2)}=\pm e_J$, the noiseless part of \eqref{eq:gram2} is proportional to $e_Je_J^{\top}$ and $e_J=\pm b^{(2)}$ is an eigenvector. If $b^{(2)}\perp e_J$, the noiseless part acts on the plane spanned by $e_J$ and $b^{(2)}$ as the $2\times2$ matrix with diagonal $\|a^{0}\|^{2}$, $\|a\|^{2}$ and off-diagonal $\langle a,a^{0}\rangle\langle e_Q,b^{(3)}\rangle$; its eigenvectors are rotated from $e_J$ and $b^{(2)}$ by an angle at most $\varepsilon_{\mathrm{mix}}$, which tends to zero by Assumption~\ref{as:mix}. The noise part of $X^{(2)}X^{(2)\top}$ is $NQdI_J$, which does not change eigenvectors, plus fluctuations of order $d\sqrt{NQJ}=d\sqrt{NM}$; relative to the eigenvalue $\|a\|^{2}$ these are of order $\varepsilon_{\mathrm{side}}$, and by (B4) and the Davis--Kahan theorem the eigenvector stays close to $b^{(2)}$. Mode~3 is treated in the same way. Without (B1), the singular vectors of the sample modes would be the factors of the singular value decomposition of $b$ arranged as a $J\times Q$ array, and $|G|$ would be spread over several pairs $(\ell_2,\ell_3)$. Without (B2), e.g.\ if $b^{(2)}$ is neither orthogonal nor equal to $e_J$, the eigenvectors would be $e_J$ and the component of $b^{(2)}$ orthogonal to $e_J$, and again $|G|$ would be split between two pairs.

\emph{Selection.} By \eqref{eq:vhat} and the orthonormality of the sample-side vectors, $G(\ell_1\ell_2^{*}\ell_3^{*})=s_{\ell_1}\langle v_{\ell_1},u_{\ell_2^{*}\cdot}\otimes u_{\ell_3^{*}\cdot}\rangle$, and $u_{\ell_2^{*}\cdot}\otimes u_{\ell_3^{*}\cdot}\approx\pm b$. For the component $\ell_1^{*}$ of Theorem~\ref{thm:main}, $|G(\ell_1^{*}\ell_2^{*}\ell_3^{*})|\approx s_{\ell_1^{*}}$. Every other right singular vector is orthogonal to $v_{\ell_1^{*}}$, hence has overlap with $b$ at most $\|b-v_{\ell_1^{*}}\|\to0$. For the components produced by noise, $s_{\ell_1}\le\sqrt{d}(\sqrt N+\sqrt M)\approx s_{\ell_1^{*}}\sqrt{\varepsilon_{\mathrm{spk}}}$, so their $|G|$ is negligible. For the mean component, $s\approx\|a^{0}\|$ and its overlap with $b$ is about the rotation angle $|\langle a,a^{0}\rangle|/\|a^{0}\|^{2}$, so the ratio of its $|G|$ to that of $\ell_1^{*}$ is about $(\|a^{0}\|/\|a\|)\,|\langle a,a^{0}\rangle|/\|a^{0}\|^{2}=\rho$, which tends to zero by (B3).
\end{proof}

Condition (B3) is stronger than Assumption~\ref{as:mix} when $\|a^{0}\|\gg\|a\|$, because the mean component, having a much larger singular value, can win the maximization of $|G|$ even when its overlap with $b$ is small. The ratio $|G(\ell_1^{\mathrm{mean}}\ell_2^{*}\ell_3^{*})|/|G(\ell_1^{*}\ell_2^{*}\ell_3^{*})|$, with $\ell_1^{\mathrm{mean}}$ the mean component, estimates $\rho$ and can be read off the core tensor. Condition (B2) agrees with the practice of TD-based UFE: a mode of replicates is represented by the constant vector, and a contrast between groups by a vector orthogonal to it. Finally, the proposition guarantees that sample-side singular vectors reproducing the design exist; in practice they are chosen by inspecting the patterns of the sample-side singular vectors, and the proposition does not address that choice.

\begin{remark}[relation between $\varepsilon_{\mathrm{side}}$ and $\varepsilon_{\mathrm{spk}}$]\label{rem:side}
Condition (B4) essentially follows from Assumption~\ref{as:spike}. With $x=(1-\pi_0)\bar\mu^{2}$, Table~\ref{tab:epsdep} gives
\begin{equation}
\frac{\varepsilon_{\mathrm{side}}}{\varepsilon_{\mathrm{spk}}}\approx\frac{\sqrt{M/N}\,(1+x+M/N)}{x\,(1+\sqrt{M/N})^{2}} .
\end{equation}
Assumption~\ref{as:spike}, $\varepsilon_{\mathrm{spk}}\ll1$, requires $x\gg(1+\sqrt{M/N})^{2}$, and then the ratio is about $\sqrt{M/N}/(1+\sqrt{M/N})^{2}\le1/4$. Hence whenever the signal component is captured in the gene-mode unfolding, $\varepsilon_{\mathrm{side}}$ is at most about a quarter of $\varepsilon_{\mathrm{spk}}$. Conversely, if $\varepsilon_{\mathrm{side}}$ is not small, then $x$ is small, $\varepsilon_{\mathrm{spk}}$ is not small either, and the right singular vector $v_{\ell_1^{*}}$ is itself dominated by noise; looking at \eqref{eq:vhat} does not help. The reason is that the quantity $d\sqrt{NM}$ in $\varepsilon_{\mathrm{side}}$ is the threshold at which a rank-one signal becomes detectable in an $N\times M$ noise matrix at all, whereas $\varepsilon_{\mathrm{spk}}$ expresses the stronger requirement that the singular vector be accurately aligned with the signal. The sample-side singular vectors, which are obtained from Gram matrices that sum over all genes, are therefore at least as reliable as $v_{\ell_1^{*}}$.
\end{remark}

\subsubsection{Interaction with the mean expression: contamination and cancellation}\label{sec:contam}
\paragraph{Mechanism.} Because genes are not centred, the mean expression $a^{0}$ forms the dominant component of $X$, and the selected component is orthogonal to it. When $\|a^{0}\|\gg\|a\|$, the part of the signal along the mean expression, $c\,a^{0}$ with $c=\langle a,a^{0}\rangle/\|a^{0}\|^{2}\approx\sin\theta$, is therefore removed together with the mean component, and the effective signal \eqref{eq:mueff} is approximately
\begin{equation}\label{eq:mueffc}
\mu_i^{\mathrm{eff}}\approx\frac{a_i-c\,a^{0}_i}{\sqrt d}.
\end{equation}
This single mechanism has three faces, treated in turn below: the right singular vector is rotated away from $b$ (the mixing angle $\theta$, Assumption~\ref{as:mix}); background genes, for which $a_i=0$, are shifted in proportion to their mean expression (contamination, Assumption~\ref{as:contam}); and signal genes whose ratio $a_i/a^{0}_i$ is close to $c$ lose their signal (cancellation, the reason why Assumption~\ref{as:weak} is stated for $\mu_i^{\mathrm{eff}}$).

\paragraph{Rotation and contamination.} If the patterns $a$ and $a^{0}$ overlap over genes, this has two effects. Over the columns of $X$ (the sample-side modes), the right singular vector is rotated away from $b$ by an angle of at most $\varepsilon_{\mathrm{mix}}$ (proof of Theorem~\ref{thm:main}); this is controlled by Assumption~\ref{as:mix}. Over the genes (mode~1), part of the mean expression is transferred into the scores of background genes, as described next; this is controlled by Assumption~\ref{as:contam}. When $\|a^{0}\|\gg\|a\|$, $\varepsilon_{\mathrm{mix}}\approx|\langle a,a^{0}\rangle|/\|a^{0}\|^{2}$, and both effects are small if the overlap is small. The rotation can be read off the core tensor. With mixing, $v_{\ell_1^{*}}\approx\cos\theta\,b-\sin\theta\,e_M$, and $e_M=e_J\otimes e_Q$ is the product of the constant vectors of modes~2 and 3, whose indices are $\ell_2^{\mathrm{mean}}$ and $\ell_3^{\mathrm{mean}}$ (Section~\ref{sec:method}). By \eqref{eq:vhat}, $G(\ell_1^{*}\ell_2^{*}\ell_3^{*})\approx s_{\ell_1^{*}}\cos\theta$ and $G(\ell_1^{*}\ell_2^{\mathrm{mean}}\ell_3^{\mathrm{mean}})\approx s_{\ell_1^{*}}\sin\theta$, so that
\begin{equation}\label{eq:tantheta}
\tan\theta\approx\frac{|G(\ell_1^{*}\ell_2^{\mathrm{mean}}\ell_3^{\mathrm{mean}})|}{|G(\ell_1^{*}\ell_2^{*}\ell_3^{*})|} .
\end{equation}

\begin{approximation}\label{prop:contam}
Suppose $\|a^{0}\|^{2}\gg\|a\|^{2}$. Then the left singular vector associated with the contrast is, up to normalization, $a-\langle a,\hat a^{0}\rangle\hat a^{0}$ with $\hat a^{0}=a^{0}/\|a^{0}\|$, and background gene $i$ acquires the deterministic shift
\begin{equation}\label{eq:shift}
\mathrm{shift}_i\approx-\frac{\langle a,a^{0}\rangle}{\|a^{0}\|^{2}}\,a^{0}_i
\end{equation}
in $s_{\ell_1^{*}}u_{\ell_1^{*} i}$.
\end{approximation}
\begin{proof}[Derivation]
The left singular vectors of the noiseless part are obtained by orthonormalizing $a^{0}$ and $a$; the first is essentially $\hat a^{0}$, and the second is proportional to $a-\langle a,\hat a^{0}\rangle\hat a^{0}$. Since $a_i=0$ for background genes, their components are $-\langle a,\hat a^{0}\rangle\hat a^{0}_i$.
\end{proof}
The shift is proportional to the mean expression $a^{0}_i$ of the background gene, so highly expressed background genes are shifted most. Its size depends on whether the effects are balanced. With $\bar\mu$ as in \eqref{eq:mubar} and $\mathrm{rms}(a^{0})=\|a^{0}\|/\sqrt{N}$:
\begin{itemize}
\item If the effects are mostly in one direction and proportional to expression (e.g.\ most signal genes are up-regulated by a common factor), $\langle a,a^{0}\rangle$ is of order $\|a\|\,\|a^{0}\|$, and the shift is proportional to $\sqrt{M}$ for a given effect size. Normality fails and highly expressed background genes are falsely selected.
\item If up- and down-regulation balance, $\langle a,a^{0}\rangle$ is a sum of terms with random signs, of size about $\sqrt{d}\,\bar\mu\sqrt{(1-\pi_0)N}\,\mathrm{rms}(a^{0})$. The shift of background gene $i$ in noise units is then about $\bar\mu\sqrt{1-\pi_0}\,(a^{0}_i/\mathrm{rms}(a^{0}))/\sqrt{N}$, and Assumption~\ref{as:contam} requires
\begin{equation}
\frac{\bar\mu\,R}{\sqrt{N}}\to0,\qquad R=\max_i|R_i|,
\end{equation}
in agreement with Table~\ref{tab:epsdep} for $\rho$ of order $N^{-1/2}$.
This is restrictive for untransformed data, in which a few genes have very large mean expression.
\end{itemize}

\paragraph{Cancellation of signals.} For a signal gene, \eqref{eq:mueffc} subtracts $c\,a^{0}_i$ from $a_i$. The signal is reduced when $a_i$ and $a^{0}_i$ have the same sign and cancelled when $a_i/a^{0}_i$ is close to $c$; a large mean expression alone does not cause cancellation. Small contamination of the background does not exclude cancellation: in the following example the background is not shifted at all while some signals are cancelled completely.

\begin{remark}[cancellation of a signal by the mean expression]\label{rem:cancel}
Assumption~\ref{as:weak} must be imposed on the effective signal, not on $\mu_i$. Let $d=1$, $M=2$, $e_M\perp b$, and divide the genes into three groups: background genes (one half, $a_i=a^{0}_i=0$), signal genes A (one quarter, $a_i=L$, $a^{0}_i=L^{2}$) and signal genes B (one quarter, $a_i=L$, $a^{0}_i=0$), with $L\to\infty$. All signal genes have $\mu_i=L$, and Assumptions~\ref{as:spike}--\ref{as:ng} hold ($\varepsilon_{\mathrm{mix}}=L/(L^{2}-2)\to0$, $\varepsilon_{\mathrm{con}}=0$). The noiseless right singular vector is $\bar v_{\ell_1^{*}}\approx b-e_M/L$, so $\langle b,\bar v_{\ell_1^{*}}\rangle\to1$. The example is built so that (1) wherever the mean expression is non-zero, that is in group A, it is a constant multiple of the signal, $a^{0}_i=L\,a_i$, and (2) the mean expression vanishes for all other genes; the vector $a$ as a whole is not parallel to $a^{0}$ ($\rho\approx0.71$). The selected component is orthogonal to the mean component, so the part of $a$ along $a^{0}$, $c\,a^{0}$ with $c=\langle a,a^{0}\rangle/\|a^{0}\|^{2}=1/L$, is removed; it equals $a_i$ for genes A and $0$ for genes B. Accordingly, with $\tan2\theta=2L/(L^{2}-2)$, i.e.\ $\sin\theta\approx1/L+1/(2L^{3})$ and $\cos\theta\approx1-1/(2L^{2})$, the effective signal of genes A is
\[
\mu_i^{\mathrm{eff}}\approx L\cos\theta-L^{2}\sin\theta=-\frac1L+O(L^{-3}):
\]
the terms of order $L$ cancel exactly and $\mu_i^{\mathrm{eff}}\to0$. In simulations with $N=4\times10^{3}$, $4\times10^{4}$ and $4\times10^{5}$ ($L\approx8$, $14$, $25$), $\langle b,v_{\ell_1^{*}}\rangle\ge0.99$ and the median absolute score of genes B was close to $L$, whereas that of genes A was $0.67$--$0.69$, the median of $|N(0,1)|$ ($0.674$), indistinguishable from the background ($0.67$--$0.68$). In general, the signal of gene $i$ is cancelled not because its mean expression is large but when its ratio $a_i/a^{0}_i$ is close to the overall ratio $c$. In real data, where many genes, including background genes, have large mean expression and up- and down-regulation are mixed, $c$ is very small and such cancellation is rare, as discussed next. With $\mu_i$ in place of $\mu_i^{\mathrm{eff}}$, Assumption~\ref{as:weak} would hold here while the conclusion (iii) fails. In terms of the model, $\mu_i^{\mathrm{eff}}$ stays close to $\mu_i$ when $|a^{0}_i\sin\theta|\ll|a_i|$ for nearly all signal genes; condition (B3), $\rho\to0$, makes $\sin\theta$ small but does not by itself exclude a few signal genes with very large $a^{0}_i$.
\end{remark}

\paragraph{Size in real data.} In most applications both effects are negligible. For log-transformed data with up- and down-regulation mixed, $\langle a,a^{0}\rangle=\sum_{i\in\mathcal{I}_1}a_ia^{0}_i$ is a sum of $n_1$ terms with random signs, which largely cancel, so that the correlation $\rho$ of \eqref{eq:rho} is of order $N^{-1/2}$. For uncentred data $\|a\|\ll\|a^{0}\|$, and the mixing angle is smaller still, $\sin\theta\approx\rho\|a\|/\|a^{0}\|$. The relative difference between $\mu_i^{\mathrm{eff}}$ and $\mu_i$ is then of order $R_i/\sqrt N$, with $R_i=a^{0}_i/\mathrm{rms}(a^{0})$ the relative mean expression of Section~\ref{sec:eps}, so that $\mu_i^{\mathrm{eff}}\approx\mu_i$ except for genes with extremely large mean expression. The effects can be substantial when $a$ is roughly proportional to $a^{0}$ for some signal genes, for example for untransformed data with multiplicative, mostly one-directional effects, or for artefacts such as group-dependent normalization or library size. When all genes are shifted in proportion to $a^{0}$, as for a normalization artefact, the shift is absorbed into the mean component and does not appear in the signal component, which is the desired behaviour. The diagnostics of Table~\ref{tab:diag} detect contamination of the background and a large correlation $\rho$, but not cancellation by itself. A possible check, which we have not validated, is whether the genes expected to carry a signal but not selected have conspicuously large mean expression.

\subsubsection{Numerical illustration}
Table~\ref{tab:sim} shows simulations of the model \eqref{eq:model} with $N=5000$, $\pi_0=0.7$, $d=1$, a contrast $b_r=\pm1/\sqrt{M}$ that divides the columns into two sets of equal size, uncentred genes, BH level $0.01$ and $K=100$, with $10$ replicates per setting. The signal strength is set through the per-entry effect size $\delta$ of Remark~\ref{rem:delta}, so that $\mu=\delta\sqrt{M}$. In the balanced settings the signs of $a_i$ are random; in the one-directional settings $a_i\propto a^{0}_i$ for signal genes. The mean expression is either $a^{0}_i\sim N(8,1)$ (nearly constant across genes) or log-normal, $a^{0}_i=e^{\xi_i}$ with $\xi_i\sim N(2,1)$ (heavy-tailed). The signal component $\ell_1^{*}$ is the component whose right singular vector has the largest overlap with $b$. $\hat\sigma$ is obtained by the maximum-drop rule (Section~\ref{sec:drop}) on a grid of $\tau\in[0.3,1.5]$ with step $0.005$. Normality is tested by a Kolmogorov--Smirnov test of the centred and standardized background scores at level $0.01$. The last column gives the standard deviation of all components of $u_{\ell_1^{*}\cdot}$, i.e.\ the estimate that would be used without optimization (Corollary~\ref{cor:local}). If the theory holds, the background scores multiplied by $s_{\ell_1^{*}}$ have mean $0$ and standard deviation $1$ and pass the test, and $\hat\sigma s_{\ell_1^{*}}\approx1$.

\begin{table}[!htbp]
\centering
\caption{Background scores and estimates of $\sigma_0$ in simulations}\label{tab:sim}
\small
\begin{tabular}{p{0.3\textwidth}cccc>{\centering\arraybackslash}p{0.15\textwidth}c}
\toprule
Setting & $\mu$ & SD & mean & KS rejected & $\hat\sigma s_{\ell_1^{*}}$ & $\sigma_{\mathrm{all}}s_{\ell_1^{*}}$\\
\midrule
balanced, $a^{0}\sim N(8,1)$, $M=200$, $\delta=1$ & 14 & 1.00 & 0.0 & 0/10 & 1.00 (0.93--1.05) & 7.8\\
balanced, $a^{0}\sim N(8,1)$, $M=40$, $\delta=0.5$ & 3.2 & 1.01 & 0.0 & 0/10 & 1.50$^{\dagger}$ & 2.0\\
one-directional, log-normal $a^{0}$, $M=200$, $\delta=0.5$ & 7.1$^{\ddagger}$ & 2.76 & $-2.0$ & 10/10 & 1.49 (1.48--1.50)$^{\dagger}$ & 5.3\\
one-directional, $a^{0}\sim N(8,1)$, $M=200$, $\delta=0.5$ & 7.1$^{\ddagger}$ & 1.04 & $-2.1$ & 0/10 & 1.33 (1.25--1.41) & 3.4\\
balanced, log-normal $a^{0}$, $M=200$, $\delta=0.5$ & 7.1 & 1.00 & 0.0 & 0/10 & 0.99 (0.96--1.03) & 4.0\\
\bottomrule
\multicolumn{7}{p{0.95\textwidth}}{\footnotesize $^{\dagger}$At or next to the upper end of the search interval. Restricting the search to $\tau\le1.2$ gives $0.73$ (0.71--0.77) in the second row. $^{\ddagger}$Value of $\delta\sqrt{M}$; the strength of each gene is proportional to its mean expression.}
\end{tabular}
\par\smallskip{\raggedright\footnotesize $N=5000$, $\pi_0=0.7$, $10$ replicates; means, with ranges for $\hat\sigma s_{\ell_1^{*}}$. All quantities are multiplied by $s_{\ell_1^{*}}$, so that the true value $\sigma_0s_{\ell_1^{*}}\approx1$.\par}
\end{table}

In the first row the theory holds: $\hat\sigma$ is within $\pm7\%$ of $\sigma_0$, whereas the standard deviation of all components overestimates it by a factor of $7.8$, in agreement with Corollary~\ref{cor:local} ($\sqrt{1+(1-\pi_0)\mu^{2}}\approx7.8$). In the second row the background is normal, but $\mu\approx3$ is close to the $z$-value of the BH threshold, so signal and background overlap (Remark~\ref{rem:weak}). The objective then has no clear valley at the true value. The maximum-drop rule picks the spurious valley of Remark~\ref{rem:spurious} at the upper end of the search interval, because $\mu\sigma_0/3\approx1.05\,\sigma_0$ lies below $\sigma_0$ and no admissible search interval exists. Restricting the search gives $\hat\sigma s_{\ell_1^{*}}\approx0.73$, an underestimate by about $27\%$. This is still much closer to the true value than $\sigma_{\mathrm{all}}$, which overestimates by a factor of $2.0$, so estimating $\sigma$ remains preferable. The two errors, however, have opposite consequences: an overestimate makes the selection conservative, whereas an underestimate inflates the scores by $1/0.73\approx1.37$ and increases the number of background genes selected by BH. In this setting the false discovery rate is therefore not controlled, which is why such data lie outside the theory; the criterion $|\mu_i|\gtrsim6$ of Table~\ref{tab:diag} detects this. The third and fourth rows show contamination by the mean expression. With heavy-tailed $a^{0}$ (third row), the shift is proportional to the mean expression of each background gene (correlation $-0.93$ with $a^{0}_i$), the background is no longer normal and $\hat\sigma$ fails. With nearly constant $a^{0}$ (fourth row), the shift is almost the same for all background genes: the background remains normal in shape but has mean $-2.1$ instead of $0$, and $\hat\sigma$ overestimates $\sigma_0$ by about $33\%$ because the $P$-values assume mean zero. In the fifth row, with balanced effects, the overlap $\langle a,a^{0}\rangle$ is only a random fluctuation, and even heavy-tailed mean expression causes no detectable contamination at this size.

\subsubsection{Diagnostics for a given data set}\label{sec:diag}
Table~\ref{tab:diag} lists, for each quantity of Section~\ref{sec:eps}, the value required and a guide for estimating it from an existing analysis; the estimates rest on idealized conditions and are rough guides rather than exact checks. Only the singular values, the optimized $\hat\sigma$ (which approximates $\sigma_0$ within the bias of Theorem~\ref{thm:cap}) and the residual variance
\begin{equation}\label{eq:dhat}
\hat d=\frac{\sum_{\ell>r_0}s_{\ell}^{2}}{(N-r_0)(M-r_0)}
\end{equation}
are needed, where $s_\ell$ are the singular values of the unfolding $X$ and $r_0$ is the number of leading components removed (the mean component, the signal components and any other systematic components); $\hat d$ is the residual variance per entry, discussed further with Prediction~1 in Section~\ref{sec:pred}. For typical data, the bound \eqref{eq:loobound} makes $\varepsilon_{\mathrm{loo}}$ small, so the normality of the background is limited mainly by contamination and by heterogeneous noise variances rather than by the sizes of $N$ and $M$, and the separation of signal genes is limited by $\mu_i$.

\begin{table}[!htbp]
\centering
\caption{Quantities controlling the approximations for a given data set}\label{tab:diag}
\small
\begin{tabular}{p{0.17\textwidth}p{0.2\textwidth}p{0.09\textwidth}p{0.40\textwidth}}
\toprule
Quantity & Definition & Required & How to compute\\
\midrule
signal strength & $\mu_i^{\mathrm{eff}}$, Eq.~\eqref{eq:mueff} & $\gtrsim6$ & $|u_{\ell_1^{*} i}|/\hat\sigma$ for selected genes; this estimate is biased upwards by the selection, and weak signal genes that are not selected are not seen; many values just above $3$ indicate weak signals\\
leave-one-out error & $\varepsilon_{\mathrm{loo}}=Md/s_{\ell_1^{*}}^{2}$ & $\ll1$ & $M\hat\sigma^{2}$; at most about $M/N$\\
spike separation & $\varepsilon_{\mathrm{spk}}$, Eq.~\eqref{eq:eps2} & $\ll1$ & $\hat d(\sqrt N+\sqrt M)^{2}/s_{\ell_1^{*}}^{2}$\\
mean--signal gap & $\varepsilon_{\mathrm{gap}}$, Eq.~\eqref{eq:epsgap} & $\ll1$ & $\hat d(\sqrt N+\sqrt M)^{2}/(s_{\ell_1^{\mathrm{mean}}}-s_{\ell_1^{*}})^{2}$, with $s_{\ell_1^{\mathrm{mean}}}$ the singular value of the mean component; the absolute value of the difference is used if the order is reversed\\
sample-mode spike & $\varepsilon_{\mathrm{side}}$, Eq.~\eqref{eq:eps4} & $\ll1$ & $\hat d\sqrt{NM}/\left(s_{\ell_1^{*}}^{2}-\hat d(N+M)\right)$, using $\|a\|^{2}\approx s_{\ell_1^{*}}^{2}-d(N+M)$ from \eqref{eq:s2}\\
contamination & $\varepsilon_{\mathrm{con}}$, Eq.~\eqref{eq:eps2} & $\ll1$ & correlation between $u_{\ell_1^{*} i}$ and mean expression over non-selected genes (Prediction~2)\\
mixing & $\varepsilon_{\mathrm{mix}}$, Eq.~\eqref{eq:eps3} & $\ll1$ & $\tan\theta\approx|G(\hat\ell_1\ell_2^{\mathrm{mean}}\ell_3^{\mathrm{mean}})|/|G(\hat\ell_1\hat\ell_2\hat\ell_3)|$ from the core tensor (Section~\ref{sec:contam}); $|\theta|\le\varepsilon_{\mathrm{mix}}$\\
variance homogeneity & $\varepsilon_{\mathrm{het}}$, Eq.~\eqref{eq:epshet} & $\ll1$ & from the residual variances $\hat d_i$ of non-selected genes after removing the leading components: $\hat\varepsilon_{\mathrm{het}}^{2}\approx\left(\mathrm{Var}(\hat d_i)/\overline{\hat d}^{\,2}-2/k\right)/(1+2/k)$ with $k=M-r_0$, where the variance and the mean $\overline{\hat d}$ are taken over the same set of genes, which holds in the idealized case $\hat d_i=d_i\chi^{2}_k/k$ (Gaussian noise, known projection); a trend of $\hat d_i$ with mean expression indicates a mean--variance relation\\
non-normality & $\varepsilon_{\mathrm{ng}}$, Eq.~\eqref{eq:eps3} & $\ll1$ & $\hat\kappa_3\max_r|v_{\ell_1^{*}r}|$, with $\hat\kappa_3$ the standardized third absolute moment of the residuals; not needed for Gaussian noise\\
correlation with mean & $\rho$, Eq.~\eqref{eq:rho} & $\ll1$ & $|G(\ell_1^{\mathrm{mean}}\hat\ell_2\hat\ell_3)|/|G(\hat\ell_1\hat\ell_2\hat\ell_3)|$ from the core tensor, with $\ell_1^{\mathrm{mean}}$ the mean component\\
sampling error & SD of $\hat\sigma/\sigma_0$ & -- & Table~\ref{tab:finiteN}, or a simulation with the same $N$\\
\bottomrule
\end{tabular}
\par\smallskip{\raggedright\footnotesize The estimates are guides based on idealized conditions: the projections and the selection of genes are estimated from the same data.\par}
\end{table}

\subsection{The limiting objective and the uniqueness of its minimizer}\label{sec:limit}
From here on we work under two conditions on the scores, which Theorem~\ref{thm:main} derives from the model but which are all that the following results use:
\begin{itemize}
\item[(N)] \emph{Normal background.} The empirical distribution of $u_{\ell_1^{*} i}/\sigma_0$ over background genes converges in probability to $N(0,1)$, uniformly, as $\nu\to\infty$.
\item[(S)] \emph{Separated signal.} For every $c>0$, the fraction of genes that are signal genes with $|u_{\ell_1^{*} i}|/\sigma_0\le c$ tends to zero in probability as $\nu\to\infty$.
\end{itemize}
In this section we take these conditions as exact in the limit, compute what the objective function \eqref{eq:obj} becomes, and study its minimizer. Section~\ref{sec:sep} shows that the empirical objective approaches this limit. Because Sections~\ref{sec:limit}--\ref{sec:drop} use only (N) and (S), their conclusions hold whatever the origin of these conditions.

\subsubsection{Limiting objective}
Under \emph{perfect separation}, all signal genes have $P$-values equal to zero and are selected by BH. Among the background genes, those with the smallest $P$-values are also selected (false positives). As $N\to\infty$ the BH threshold converges to $t^{*}(\tau)$ defined by
\begin{equation}\label{eq:tstar}
t^{*}(\tau)=\alpha\left[(1-\pi_0)+\pi_0F_\tau\left(t^{*}(\tau)\right)\right]
\end{equation}
\cite{GenoveseWasserman2004}, taking the largest solution in $(0,1]$ (the solution is in fact unique, as shown in the proof of Theorem~\ref{thm:consist}); the right-hand side is $\alpha$ times the fraction of genes with $P$-value at most $t$ (all signal genes plus the background genes with $P\le t$). Since the right-hand side is at most $\alpha$, we have $t^{*}\le\alpha$: BH never selects a gene with $P>\alpha$. In the setting used in practice, $\alpha=0.01=1/K$, so the selected genes all lie in the first bin $[0,1/K]$, and only the first bin is affected by the exclusion.

We now compute the limit of the bin counts. Let
\begin{equation}
\begin{gathered}
z_k=z(k/K),\quad k=0,\dots,K\ (z_0=\infty),\\
\beta_k(\tau)=F_\tau\left(\frac kK\right)-F_\tau\left(\frac{k-1}{K}\right),\qquad
E(\tau)=F_\tau\left(t^{*}(\tau)\right).
\end{gathered}
\end{equation}
$\beta_k$ is the probability that a background $P$-value falls into bin $k$, and $E$ is the fraction of background genes excluded by BH. The latter holds because, in the limit, BH selects exactly the genes whose $P$-values are at most the threshold $t^{*}$ and no others; the excluded background genes are therefore those with $P\le t^{*}$, and since $F_\tau$ is the distribution function of background $P$-values, their fraction is $F_\tau(t^{*})$. Dividing the counts by $N$, the law of large numbers gives $h_k/N\to\pi_0m_k$ with
\begin{equation}
m_1=\beta_1-E,\qquad m_k=\beta_k\quad(k=2,\dots,K):
\end{equation}
bin 1 loses the excluded background genes, and the other bins are unaffected. Since $\SD_k(c\,x_k)=c\,\SD_k(x_k)$ for a constant $c>0$, the objective \eqref{eq:obj} at $\sigma=\tau\sigma_0$ satisfies $L_N=\SD_k(h_k/N)\to\SD_k(\pi_0m_k)=\pi_0\SD_k(m_k)$: the counts $h_k$ of the data are replaced by the limiting bin masses $m_k$ per background gene, and the factor $\pi_0$ comes out of the standard deviation. The limiting objective is therefore
\begin{equation}
L(\tau)=\pi_0\,\SD_k(m_k).
\end{equation}
Since $\pi_0$ is a constant factor, the shape of $L$ in $\tau$ is that of $\SD_k(m_k)$. In what follows, $\tau_{\max}$ denotes the local maximum of $L$ for $\tau<1$ and $\tau^{*}$ the minimizer of $L$ to the right of $\tau_{\max}$. That each of them exists and is unique is shown in Theorem~\ref{thm:cap}; until then the names refer to values found by numerical computation. We further define
\begin{equation}
S(\tau)=\sum_{k=1}^{K}\left(\beta_k-\frac1K\right)^{2},\qquad \eta_\tau(z)=z\,\phi(\tau z);
\end{equation}
$S$ measures the unevenness of the histogram of background $P$-values without exclusion, and $\eta_\tau$ appears in the derivative of $F_\tau$.

\subsubsection{The objective without exclusion}
We first ignore the exclusion ($E=0$). Then the objective is a function of $S$ alone.
\begin{lemma}\label{lem:unimodal}
For every $K\ge2$, $S$ is strictly decreasing on $(0,1)$, strictly increasing on $(1,\infty)$, and $S(1)=0$.
\end{lemma}
\begin{proof}
By \eqref{eq:ftau}, and since $z(t)$ decreases in $t$, the density $f_\tau$ is strictly decreasing in $t$ for $\tau<1$ and strictly increasing for $\tau>1$; hence $\beta_k$ is strictly monotone in $k$ in the same direction. From \eqref{eq:Ftau}, $\partial_\tau F_\tau(t)=-2\eta_\tau(z(t))$, so
\begin{equation}\label{eq:betap}
\beta_k'=2\left[\eta_\tau(z_{k-1})-\eta_\tau(z_k)\right],\qquad\sum_{k=1}^{K}\beta_k'=0 ,
\end{equation}
where $'$ denotes $d/d\tau$. As $\eta_\tau$ increases up to $z=1/\tau$ and decreases afterwards, and $z_k$ decreases in $k$, the sequence $\beta_k'$ changes sign exactly once, from negative to positive. We now use an elementary fact: if a sequence $\zeta_k$ changes sign once at $k_0$ from negative to positive, $\sum_k\zeta_k=0$, and $\xi_k$ is decreasing, then $\sum_k\xi_k\zeta_k=\sum_k(\xi_k-\xi_{k_0})\zeta_k\le0$, because each term is a product of a non-negative and a non-positive factor, with strict inequality if $\xi$ is strictly decreasing. Since $S'=2\sum_k(\beta_k-1/K)\beta_k'$ and $\sum_k\beta_k'=0$ by \eqref{eq:betap}, the term with $1/K$ vanishes and $S'=2\sum_k\beta_k\beta_k'$. Applying this to $S'$ gives $S'<0$ for $\tau<1$ and $S'>0$ for $\tau>1$.
\end{proof}
Without exclusion, therefore, the objective decreases steadily towards the correct value $\tau=1$ and increases beyond it; it has no other dip. In practice one observes that the objective decreases again at very small $\sigma$; the lemma shows that this is caused solely by the BH exclusion, which removes almost all genes when $\sigma$ is very small.

\subsubsection{Local analysis near \texorpdfstring{$\tau=1$}{tau=1}}
With exclusion, the variance of the bin masses can be computed directly from the definition of $m_k$. Since $\sum_{k=1}^{K}\beta_k=F_\tau(1)-F_\tau(0)=1$ and only $m_1$ differs from $\beta_1$, by $-E$,
\begin{equation}
\sum_{k=1}^{K}m_k=1-E,\qquad \sum_{k=1}^{K}m_k^{2}=\sum_{k=1}^{K}\beta_k^{2}-2E\beta_1+E^{2}.
\end{equation}
Moreover $S=\sum_k(\beta_k-1/K)^{2}=\sum_k\beta_k^{2}-1/K$. Substituting these into $K\,\mathrm{Var}_k(m)=\sum_km_k^{2}-\frac1K\left(\sum_km_k\right)^{2}$ gives
\begin{equation*}
K\,\mathrm{Var}_k(m)=\left(S+\frac1K\right)-2E\beta_1+E^{2}-\frac{(1-E)^{2}}{K},
\end{equation*}
and collecting the terms in $E$,
\begin{equation}\label{eq:var}
K\,\mathrm{Var}_k(m)=S(\tau)-2E\left(\beta_1-\frac1K\right)+E^{2}\left(1-\frac1K\right).
\end{equation}
At $\tau=1$ the background $P$-values are uniform, $F_1(t)=t$ (Lemma~\ref{lem:pit}). Hence $\beta_1=1/K$, $S=S'=0$ (Lemma~\ref{lem:unimodal}), and $E(1)=F_1(t^{*}(1))=t^{*}(1)$: the fraction of background genes with $P\le t^{*}$ equals $t^{*}$ itself. Substituting $F_1(t)=t$ into \eqref{eq:tstar} gives the linear equation $t=\alpha(1-\pi_0)+\alpha\pi_0t$, whose only solution is
\begin{equation}
E(1)=t^{*}(1)=\frac{\alpha(1-\pi_0)}{1-\alpha\pi_0}.
\end{equation}
Finally, $\beta_1'(1)=-2\eta_1(z_1)$ by the formula for $\beta_k'$ in the proof of Lemma~\ref{lem:unimodal}, with $\eta_1(z_0)=0$. Hence, to leading order in $t^{*}(1)$,
\begin{equation}
\left.\frac{\mathrm{d}}{\mathrm{d}\tau}K\,\mathrm{Var}_k(m)\right|_{\tau=1}\approx4\,t^{*}(1)\,\eta_1(z_1)>0,\qquad
S''(1)=8\sum_{k=1}^{K}\left[\eta_1(z_{k-1})-\eta_1(z_k)\right]^{2}>0 .
\end{equation}
The first expression follows by differentiating \eqref{eq:var}: at $\tau=1$ the terms with $S'$ and with $\beta_1-1/K$ vanish, and what remains is $-2E\beta_1'=4t^{*}(1)\eta_1(z_1)$, up to the term $2EE'(1-1/K)$, which is of second order in $t^{*}(1)$. For the second, differentiating $S=\sum_k(\beta_k-1/K)^{2}$ twice gives
\begin{equation*}
S'=2\sum_{k=1}^{K}\left(\beta_k-\frac1K\right)\beta_k',\qquad
S''=2\sum_{k=1}^{K}\left[(\beta_k')^{2}+\left(\beta_k-\frac1K\right)\beta_k''\right].
\end{equation*}
At $\tau=1$, $\beta_k=1/K$, so the second term vanishes and $S''(1)=2\sum_k\beta_k'(1)^{2}$; substituting $\beta_k'=2[\eta_\tau(z_{k-1})-\eta_\tau(z_k)]$ from the proof of Lemma~\ref{lem:unimodal} gives the stated form. It is positive because not all the differences vanish: for $k=1$, $\eta_1(z_0)=0$ (since $z_0=\infty$) while $\eta_1(z_1)>0$.
The objective is thus still increasing at $\tau=1$, and its minimum lies slightly to the left.
\begin{approximation}\label{prop:local}
For sufficiently small $t^{*}(1)$ the limiting objective is strictly convex in a neighbourhood of $\tau=1$ and has a unique minimizer there, located at
\begin{equation}\label{eq:bias}
\tau^{*}-1\approx-\frac{t^{*}(1)\,\eta_1(z_1)}{2\sum_{k=1}^{K}\left[\eta_1(z_{k-1})-\eta_1(z_k)\right]^{2}} .
\end{equation}
\end{approximation}
\begin{proof}[Derivation]
Let $f(\tau)=K\,\mathrm{Var}_k(m)$. By \eqref{eq:var}, $f''(1)=S''(1)+O(t^{*}(1))$, because the terms containing $E$ are of order $t^{*}(1)$ near $\tau=1$. $S''(1)$ is positive and does not depend on $t^{*}(1)$, so for sufficiently small $t^{*}(1)$, $f''>0$ in a neighbourhood of $\tau=1$, i.e.\ $f$ is strictly convex there and has at most one minimizer in it. At the minimizer, $f'(\tau^{*})=0$. Expanding around $\tau=1$ to first order (one Newton step),
\begin{equation*}
0=f'(\tau^{*})\approx f'(1)+f''(1)(\tau^{*}-1),\qquad\text{so}\qquad \tau^{*}-1\approx-\frac{f'(1)}{f''(1)} .
\end{equation*}
Substituting $f'(1)\approx4t^{*}(1)\eta_1(z_1)$ and $f''(1)\approx S''(1)=8\sum_k[\eta_1(z_{k-1})-\eta_1(z_k)]^{2}$ gives \eqref{eq:bias}. Since $f'(1)>0$ and $f''>0$, the minimizer lies to the left of $\tau=1$ at a distance of order $t^{*}(1)$, hence inside the neighbourhood, which shows its existence.
\end{proof}
The approximation is of first order in $t^{*}(1)$; as Table~\ref{tab:bias} shows, it deviates from the numerical minimizer when $\pi_0$ is small and $t^{*}(1)$ correspondingly large. The rigorous counterpart of this approximation is Theorem~\ref{thm:cap}, which locates $\tau^{*}$ by interval arithmetic.
The bias has a simple origin. The background genes falsely selected by BH are removed from the first bin, which therefore looks depleted. To refill it, the optimization chooses a slightly smaller $\sigma$, which moves background $P$-values towards zero. The bias is proportional to the fraction $t^{*}(1)$ of excluded background genes, underestimates $\sigma$, and vanishes as $\alpha\to0$ or $\pi_0\to1$. Table~\ref{tab:bias} compares \eqref{eq:bias} with the numerical minimizer.

\begin{table}[!htbp]
\centering
\caption{Minimizer and local maximum of the limiting objective}\label{tab:bias}
\begin{tabular}{cccc}
\toprule
$\pi_0$ & $\tau^{*}$, Eq.~\eqref{eq:bias} & $\tau^{*}$, numerical & local maximum $\tau_{\max}$ (numerical)\\
\midrule
0.95 & 0.9981 & 0.998 & 0.48\\
0.9 & 0.9962 & 0.996 & 0.47\\
0.8 & 0.9924 & 0.994 & 0.45\\
0.7 & 0.9885 & 0.991 & 0.44\\
0.5 & 0.9809 & 0.989 & 0.42\\
\bottomrule
\end{tabular}
\par\smallskip{\raggedright\footnotesize $K=100$, $\alpha=0.01$.\par}
\end{table}

\subsubsection{An exact expression for the derivative}
Approximation~\ref{prop:local} is a local statement. To show that there is no other minimizer anywhere, we need the sign of $L'(\tau)$ for all $\tau$. Differentiating \eqref{eq:var} gives
\begin{equation}\label{eq:D}
D(\tau)\equiv\frac K2\frac{\mathrm{d}}{\mathrm{d}\tau}\mathrm{Var}_k(m)=\frac{S'(\tau)}{2}-E\beta_1'-E'\left(\beta_1-\frac1K\right)+EE'\left(1-\frac1K\right),
\end{equation}
where, by implicit differentiation of \eqref{eq:tstar},
\begin{equation}\label{eq:Eprime}
E'=\frac{\partial_\tau F_\tau(t^{*})}{1-\alpha\pi_0 f_\tau(t^{*})} .
\end{equation}
Since $L=\pi_0\sqrt{\mathrm{Var}_k(m)}$, the sign of $D(\tau)$ is the sign of $L'(\tau)$: $D>0$ means that the objective increases with $\tau$, and $D<0$ that it decreases. All terms of $D$ are explicit functions of $\tau$, $\pi_0$ and $t^{*}$.

For $\tau\ge1$ the dependence on $\pi_0$ can be removed by a worst-case bound, so that one function of $\tau$ covers all $\pi_0$ at once.
\begin{proposition}\label{prop:lambda}
Let $\alpha\le1/K$ and $\tau\ge1$. Then $D(\tau)\ge\Lambda(\tau)$, where
\begin{equation}
\Lambda(\tau)=\frac{S'(\tau)}{2}-\frac{|\beta_1'(\tau)|}{1-\alpha}\left(\frac1K-\beta_1(\tau)\right)
\end{equation}
does not depend on $\pi_0$.
\end{proposition}
\begin{proof}
For $\tau\ge1$, $F_\tau(t)\le t$ (Lemma~\ref{lem:pit}), so $t^{*}\le\alpha(1-\pi_0)/(1-\alpha\pi_0)$ and $E\le t^{*}$. Let $w^{*}=z(t^{*})\ge z_1$. Since $z_1\ge1\ge1/\tau$ and $\eta_\tau$ decreases beyond $1/\tau$, $|\partial_\tau F_\tau(t^{*})|=2\eta_\tau(w^{*})\le2\eta_\tau(z_1)=|\beta_1'|$. Moreover $f_\tau(t^{*})\le1$, because $\log\tau\le(\tau^{2}-1)/2\le(\tau^{2}-1)w^{*2}/2$. Hence $|E'|\le|\beta_1'|/(1-\alpha\pi_0)$. The terms of \eqref{eq:D} that contain $E$ satisfy $E|\beta_1'|-E|E'|(1-1/K)\ge E|\beta_1'|[1-(1-1/K)/(1-\alpha\pi_0)]\ge0$ for $\alpha\le1/K$. The remaining term is bounded using $|E'|\le|\beta_1'|/(1-\alpha)$.
\end{proof}
$\Lambda$ is a closed-form function of $\tau$ alone. For $K=100$ and $\alpha=0.01$ we evaluated $\Lambda$ by ordinary floating-point arithmetic (not by interval arithmetic) on a grid of $5300$ points in $(1,20]$ ($300$ logarithmically spaced points in $(1,1.1]$ and $5000$ equally spaced points in $[1.1,20]$); $\Lambda(\tau)>0$ at every grid point, the minimum of $\Lambda(\tau)/(\tau-1)$ being $1.8\times10^{-4}$ (at $\tau=20$; script \texttt{lambda\_check.py} in the supplementary code). Moreover $\Lambda'(1)=S''(1)/2-|\beta_1'(1)|^{2}/(1-\alpha)=0.0196-0.0056>0$. $\Lambda$ is a lower bound of $D$ only for $\tau\ge1$: its proof uses $F_\tau(t)\le t$ and the decrease of $\eta_\tau$ beyond $1/\tau$, both of which fail for $\tau<1$. For $\tau<1$ no bound of this kind can be positive, because $D<0$ on $(\tau_{\max},\tau^{*})$, where the objective decreases, and the location of $\tau^{*}$ depends on $\pi_0$. The sign of $D$ for $\tau<1$ is therefore established $\pi_0$ by $\pi_0$ in Theorem~\ref{thm:cap}, by interval arithmetic (Tables~S1 and S2 in Appendix~\ref{app:boxes}); Theorem~\ref{thm:cap} also covers $(1,20]$ rigorously, so that the numerical check of $\Lambda$ serves only as a supplementary argument independent of $\pi_0$.

\subsubsection{Computer-assisted proof of uniqueness}
\begin{theorem}[computer-assisted]\label{thm:cap}
Let $K=100$, $\alpha=0.01$ and $\pi_0\in[0.1,0.99]$. On $\tau\in[0.05,20]$ the limiting objective $L$ has exactly two critical points: a local maximum $\tau_{\max}(\pi_0)\in[0.4078,0.4985]$ and a local minimum $\tau^{*}(\pi_0)\in[0.98923,0.99963]$. $L$ is strictly increasing on $[0.05,\tau_{\max}]$, strictly decreasing on $[\tau_{\max},\tau^{*}]$ and strictly increasing on $[\tau^{*},20]$. In particular $\tau^{*}$ is the unique minimizer of $L$ on $[\tau_{\max},20]$, and in the limit of large data sets
\begin{equation}
0.98923\le\frac{\hat\sigma}{\sigma_0}\le0.99963 .
\end{equation}
\end{theorem}
The objective therefore has the shape ``rise, peak at $\tau_{\max}\approx0.45$, fall, valley at $\tau^{*}\approx0.99$, rise''. The fall to the left of the peak is the effect of the exclusion discussed after Lemma~\ref{lem:unimodal}; the valley is at the correct value up to a bias of at most $1.1\%$.

\paragraph{Method.} The proof uses interval arithmetic \cite{Tucker2011} with the Arb library \cite{Johansson2017} (python-flint 0.9, 96-bit precision). In interval arithmetic each number is replaced by an interval (a ``ball'') that is guaranteed to contain its exact value, and every operation returns an interval guaranteed to contain all possible results; rounding errors are included. If the interval computed for $D$ on a set of parameter values does not contain zero, the sign of $D$ on that whole set is proved. The parameter plane $(\tau,\pi_0)$ is covered by small rectangles (``boxes''), and the sign of $D$ is established on each box as follows.
\begin{enumerate}
\item The bin edges $z_k=\sqrt2\,\mathrm{erfc}^{-1}(k/K)$ are computed as rigorous enclosures.
\item The BH threshold is handled in $z$-units: $w^{*}=z(t^{*})$ is the smallest root, on $w\ge z(\alpha)$, of
\begin{equation}
\gamma(w)=\alpha\left[\frac{1-\pi_0}{2}+\pi_0\Phi(-\tau w)\right]-\Phi(-w),
\end{equation}
which is \eqref{eq:tstar} rewritten with $t=2\Phi(-w)$. There are no roots below $z(\alpha)$, because $t>\alpha$ implies $t>\alpha[(1-\pi_0)+\pi_0F_\tau(t)]$. For each endpoint of a box we prove $\gamma(w_{\mathrm{hi}})>0$ and $\gamma<0$ on $[z(\alpha),w_{\mathrm{lo}}]$ (by bisection into intervals), which shows $w^{*}\in[w_{\mathrm{lo}},w_{\mathrm{hi}}]$.
\item By \eqref{eq:tstar}, $t^{*}$ is non-increasing both in $\tau$ (because $F_\tau$ decreases in $\tau$) and in $\pi_0$ (because $(1-\pi_0)+\pi_0F_\tau=1-\pi_0(1-F_\tau)$ decreases in $\pi_0$). Hence $w^{*}$ is non-decreasing in both, and its range on a box $[\tau_a,\tau_b]\times[\pi_a,\pi_b]$ is $[w^{*}(\tau_a,\pi_a),w^{*}(\tau_b,\pi_b)]$; only the two corners need to be computed.
\item Implicit differentiation gives
\begin{equation}
w^{*\prime}=\frac{\alpha\pi_0\,\phi(\tau w^{*})\,w^{*}}{\phi(w^{*})-\alpha\pi_0\tau\,\phi(\tau w^{*})},\qquad
E=2\Phi(-\tau w^{*}),\qquad E'=-2\phi(\tau w^{*})\left(w^{*}+\tau w^{*\prime}\right),
\end{equation}
so $D$ in \eqref{eq:D} is an explicit function of $(\tau,\pi_0,w^{*},w^{*\prime})$ and can be evaluated in interval arithmetic on each box.
\item Near a zero of $D$ the sign of $D$ cannot be decided. There we decide the sign of the derivative $D'$ instead, computed exactly by forward-mode automatic differentiation (arithmetic with pairs ``value, derivative'' propagated through every operation, again in intervals; $w^{*\prime\prime}$ is obtained by the chain rule). If neither sign can be decided, the box is bisected.
\item On a run of consecutive boxes on which $D'$ has a constant sign, $D$ is strictly monotone and has exactly one zero if the signs of $D$ at the two ends of the run differ, and none otherwise. Counting these zeros counts the critical points of $L$. Finally, each zero is enclosed by bisection with point evaluations on sub-boxes of $\pi_0$ of width $0.001$.
\end{enumerate}
The range $\tau\in[0.3,20]$ was covered with $122$ boxes in $\pi_0$ (width $0.01$ for $\pi_0<0.84$, $0.005$ for $0.84\le\pi_0<0.90$ and $0.0025$ for $\pi_0\ge0.90$) and a total of $387{,}532$ boxes in $\tau$; the range $\tau\in[0.05,0.3]$ with $89$ boxes in $\pi_0$ and $122{,}870$ boxes in $\tau$. The sign was decided on every box. Tables~S1 and S2 in Appendix~\ref{app:boxes} list the results box by box. As a by-product, the computation verifies that the denominator of $w^{*\prime}$, i.e.\ the slope $1-\alpha\pi_0f_\tau(t^{*})$ with which the two sides of \eqref{eq:tstar} cross, is positive; this is used in Section~\ref{sec:sep}.

\begin{remark}
A seemingly natural approach, combining the monotonicity $m_1\ge m_2$ with the inequality $\sum_k\xi_k\zeta_k\le\xi_{k_0}\sum_k\zeta_k$ used in Lemma~\ref{lem:unimodal}, does not work. Near $\tau=1$ the exclusion removes about $30\%$ of the first bin (for $\pi_0=0.7$), so that $m_1<m_2$, and the resulting sufficient condition $m_{k_0}\le\bar m$ fails for $\tau<1$. This is why the computer-assisted argument is used.
\end{remark}

\subsection{Separation of signal genes and consistency}\label{sec:sep}
Section~\ref{sec:limit} assumed perfect separation. We now show that the empirical objective computed from data approaches the limiting objective, so that its minimizer approaches $\tau^{*}$, and we quantify what happens for finite $N$ and finite signal strength.

\begin{theorem}\label{thm:consist}
Consider a sequence of data sets satisfying \eqref{eq:seq}, $n_0/N\to\pi_0$, and conditions (N) and (S) (for instance under the assumptions of Theorem~\ref{thm:main}), and let $K=100$, $\alpha=0.01$ and $\pi_0\in[0.1,0.99]$. Then, as $\nu\to\infty$, $\sup_{\tau\in[\tau_L,\tau_U]}|L_N(\tau\sigma_0)-L(\tau)|\to0$ in probability for every $[\tau_L,\tau_U]\subset[0.05,20]$, and the minimizer $\hat\sigma$ of $L_N$ over $[\tau_{\max}\sigma_0,\tau_U\sigma_0]$, $\tau_U\in(1,20]$, satisfies $\hat\sigma/\sigma_0\to\tau^{*}$ in probability.
\end{theorem}
\begin{proof}
Write $x_i=s_{\ell_1^{*}}u_{\ell_1^{*} i}/\sqrt d=u_{\ell_1^{*} i}/\sigma_0$ for the standardized scores, so that at $\sigma=\tau\sigma_0$ the $P$-value of gene $i$ is $P_i=2\Phi(-|x_i|/\tau)$, and $P_i\le t$ if and only if $|x_i|\ge\tau z(t)$. Let $n_1=N-n_0$ and $\pi_{0N}=n_0/N\to\pi_0$. Throughout, $\tau$ ranges over $[\tau_L,\tau_U]\subset[0.05,20]$, and ``uniformly'' means uniformly in $\tau$ and, where it appears, in $t$.

\emph{Step 1: empirical distribution functions.} Let $F^{0}_N(t;\tau)$ and $H_N(t;\tau)$ be the fractions of background genes and of signal genes with $P_i\le t$. Condition (N) states that the empirical distribution function of $x_i$ over background genes converges uniformly to $\Phi$; hence that of $|x_i|$ converges uniformly to $2\Phi(y)-1$, and since $F^{0}_N(t;\tau)$ depends on $(t,\tau)$ only through $y=\tau z(t)$,
\[
\sup_{t\in[0,1],\,\tau}\left|F^{0}_N(t;\tau)-F_\tau(t)\right|\to0
\]
in probability, with $F_\tau(t)=2\Phi(-\tau z(t))$ as in \eqref{eq:Ftau}. For signal genes, fix $t_0>0$. For $t\ge t_0$, a signal gene with $P_i>t$ has $|x_i|<\tau z(t)\le\tau_Uz(t_0)$; by condition (S) the fraction of all genes that are such signal genes tends to zero, and since $n_1/N\to1-\pi_0\ge0.01$,
\[
\sup_{t\ge t_0,\,\tau}\left(1-H_N(t;\tau)\right)\to0 .
\]
Hence the empirical distribution function of all $P$-values, $\Gamma_N=\pi_{0N}F^{0}_N+(1-\pi_{0N})H_N$, satisfies $\sup_{t\ge t_0,\tau}|\Gamma_N(t;\tau)-\Gamma(t;\tau)|\to0$ with $\Gamma(t;\tau)=\pi_0F_\tau(t)+1-\pi_0$.

\emph{Step 2: the BH threshold.} The BH procedure applied to $N$ $P$-values with empirical distribution function $\Gamma_N$ selects exactly the genes with $P_i\le\hat t$, where $\hat t=\sup\{t\in(0,1]:\Gamma_N(t)\ge t/\alpha\}$, with $\hat t=0$ if no such $t$ exists (the standard characterization of the BH threshold \cite{GenoveseWasserman2002}). Let $g(t;\tau)=\Gamma(t;\tau)-t/\alpha$; its largest zero is $t^{*}(\tau)$ of \eqref{eq:tstar}. Since $\Gamma\ge1-\pi_0$, $t^{*}\ge\underline t:=\alpha(1-\pi_0)\ge10^{-4}$; since $\Gamma\le1$, $t^{*}\le\alpha$. We take $t_0=\underline t/2$.

The crossing at $t^{*}$ is transversal: $\partial_tg(t^{*};\tau)=\pi_0f_\tau(t^{*})-1/\alpha<0$. With $w=z(t^{*})$, $f_\tau(t^{*})=\tau\phi(\tau w)/\phi(w)$ by \eqref{eq:ftau}. If $\tau\ge1$, $\pi_0f_\tau(t^{*})\le\pi_0\tau\le20<1/\alpha$. If $\tau<1$, the defining equation gives $t^{*}\ge\alpha\pi_0F_\tau(t^{*})$, i.e.\ $\Phi(-w)\ge\alpha\pi_0\Phi(-\tau w)$; the Mills ratio $R(x)=\Phi(-x)/\phi(x)$ is decreasing for $x\ge0$, so $R(\tau w)>R(w)>\tau R(w)$, i.e.\ $\tau\phi(\tau w)\Phi(-w)<\phi(w)\Phi(-\tau w)$, and therefore $\pi_0f_\tau(t^{*})=\pi_0\tau\phi(\tau w)/\phi(w)<\pi_0\Phi(-\tau w)/\Phi(-w)\le1/\alpha$. This argument uses only that $t^{*}$ is a zero of $g(\cdot;\tau)$, so $\partial_tg<0$ at every zero in $(0,1]$. Since $g(t;\tau)\to1-\pi_0>0$ as $t\to0$, $g$ can cross zero only downwards and therefore has exactly one zero, which is $t^{*}(\tau)$. The implicit function theorem thus applies to $t^{*}(\tau)$ itself and shows that it is continuous in $\tau$.

Fix $0<\delta<t_0$. Transversality gives two margins, uniform in $\tau$: (a) $\sup\{g(t;\tau):t\ge t^{*}(\tau)+\delta\}<0$; (b) $\inf_\tau\max\{g(t;\tau):t\in[t^{*}(\tau)-\delta,t^{*}(\tau)]\}>0$. Both follow by compactness from the continuity of $g$ in $(t,\tau)$: (a) for each $\tau$ because $t^{*}$ is the largest zero and $g(1;\tau)=1-1/\alpha<0$, and (b) because $g$ changes sign at $t^{*}$; the continuity of $t^{*}(\tau)$ shown above makes the margins uniform. If $\sup_{t\ge t_0,\tau}|\Gamma_N-\Gamma|$ is smaller than both margins, then $\hat t(\tau)\in[t^{*}(\tau)-\delta,t^{*}(\tau)+\delta]$ for all $\tau$. By Step~1 this holds with probability tending to one, and since $\delta$ can be taken arbitrarily small, $\sup_\tau|\hat t(\tau)-t^{*}(\tau)|\to0$ in probability. Since $t^{*}\ge2t_0$ and $\delta<t_0$, also $\hat t>t_0$ with probability tending to one.

\emph{Step 3: the histogram.} Since $\hat t\le\alpha=1/K$, the selected genes lie in the first bin, and the counts of non-selected genes are, for $k\ge2$,
\[
\frac{h_k}{N}=\pi_{0N}\left[F^{0}_N\!\left(\tfrac kK\right)-F^{0}_N\!\left(\tfrac{k-1}K\right)\right]+(1-\pi_{0N})\left[H_N\!\left(\tfrac kK\right)-H_N\!\left(\tfrac{k-1}K\right)\right],
\]
and for $k=1$ the same expression with $F^{0}_N(0)$ and $H_N(0)$ replaced by $F^{0}_N(\hat t)$ and $H_N(\hat t)$. The signal parts are at most $1-H_N(t_0)\to0$ uniformly, by Step~1, on the event $\hat t\ge t_0$. For the background parts, Step~1 and the uniform continuity of $F_\tau(t)$ in $(t,\tau)$ give $F^{0}_N(\hat t)\to F_\tau(t^{*})=E(\tau)$ uniformly. Hence $h_k/N\to\pi_0m_k(\tau)$ uniformly in $\tau$ for each $k$, with $m_k$ as in Section~\ref{sec:limit}. Since the standard deviation over $K$ values satisfies $|\SD_k(x_k)-\SD_k(y_k)|\le\max_k|x_k-y_k|$,
\[
\sup_\tau\left|L_N(\tau\sigma_0)-L(\tau)\right|\le\max_k\sup_\tau\left|\frac{h_k}N-\pi_0m_k\right|\to0
\]
in probability.

\emph{Step 4: the minimizer.} By Theorem~\ref{thm:cap}, on $[\tau_{\max},\tau_U]$ the continuous function $L$ has the unique minimizer $\tau^{*}$. For $\delta>0$, compactness gives $\epsilon=\min\{L(\tau):\tau\in[\tau_{\max},\tau_U],|\tau-\tau^{*}|\ge\delta\}-L(\tau^{*})>0$. On the event $\sup_\tau|L_N(\tau\sigma_0)-L(\tau)|<\epsilon/2$, which has probability tending to one, every minimizer of $L_N(\tau\sigma_0)$ over $[\tau_{\max},\tau_U]$ (one exists because $L_N$ takes finitely many values) lies within $\delta$ of $\tau^{*}$. Hence $\hat\sigma/\sigma_0\to\tau^{*}$ in probability \cite{vanderVaart1998}.
\end{proof}

Theorem~\ref{thm:consist} does not say that $\hat\sigma$ is a consistent estimator of $\sigma_0$: it converges to $\tau^{*}\sigma_0$ with $\tau^{*}<1$, i.e.\ it has an asymptotic relative bias of at most $1.1\%$. Because $\tau^{*}(\pi_0)$ is known (Theorem~\ref{thm:cap} and Table~S1 in Appendix~\ref{app:boxes}), the bias can be removed.
\begin{corollary}[bias correction]\label{cor:biascorr}
Under the assumptions of Theorem~\ref{thm:consist}, if $\hat\pi_0\to\pi_0$ in probability as $\nu\to\infty$, then $\hat\sigma/\tau^{*}(\hat\pi_0)\to\sigma_0$ in the sense that $\hat\sigma/(\tau^{*}(\hat\pi_0)\sigma_0)\to1$ in probability.
\end{corollary}
\begin{proof}
$\tau^{*}(\pi_0)$ is the zero of $D$, which depends smoothly on $\pi_0$, and $D'>0$ there (Theorem~\ref{thm:cap}); by the implicit function theorem $\tau^{*}$ is continuous in $\pi_0$, and the claim follows from Theorem~\ref{thm:consist}.
\end{proof}
A simple choice is $\hat\pi_0=1-|\mathcal{S}(\hat\sigma)|/N$. It is not exactly consistent, because BH also selects a fraction of order $\alpha$ of background genes, but since $\tau^{*}$ varies only between $0.989$ and $1.000$ over $\pi_0\in[0.1,0.99]$ (Table~S1 in Appendix~\ref{app:boxes}), the resulting error in $\tau^{*}$ is negligible.

\paragraph{Finite $N$.} For a given data set, $\hat\sigma/\sigma_0$ differs from $\tau^{*}$ by the random fluctuation of the histogram counts. Table~\ref{tab:finiteN} shows simulations in which background scores are $N(0,1)$, signal genes have strength $\mu=8$ with random signs, $\hat\sigma$ is chosen by the maximum-drop rule (Section~\ref{sec:drop}) on a grid of $\tau\in[0.3,1.5]$ with step $0.0025$, $K=100$ and $\alpha=0.01$. The mean agrees with $\tau^{*}$ of Theorem~\ref{thm:cap}, and in these simulations the standard deviation decreases roughly as $N^{-1/2}$. Conditions (N) and (S) do not specify rates of convergence, so this rate is an observation for this setting, not a consequence of the theorems. For a given data set the error therefore decomposes as
\begin{equation}
\frac{\hat\sigma}{\sigma_0}-1=\underbrace{(\tau^{*}-1)}_{\text{deterministic bias},\ |\cdot|\le1.1\%}+\underbrace{e_N}_{\text{sampling fluctuation},\ \mathrm{SD}\approx0.5\text{--}3.6\%} .
\end{equation}
\begin{table}[!htbp]
\centering
\caption{Mean $\pm$ standard deviation of $\hat\sigma/\sigma_0$ for finite $N$}\label{tab:finiteN}
\begin{tabular}{cccc}
\toprule
$N$ & $\pi_0=0.9$ & $\pi_0=0.7$ & replicates\\
\midrule
$2\times10^{3}$ & $0.9960\pm0.0251$ & $0.9869\pm0.0359$ & 100\\
$5\times10^{3}$ & $0.9958\pm0.0189$ & $0.9942\pm0.0209$ & 60\\
$2\times10^{4}$ & $0.9973\pm0.0077$ & $0.9907\pm0.0125$ & 40\\
$10^{5}$ & $0.9973\pm0.0056$ & $0.9925\pm0.0053$ & 14\\
\midrule
$\infty$ ($\tau^{*}$) & $0.9965$ & $0.9913$ & --\\
\bottomrule
\end{tabular}
\par\smallskip{\raggedright\footnotesize Signal strength $\mu=8$; $K=100$, $\alpha=0.01$.\par}
\end{table}

\paragraph{Finite signal strength.} How strong must the signals be for the limit to apply? Let all signal genes have standardized score $\mu s_i+g_i$ with random signs $s_i=\pm1$ and $g_i\sim N(0,1)$, and compute the minimizer of the exact limiting objective, now including the signal genes that are not selected and therefore remain in the histogram (Table~\ref{tab:finite}).
\begin{table}[!htbp]
\centering
\caption{Minimizer $\tau^{*}=\hat\sigma/\sigma_0$ of the limiting objective for finite signal strength}\label{tab:finite}
\begin{tabular}{ccccc}
\toprule
$\mu$ & $\pi_0=0.5$ & $\pi_0=0.7$ & $\pi_0=0.9$ & signal genes not selected ($\pi_0=0.7$)\\
\midrule
3 & fails & 0.712 & 1.420 & 19\%\\
4 & 0.856 & 0.911 & 0.984 & 10\%\\
5 & 1.013 & 1.009 & 1.009 & 2.3\%\\
6 & 0.992 & 0.993 & 0.998 & 0.11\%\\
8 & 0.989 & 0.991 & 0.996 & $2\times10^{-5}\%$\\
$\infty$ & 0.989 & 0.991 & 0.996 & 0\\
\bottomrule
\end{tabular}
\par\smallskip{\raggedright\footnotesize $K=100$, $\alpha=0.01$. ``fails'': the minimizer lies at the boundary of the search interval $[0.6,1.5]$.\par}
\end{table}
``Fails'' means that the minimizer reached the boundary of the search interval $[0.6,1.5]$. For $\mu\ge6$ the minimizer agrees with the perfectly separated value within $0.3\%$. A practical guideline is therefore $|\mu_i|\gtrsim6$, which can be checked with the estimates $|u_{\ell_1^{*} i}|/\hat\sigma$.

\paragraph{Why finite signal strength biases $\hat\sigma$ in either direction.} Table~\ref{tab:finite} shows both underestimates and overestimates of $\sigma_0$. Two steps explain where the overestimates come from. First, a signal gene can have a score below the selection threshold. Its standardized score is about $\mu+g$ with noise $g\sim N(0,1)$; for $\mu=5$, a noise value $g\approx-2.5$ gives a score of about $2.5$ and a $P$-value of about $0.012$, above the BH threshold $t^{*}\approx0.003$, so the gene is not selected although it carries a signal (a false negative). The probability of this is that of $g<z(t^{*})-\mu\approx2.97-\mu$: about $2.1\%$ for $\mu=5$, about $15\%$ for $\mu=4$, and zero for $\mu=\infty$. Second, the histogram is built from all genes that are not selected (Section~\ref{sec:method}), because it is not known which genes are background. The false-negative signal genes are therefore counted together with the background genes. Their $P$-values are mostly just above the threshold, so they accumulate in the first few bins and create an excess near $P=0$. By Lemma~\ref{lem:pit}, such an excess is reduced by a larger $\sigma$, which moves $P$-values towards $1$; this pushes $\hat\sigma$ upwards.

Two effects push in the opposite direction. One is the depletion of the first bin by the background genes excluded by BH (Approximation~\ref{prop:local}). The other is that a smaller $\sigma$ makes the scores of signal genes relatively larger, so that more of them are selected by BH and removed from the histogram. Table~\ref{tab:leak} shows, for $\pi_0=0.7$, the fraction of signal genes that remain in the histogram, computed from the limiting objective.

\begin{table}[!htbp]
\centering
\caption{Fraction of signal genes left in the histogram}\label{tab:leak}
\begin{tabular}{ccccc}
\toprule
$\mu$ & $\tau=0.90$ & $\tau=0.95$ & $\tau=1.00$ & $\tau=1.05$\\
\midrule
4 & 9.6\% & 12.6\% & 16.4\% & 20.9\%\\
5 & 1.0\% & 1.5\% & 2.1\% & 3.0\%\\
\bottomrule
\end{tabular}
\par\smallskip{\raggedright\footnotesize Signal genes not selected by BH, as a function of $\tau=\sigma/\sigma_0$ ($\pi_0=0.7$, $K=100$, $\alpha=0.01$, limiting objective).\par}
\end{table}

The balance of these effects explains Table~\ref{tab:finite}. For $\mu\ge6$ almost no signal genes remain, only the exclusion of background genes acts, and $\sigma$ is slightly underestimated ($\tau^{*}\approx0.99$). For $\mu=5$ the remaining signal genes make up about $0.6\%$ of all genes, more than the excluded background genes (about $0.2\%$), and their number changes little when $\sigma$ is varied; offsetting the excess by a larger $\sigma$ wins, and $\sigma$ is slightly overestimated ($\tau^{*}\approx1.01$). For $\mu=4$ the remaining signal genes make up about $5\%$ of all genes, and lowering $\tau$ from $1$ to $0.9$ removes about $40\%$ of them; absorbing them into the BH selection lowers the objective more than offsetting the excess, and $\sigma$ is underestimated ($\tau^{*}\approx0.91$). For $\mu=3$ and $\pi_0=0.9$, BH selects almost nothing and all signal genes remain; with no possibility of absorption, the excess is offset by a much larger $\sigma$ ($\tau^{*}\approx1.42$). Thus a few false negatives bias $\hat\sigma$ upwards and many bias it downwards, depending on the signal strength and on $\pi_0$.

\begin{remark}[spurious valley]\label{rem:spurious}
For finite $\mu$ a second local maximum and a second local minimum may appear near $\tau\approx\mu/3$. The reason is that for large $\sigma$ the $P$-values of signal genes exceed the BH threshold, so these genes are no longer selected and enter the histogram. For $\mu=4$--$6$ the second minimum was $2$ to $150$ times higher than the true minimum, so the global minimizer was unchanged. Nevertheless the upper end of the search interval should be kept below about $\mu\sigma_0/3$. As $\mu\to\infty$ this structure moves to infinity, in agreement with Theorem~\ref{thm:consist}. Two consequences should be noted. First, Theorem~\ref{thm:drop} requires $\sigma_U\ge\sigma_0$, so an admissible search interval exists only if $\mu/3>1$, and in practice with some margin; for $\mu\lesssim3$ no interval satisfies both requirements. Second, the maximum-drop rule does not automatically avoid the spurious valley. That the second minimum lies above the true minimum protects the global minimizer, but the rule compares drops, and the drop after the second maximum can exceed that into the true valley, because signal genes, a sizeable fraction of all genes, enter the histogram there together, whereas the true valley is only a shallow distortion caused by excluded background genes. In the second row of Table~\ref{tab:sim} ($\mu\approx3.2$), averaged over the $10$ replicates, the drop beyond the second maximum ($\tau\ge1$) was $4.8\times10^{-3}$, about five times that into the true valley ($\tau\le1$; $1.0\times10^{-3}$).
\end{remark}

\subsection{The maximum-drop rule}\label{sec:drop}
In practice $\hat\sigma$ is taken at the valley to the right of the peak of $L_N$, and the region where $L_N$ decreases again at small $\sigma$ is ignored. Theorem~\ref{thm:consist} assumed that the search starts at the peak $\tau_{\max}\sigma_0$, which is unknown. The maximum-drop rule weakens this assumption. The search no longer has to start at the peak; the ends of the search interval may lie anywhere in wide ranges measured in units of $\sigma_0$ (Theorem~\ref{thm:drop}). The value of $\sigma_0$ need not be known, but a rough idea of its magnitude is still required.
\begin{definition}[maximum-drop rule]
For two points $\sigma_L\le\sigma'\le\sigma''\le\sigma_U$ of a search interval $[\sigma_L,\sigma_U]$, let
\[
\Delta_N(\sigma',\sigma'')=L_N(\sigma')-L_N(\sigma'')
\]
be the drop of the objective from the left point to the right point. The left point of the pair maximizing $\Delta_N$ is denoted by $\hat\sigma_{\max}$ (the peak), and the right point by $\hat\sigma$ (the bottom of the valley, the estimate of $\sigma$):
\begin{equation}
(\hat\sigma_{\max},\hat\sigma)=\argmax_{\sigma_L\le\sigma'\le\sigma''\le\sigma_U}\Delta_N(\sigma',\sigma'') .
\end{equation}
Since $L_N$ is a step function, any pair attaining the supremum up to $o_P(1)$ may be taken.
\end{definition}
The rule looks for the largest drop of the objective when moving from left to right. For the maximizing pair, $\hat\sigma_{\max}$ maximizes $L_N$ on $[\sigma_L,\hat\sigma]$ and $\hat\sigma$ minimizes $L_N$ on $[\hat\sigma_{\max},\sigma_U]$, so the rule is exactly ``find the peak and take the deepest valley to its right''. The decrease of $L_N$ towards small $\sigma$ is, read from left to right, an increase, and is therefore never selected.

\begin{theorem}\label{thm:drop}
Under the assumptions of Theorem~\ref{thm:consist}, let $0.05\,\sigma_0\le\sigma_L<0.9892\,\sigma_0$ and $\sigma_0\le\sigma_U\le20\,\sigma_0$ for all sufficiently large $\nu$, with probability tending to one. If $\sigma_L<\tau_{\max}\sigma_0$, then, as $\nu\to\infty$,
\begin{equation}
\left(\frac{\hat\sigma_{\max}}{\sigma_0},\ \frac{\hat\sigma}{\sigma_0}\right)\xrightarrow{P}\left(\tau_{\max},\ \tau^{*}\right);
\end{equation}
if $\sigma_L\ge\tau_{\max}\sigma_0$, then $\hat\sigma_{\max}/\sigma_L\to1$ in probability and still $\hat\sigma/\sigma_0\to\tau^{*}$.
\end{theorem}
The conditions on $\sigma_L$ and $\sigma_U$ are mild: the search interval must start below the valley and end above the true value, and may extend from $0.05\sigma_0$ to $20\sigma_0$. They are, however, stated in units of $\sigma_0$, so they are not guaranteed to hold without any knowledge of $\sigma_0$. What the theorem establishes is robustness: the lower end may be off by a factor of about $20$ below $\sigma_0$ and the upper end by a factor of about $20$ above it, so the search interval may be chosen wide as long as the order of magnitude of $\sigma_0$ is known. We do not show that an interval computed from the data satisfies these conditions.
\begin{proof}
Write $\tau_L=\sigma_L/\sigma_0$, $\tau_U=\sigma_U/\sigma_0$ and $\Delta(\tau',\tau'')=L(\tau')-L(\tau'')$ on $T=\{\tau_L\le\tau'\le\tau''\le\tau_U\}$, the limit of the drop $\Delta_N(\tau'\sigma_0,\tau''\sigma_0)$, which we write briefly as $\Delta_N(\tau',\tau'')$; consider first $\tau_L<\tau_{\max}$. By Theorem~\ref{thm:cap}, $L$ increases on $[\tau_L,\tau_{\max}]$, decreases on $[\tau_{\max},\tau^{*}]$ and increases on $[\tau^{*},\tau_U]$, and $\tau^{*}<1\le\tau_U$. Let $\Delta^{*}=L(\tau_{\max})-L(\tau^{*})>0$. If $\tau''\le\tau_{\max}$, then $\Delta(\tau',\tau'')\le0$ because $L$ increases there. If $\tau_{\max}<\tau''\le\tau^{*}$, then $\Delta(\tau',\tau'')\le L(\tau_{\max})-L(\tau'')\le\Delta^{*}$, with equality only for $(\tau',\tau'')=(\tau_{\max},\tau^{*})$. If $\tau''>\tau^{*}$, then $\max_{\tau'\le\tau''}L(\tau')=\max\{L(\tau_{\max}),L(\tau'')\}$ and $L(\tau'')>L(\tau^{*})$, so $\Delta(\tau',\tau'')<\Delta^{*}$. Thus $(\tau_{\max},\tau^{*})$ is the unique maximizer of $\Delta$ on $T$; the case $\tau_L\ge\tau_{\max}$ is identical with $\tau_{\max}$ replaced by $\tau_L$. Next, $\sup_T|\Delta_N-\Delta|\le2\sup_{[\tau_L,\tau_U]}|L_N-L|\to0$ by Theorem~\ref{thm:consist}. Since $\Delta$ is continuous on the compact set $T$ and has a unique maximizer, for every neighbourhood $\mathcal{U}$ of the maximizer there is $\lambda>0$ with $\sup_{T\setminus\mathcal{U}}\Delta\le\Delta^{*}-\lambda$, and the (near-)maximizer of $\Delta_N$ lies in $\mathcal{U}$ with probability tending to one.
\end{proof}
To avoid the spurious valley of Remark~\ref{rem:spurious}, $\sigma_U$ should in addition be below about $\mu\sigma_0/3$.

\subsection{Predictions testable on existing analyses}\label{sec:pred}
\paragraph{Prediction 1 (independent prediction of $\sigma$).} By \eqref{eq:sigma0refined} and Theorem~\ref{thm:cap}, the optimized $\hat\sigma$ should satisfy
\begin{equation}
\hat\sigma\approx(1+M\hat\sigma^{2})\,\frac{\sqrt{\hat d}}{s_{\ell_1^{*}}},
\end{equation}
where $\hat d$ is the residual variance per entry of \eqref{eq:dhat}. Note that $\hat d$ estimates the mean noise variance over \emph{all} genes, whereas $d$ is the mean over background genes; the two agree only if the mean noise variance of signal genes is about the same as that of background genes. If, for example, background genes have variance $d$ and signal genes $2d$, then $\varepsilon_{\mathrm{het}}=0$ but $\hat d$ estimates $(2-\pi_0)d$. When this is in doubt, the mean of the residual variances $\hat d_i$ of the genes that are not selected can be used as a candidate estimate of the background $d$. This requires that the influence of the remaining signal genes and the bias caused by selecting genes from the same data be small; its consistency is not established in this paper. The factor $1+M\hat\sigma^{2}$ is the leave-one-out correction \eqref{eq:sigma0refined}, with $\varepsilon_{\mathrm{loo}}$ estimated by $M\hat\sigma^{2}$. With the estimate $\hat d$ from all genes defined above, and apart from the small leave-one-out factor, which uses $\hat\sigma$ itself, the right-hand side is computed from the singular values, without the histogram; this does not apply to the alternative estimate from non-selected genes, which depends on the selection. The comparison is therefore a consistency check rather than a fully independent estimate, and the estimation error of $\hat d$ and the approximation errors of the model also enter; a large discrepancy indicates that the approximations do not apply. Agreement would support quantitatively that flattening estimates the noise level of the background.
\paragraph{Prediction 2 (diagnosis of contamination).} By Approximation~\ref{prop:contam}, contamination shifts background genes in proportion to their mean expression. For non-selected genes, the correlation between $u_{\ell_1^{*} i}$ and the mean expression (or $u_{\ell_1^{\mathrm{mean}} i}$, the singular vector of the mean component, which essentially is the normalized mean expression) should therefore be close to zero. A clear correlation indicates a violation of Assumption~\ref{as:contam}.
\paragraph{Diagnosis of weak signals.} By Table~\ref{tab:finite}, the limit applies when $|\mu_i|\gtrsim6$. If many selected genes have $|u_{\ell_1^{*} i}|/\hat\sigma$ just above the selection threshold ($\approx3$), weak signals are present, and some of them are likely to remain in the histogram.

\section{Discussion}
We have shown that flattening the $P$-value histogram estimates the standard deviation of background scores with an asymptotic relative bias of at most $1.1\%$, which can be removed by dividing by $\tau^{*}$, provided the background scores are normal and the signal genes are separated; and we have proved that both conditions follow from the model when the effective signals $\mu_i^{\mathrm{eff}}$ are strong (Theorem~\ref{thm:main}). In the simulations of Table~\ref{tab:finiteN} the remaining error for a given data set was a sampling fluctuation that decreased roughly as $N^{-1/2}$ (the theory itself gives no rate), and whether the approximations apply can be checked with the diagnostics of Table~\ref{tab:diag}. The analysis explains three empirical observations: why the standard deviation of all components cannot be used (Corollary~\ref{cor:local}), why the objective decreases again at small $\sigma$ (Lemma~\ref{lem:unimodal}), and why the valley to the right of the peak is the correct choice (Theorem~\ref{thm:drop}). It also shows where the normal distribution used by TD-based UFE comes from: each score is a weighted sum of the noise over the $M$ columns of the unfolded data, whose projection on the deterministic vector $\bar v_{\ell_1^{*}}$ is exactly normal for Gaussian noise and approximately normal otherwise by the central limit theorem (Theorem~\ref{thm:main}(ii)).

The requirement of strong signals deserves comment, since it might seem to restrict the method severely. Two problems must be distinguished. Detecting the signal component, i.e.\ $v_{\ell_1^{*}}\approx b$, benefits from the superposition of all signal genes: its condition $(1-\pi_0)\bar\mu^{2}\gg(1+\sqrt{M/N})^{2}$ (Assumption~\ref{as:spike}) involves the signal summed over genes and is mild. Selecting individual genes, on which the flattening relies, does not benefit from this superposition: each signal gene must stand out from the background on its own, which requires $\mu_i$ of about $3$ to $6$. Consequently, when the signal is spread thinly over many genes, the component can be found while the genes cannot be selected; the flattening then does not estimate $\sigma_0$ either, and the genes ranked by $|u_{\ell_1^{*} i}|$ still carry information that significance-based selection cannot use. This situation is detected by the diagnostics of Table~\ref{tab:diag}. For the selection of genes, first, $\mu_i=\delta_i\sqrt M$ is the signal accumulated over all $M=JQ$ entries, so that a modest per-entry effect gives a large $\mu_i$ when there are many combinations of samples and conditions: $\mu_i\ge6$ requires $\delta_i\ge1.3$ for $M=20$ but only $\delta_i\ge0.42$ for $M=200$. Second, $\mu\approx3$ is the $z$-value of the BH threshold at $\alpha=0.01$; signal genes of this strength are selected only about half the time even with the correct $\sigma$, and weaker ones are hardly detectable by any method based on $z$-values, so the requirement largely coincides with the signal being detectable at all. Third, what actually limits the method is a large fraction of signal genes near this threshold ($\mu$ of about $3$ to $5$); a few such genes bias $\hat\sigma$ only in proportion to their fraction (Table~\ref{tab:leak}), and their presence can be detected from the distribution of $|u_{\ell_1^{*} i}|/\hat\sigma$ among selected genes (Table~\ref{tab:diag}). Fourth, this limit belongs to the problem rather than to the method: when signal and background overlap, the background scale is not identifiable from the data (Proposition~\ref{prop:nonid}), and any data-based estimate of the null distribution faces the same obstacle. In past applications of TD-based UFE, the scores of selected genes were usually far above the threshold, which suggests that many real data sets lie in the regime covered by the theory; this is an empirical observation, not a guarantee.

It is worth separating what in this analysis is specific to tensor decomposition. The gene singular vectors of the HOSVD are the left singular vectors of the gene-mode unfolding $X$, so the scores, $\sigma_0$ and the whole theory of the flattening (Sections~\ref{sec:limit}--\ref{sec:drop}) are statements about the singular value decomposition of $X$ and would be the same for an analysis of $X$ as a matrix. What the tensor decomposition adds concerns the sample-side modes. By \eqref{eq:vhat} it decomposes the right singular vector $v_{\ell_1^{*}}$, a vector of length $M=JQ$, into products of singular vectors of the individual modes weighted by the core tensor. This shows to which combination of sample and condition patterns the signal component corresponds, allows the component to be selected through the core tensor (Proposition~\ref{prop:select}), and reveals problems that the matrix decomposition hides. For example, mixing with the mean expression appears as a non-zero core entry $G(\ell_1^{*}\ell_2^{\mathrm{mean}}\ell_3^{\mathrm{mean}})$ next to $G(\ell_1^{*}\ell_2^{*}\ell_3^{*})$, and their ratio estimates the mixing angle (Table~\ref{tab:diag}). A strong correlation between the signal and the mean expression is, however, a problem of uncentred data in general; the tensor decomposition can detect it but does not remove it.

Several limitations remain. The proof of Proposition~\ref{prop:select}, which concerns the selection of the component rather than the standard-deviation optimization, is given in outline. Theorem~\ref{thm:main} gives no rate of convergence in terms of $N$ alone, and the leave-one-out refinement of $\sigma_0$ (Remark~\ref{rem:loo}) is heuristic. Separation is guaranteed for the effective signal \eqref{eq:mueff}, which depends on the mean expression of each signal gene; a signal parallel to the mean expression can be cancelled (Remark~\ref{rem:cancel}). The noise variances of background genes must be nearly equal (Assumption~\ref{as:het}, Remark~\ref{rem:hetero}); for count data this usually requires a variance-stabilizing transformation such as the logarithm, and the spread can be checked with the diagnostic of Table~\ref{tab:diag}. For non-Gaussian noise, normality also requires that the design pattern is not concentrated on a few columns of the unfolded data; designs in which only a few samples differ from the rest fall outside the theory. The computer-assisted part covers $K=100$, $\alpha=0.01$ and $\pi_0\in[0.1,0.99]$. Background fractions below $0.1$ and $\tau>20$ are not relevant in practice, but $\pi_0>0.99$ and other values of $K$ and $\alpha$ require further computation. The relation between Assumption~\ref{as:contam} and log transformation or sample-wise normalization remains to be studied, preferably after examining Prediction~2 on real data. The conditions of the maximum-drop rule on the search interval are stated in units of $\sigma_0$; a guarantee that an interval computed from the data satisfies them remains to be established. Finally, the results concern the estimation of $\sigma_0$; control of the false discovery rate by BH with the estimated $\sigma$ is a separate question.

\section{Conclusions}
Flattening the histogram of the $P$-values of the genes that are not selected, the standard-deviation optimization of TD-based UFE, estimates the noise level of the background genes in the gene singular vector. In the limit of large data sets with perfectly separated signals its systematic bias is at most $1.1\%$ for 100 bins, BH level $0.01$ and background fractions between $0.1$ and $0.99$, and the maximum-drop rule turns this into a consistent estimator up to that bias. The conditions are that the signals are strong enough to be detected gene by gene, that the noise variances of the background genes are nearly equal, and that the signal is not strongly aligned with the mean expression; each can be checked with the diagnostics of Table~\ref{tab:diag}.

\section*{List of abbreviations}
BH: Benjamini--Hochberg; FDR: false discovery rate; HOSVD: higher-order singular value decomposition; KS: Kolmogorov--Smirnov; SVD: singular value decomposition; TD-based UFE: tensor-decomposition-based unsupervised feature extraction.

\section*{Declarations}
\subsection*{Ethics approval and consent to participate}
Not applicable.
\subsection*{Consent for publication}
Not applicable.
\subsection*{Availability of data and materials}
No experimental data were generated or analysed in this study. The datasets supporting the conclusions of this article are included within the article and its additional files; the Python scripts and their outputs that reproduce all numerical results, including the interval-arithmetic verification of Theorem~\ref{thm:cap}, are provided as Additional file~2.
\subsection*{Competing interests}
The authors declare that they have no competing interests.
\subsection*{Funding}
The authors gratefully acknowledge the Deanship of Scientific Research (DSR) at King Abdulaziz University, Jeddah, Saudi Arabia, for funding this project under grant No. (IPP: 1083-611-2026) and for providing technical support.
\subsection*{Authors' contributions}
Y. H. T. conceived the study, derived the results, performed the computations and wrote the manuscript. T.T. evaluated the results and wrote and reviewed the manuscript. Conceptualization, data curation, and analysis were performed by Y. H. T. All the authors have read and agreed to the published version of the manuscript.
\subsection*{Acknowledgements}
Not applicable.

\section*{Supplementary material}
Additional file~1 of the journal submission is reproduced here as Appendix~\ref{app:boxes} (Tables~S1 and S2). Additional file~2, the Python scripts and their outputs that reproduce all numerical results, is provided as the ancillary files of this arXiv record (directory \texttt{anc/}); its README lists which script produces which result.

\appendix
\section{Box-by-box results of the computer-assisted proof}\label{app:boxes}
\setcounter{table}{0}
\renewcommand{\thetable}{S\arabic{table}}
Table~\ref{tab:A1} covers $\tau\in[0.3,20]$. Each row corresponds to one box of $\pi_0$ values. The second column gives the number of $\tau$ boxes that were needed to decide the sign of $D$ or $D'$ on the whole range of $\tau$. The third and fourth columns give intervals that are proved to contain the local maximum $\tau_{\max}$ and the minimizer $\tau^{*}$ for every $\pi_0$ in the box. In every box, the sign of $D$ along increasing $\tau$ followed the same pattern: positive; decreasing through zero (the local maximum); negative; increasing through zero (the minimizer); positive.

Table~\ref{tab:A2} covers $\tau\in[0.05,0.3]$. Each entry gives a box of $\pi_0$ values and the number of $\tau$ boxes used. In every box $D>0$ was proved throughout, i.e.\ $L$ is strictly increasing and there is no critical point.

{\small
\begin{longtable}{cccc}
\caption{Results on $\tau\in[0.3,20]$.}\label{tab:A1}\\
\toprule
$\pi_0$ box & $\tau$ boxes & enclosure of $\tau_{\max}$ & enclosure of $\tau^{*}$\\
\midrule
\endfirsthead
\toprule
$\pi_0$ box & $\tau$ boxes & enclosure of $\tau_{\max}$ & enclosure of $\tau^{*}$\\
\midrule
\endhead
\bottomrule
\endfoot
$[0.1000,\,0.1100]$ & 3194 & $[0.4100,\,0.4181]$ & $[0.99552,\,0.99601]$ \\
$[0.1100,\,0.1200]$ & 3200 & $[0.4100,\,0.4182]$ & $[0.99519,\,0.99568]$ \\
$[0.1200,\,0.1300]$ & 3206 & $[0.4097,\,0.4179]$ & $[0.99489,\,0.99541]$ \\
$[0.1300,\,0.1400]$ & 3212 & $[0.4094,\,0.4176]$ & $[0.99459,\,0.99506]$ \\
$[0.1400,\,0.1500]$ & 3216 & $[0.4091,\,0.4174]$ & $[0.99432,\,0.99476]$ \\
$[0.1500,\,0.1600]$ & 3222 & $[0.4089,\,0.4200]$ & $[0.99405,\,0.99447]$ \\
$[0.1600,\,0.1700]$ & 3230 & $[0.4086,\,0.4197]$ & $[0.99375,\,0.99420]$ \\
$[0.1700,\,0.1800]$ & 3236 & $[0.4086,\,0.4198]$ & $[0.99348,\,0.99396]$ \\
$[0.1800,\,0.1900]$ & 3242 & $[0.4084,\,0.4196]$ & $[0.99324,\,0.99367]$ \\
$[0.1900,\,0.2000]$ & 3250 & $[0.4081,\,0.4194]$ & $[0.99297,\,0.99338]$ \\
$[0.2000,\,0.2100]$ & 3254 & $[0.4080,\,0.4192]$ & $[0.99273,\,0.99313]$ \\
$[0.2100,\,0.2200]$ & 3262 & $[0.4078,\,0.4191]$ & $[0.99247,\,0.99289]$ \\
$[0.2200,\,0.2300]$ & 3268 & $[0.4080,\,0.4193]$ & $[0.99227,\,0.99268]$ \\
$[0.2300,\,0.2400]$ & 3274 & $[0.4107,\,0.4192]$ & $[0.99205,\,0.99243]$ \\
$[0.2400,\,0.2500]$ & 3282 & $[0.4105,\,0.4190]$ & $[0.99182,\,0.99222]$ \\
$[0.2500,\,0.2600]$ & 3288 & $[0.4104,\,0.4189]$ & $[0.99162,\,0.99199]$ \\
$[0.2600,\,0.2700]$ & 3294 & $[0.4102,\,0.4188]$ & $[0.99142,\,0.99182]$ \\
$[0.2700,\,0.2800]$ & 3300 & $[0.4100,\,0.4193]$ & $[0.99123,\,0.99157]$ \\
$[0.2800,\,0.2900]$ & 3310 & $[0.4109,\,0.4193]$ & $[0.99104,\,0.99140]$ \\
$[0.2900,\,0.3000]$ & 3318 & $[0.4106,\,0.4217]$ & $[0.99088,\,0.99120]$ \\
$[0.3000,\,0.3100]$ & 3330 & $[0.4113,\,0.4208]$ & $[0.99070,\,0.99103]$ \\
$[0.3100,\,0.3200]$ & 3352 & $[0.4118,\,0.4213]$ & $[0.99056,\,0.99087]$ \\
$[0.3200,\,0.3300]$ & 3370 & $[0.4108,\,0.4216]$ & $[0.99040,\,0.99071]$ \\
$[0.3300,\,0.3400]$ & 3394 & $[0.4114,\,0.4220]$ & $[0.99027,\,0.99057]$ \\
$[0.3400,\,0.3500]$ & 3412 & $[0.4119,\,0.4224]$ & $[0.99013,\,0.99042]$ \\
$[0.3500,\,0.3600]$ & 3436 & $[0.4127,\,0.4230]$ & $[0.99001,\,0.99029]$ \\
$[0.3600,\,0.3700]$ & 3454 & $[0.4132,\,0.4234]$ & $[0.98989,\,0.99017]$ \\
$[0.3700,\,0.3800]$ & 3474 & $[0.4105,\,0.4238]$ & $[0.98977,\,0.99005]$ \\
$[0.3800,\,0.3900]$ & 3494 & $[0.4116,\,0.4247]$ & $[0.98969,\,0.98996]$ \\
$[0.3900,\,0.4000]$ & 3512 & $[0.4122,\,0.4251]$ & $[0.98958,\,0.98986]$ \\
$[0.4000,\,0.4100]$ & 3530 & $[0.4133,\,0.4261]$ & $[0.98954,\,0.98977]$ \\
$[0.4100,\,0.4200]$ & 3550 & $[0.4144,\,0.4241]$ & $[0.98942,\,0.98971]$ \\
$[0.4200,\,0.4300]$ & 3554 & $[0.4155,\,0.4250]$ & $[0.98941,\,0.98969]$ \\
$[0.4300,\,0.4400]$ & 3560 & $[0.4151,\,0.4270]$ & $[0.98933,\,0.98959]$ \\
$[0.4400,\,0.4500]$ & 3570 & $[0.4165,\,0.4254]$ & $[0.98933,\,0.98956]$ \\
$[0.4500,\,0.4600]$ & 3578 & $[0.4137,\,0.4293]$ & $[0.98928,\,0.98949]$ \\
$[0.4600,\,0.4700]$ & 3586 & $[0.4152,\,0.4267]$ & $[0.98928,\,0.98949]$ \\
$[0.4700,\,0.4800]$ & 3598 & $[0.4167,\,0.4281]$ & $[0.98926,\,0.98942]$ \\
$[0.4800,\,0.4900]$ & 3604 & $[0.4181,\,0.4279]$ & $[0.98926,\,0.98942]$ \\
$[0.4900,\,0.5000]$ & 3612 & $[0.4187,\,0.4296]$ & $[0.98924,\,0.98941]$ \\
$[0.5000,\,0.5100]$ & 3620 & $[0.4197,\,0.4305]$ & $[0.98924,\,0.98941]$ \\
$[0.5100,\,0.5200]$ & 3630 & $[0.4207,\,0.4304]$ & $[0.98925,\,0.98945]$ \\
$[0.5200,\,0.5300]$ & 3630 & $[0.4208,\,0.4318]$ & $[0.98930,\,0.98946]$ \\
$[0.5300,\,0.5400]$ & 3634 & $[0.4214,\,0.4328]$ & $[0.98932,\,0.98952]$ \\
$[0.5400,\,0.5500]$ & 3638 & $[0.4202,\,0.4335]$ & $[0.98937,\,0.98958]$ \\
$[0.5500,\,0.5600]$ & 3640 & $[0.4213,\,0.4329]$ & $[0.98940,\,0.98964]$ \\
$[0.5600,\,0.5700]$ & 3644 & $[0.4221,\,0.4337]$ & $[0.98947,\,0.98968]$ \\
$[0.5700,\,0.5800]$ & 3654 & $[0.4226,\,0.4362]$ & $[0.98954,\,0.98978]$ \\
$[0.5800,\,0.5900]$ & 3656 & $[0.4253,\,0.4366]$ & $[0.98963,\,0.98986]$ \\
$[0.5900,\,0.6000]$ & 3662 & $[0.4261,\,0.4374]$ & $[0.98972,\,0.98995]$ \\
$[0.6000,\,0.6100]$ & 3664 & $[0.4272,\,0.4380]$ & $[0.98981,\,0.99003]$ \\
$[0.6100,\,0.6200]$ & 3668 & $[0.4286,\,0.4381]$ & $[0.98992,\,0.99016]$ \\
$[0.6200,\,0.6300]$ & 3670 & $[0.4296,\,0.4390]$ & $[0.99000,\,0.99027]$ \\
$[0.6300,\,0.6400]$ & 3674 & $[0.4302,\,0.4410]$ & $[0.99016,\,0.99040]$ \\
$[0.6400,\,0.6500]$ & 3676 & $[0.4309,\,0.4416]$ & $[0.99026,\,0.99054]$ \\
$[0.6500,\,0.6600]$ & 3676 & $[0.4319,\,0.4425]$ & $[0.99039,\,0.99067]$ \\
$[0.6600,\,0.6700]$ & 3680 & $[0.4326,\,0.4431]$ & $[0.99056,\,0.99083]$ \\
$[0.6700,\,0.6800]$ & 3684 & $[0.4345,\,0.4449]$ & $[0.99071,\,0.99101]$ \\
$[0.6800,\,0.6900]$ & 3688 & $[0.4354,\,0.4458]$ & $[0.99087,\,0.99117]$ \\
$[0.6900,\,0.7000]$ & 3700 & $[0.4357,\,0.4469]$ & $[0.99106,\,0.99135]$ \\
$[0.7000,\,0.7100]$ & 3722 & $[0.4375,\,0.4475]$ & $[0.99123,\,0.99153]$ \\
$[0.7100,\,0.7200]$ & 3744 & $[0.4383,\,0.4488]$ & $[0.99140,\,0.99174]$ \\
$[0.7200,\,0.7300]$ & 3764 & $[0.4370,\,0.4495]$ & $[0.99162,\,0.99192]$ \\
$[0.7300,\,0.7400]$ & 3788 & $[0.4387,\,0.4509]$ & $[0.99182,\,0.99215]$ \\
$[0.7400,\,0.7500]$ & 3812 & $[0.4407,\,0.4527]$ & $[0.99205,\,0.99235]$ \\
$[0.7500,\,0.7600]$ & 3840 & $[0.4421,\,0.4539]$ & $[0.99226,\,0.99258]$ \\
$[0.7600,\,0.7700]$ & 3870 & $[0.4429,\,0.4546]$ & $[0.99250,\,0.99281]$ \\
$[0.7700,\,0.7800]$ & 3904 & $[0.4440,\,0.4556]$ & $[0.99272,\,0.99305]$ \\
$[0.7800,\,0.7900]$ & 3938 & $[0.4451,\,0.4566]$ & $[0.99295,\,0.99330]$ \\
$[0.7900,\,0.8000]$ & 3982 & $[0.4462,\,0.4576]$ & $[0.99321,\,0.99357]$ \\
$[0.8000,\,0.8100]$ & 4096 & $[0.4470,\,0.4598]$ & $[0.99348,\,0.99383]$ \\
$[0.8100,\,0.8200]$ & 4230 & $[0.4484,\,0.4612]$ & $[0.99375,\,0.99410]$ \\
$[0.8200,\,0.8300]$ & 4414 & $[0.4500,\,0.4637]$ & $[0.99401,\,0.99435]$ \\
$[0.8300,\,0.8400]$ & 4810 & $[0.4530,\,0.4637]$ & $[0.99428,\,0.99464]$ \\
$[0.8400,\,0.8450]$ & 2734 & $[0.4569,\,0.4624]$ & $[0.99459,\,0.99477]$ \\
$[0.8450,\,0.8500]$ & 2732 & $[0.4575,\,0.4630]$ & $[0.99474,\,0.99492]$ \\
$[0.8500,\,0.8550]$ & 2726 & $[0.4581,\,0.4636]$ & $[0.99489,\,0.99507]$ \\
$[0.8550,\,0.8600]$ & 2726 & $[0.4587,\,0.4641]$ & $[0.99504,\,0.99521]$ \\
$[0.8600,\,0.8650]$ & 2724 & $[0.4597,\,0.4650]$ & $[0.99518,\,0.99537]$ \\
$[0.8650,\,0.8700]$ & 2718 & $[0.4603,\,0.4673]$ & $[0.99534,\,0.99552]$ \\
$[0.8700,\,0.8750]$ & 2716 & $[0.4616,\,0.4668]$ & $[0.99549,\,0.99567]$ \\
$[0.8750,\,0.8800]$ & 2716 & $[0.4625,\,0.4677]$ & $[0.99565,\,0.99582]$ \\
$[0.8800,\,0.8850]$ & 2718 & $[0.4624,\,0.4685]$ & $[0.99580,\,0.99598]$ \\
$[0.8850,\,0.8900]$ & 2718 & $[0.4633,\,0.4694]$ & $[0.99596,\,0.99614]$ \\
$[0.8900,\,0.8950]$ & 2718 & $[0.4643,\,0.4702]$ & $[0.99611,\,0.99630]$ \\
$[0.8950,\,0.9000]$ & 2712 & $[0.4656,\,0.4712]$ & $[0.99628,\,0.99646]$ \\
$[0.9000,\,0.9025]$ & 2614 & $[0.4680,\,0.4709]$ & $[0.99645,\,0.99654]$ \\
$[0.9025,\,0.9050]$ & 2614 & $[0.4677,\,0.4715]$ & $[0.99653,\,0.99662]$ \\
$[0.9050,\,0.9075]$ & 2614 & $[0.4688,\,0.4722]$ & $[0.99661,\,0.99670]$ \\
$[0.9075,\,0.9100]$ & 2614 & $[0.4690,\,0.4728]$ & $[0.99669,\,0.99679]$ \\
$[0.9100,\,0.9125]$ & 2612 & $[0.4698,\,0.4728]$ & $[0.99678,\,0.99687]$ \\
$[0.9125,\,0.9150]$ & 2598 & $[0.4696,\,0.4733]$ & $[0.99686,\,0.99696]$ \\
$[0.9150,\,0.9175]$ & 2568 & $[0.4702,\,0.4739]$ & $[0.99694,\,0.99704]$ \\
$[0.9175,\,0.9200]$ & 2548 & $[0.4713,\,0.4746]$ & $[0.99703,\,0.99712]$ \\
$[0.9200,\,0.9225]$ & 2536 & $[0.4715,\,0.4752]$ & $[0.99711,\,0.99721]$ \\
$[0.9225,\,0.9250]$ & 2526 & $[0.4721,\,0.4758]$ & $[0.99720,\,0.99729]$ \\
$[0.9250,\,0.9275]$ & 2518 & $[0.4728,\,0.4763]$ & $[0.99728,\,0.99738]$ \\
$[0.9275,\,0.9300]$ & 2510 & $[0.4736,\,0.4770]$ & $[0.99737,\,0.99747]$ \\
$[0.9300,\,0.9325]$ & 2502 & $[0.4742,\,0.4776]$ & $[0.99745,\,0.99755]$ \\
$[0.9325,\,0.9350]$ & 2496 & $[0.4747,\,0.4782]$ & $[0.99754,\,0.99764]$ \\
$[0.9350,\,0.9375]$ & 2490 & $[0.4753,\,0.4788]$ & $[0.99763,\,0.99772]$ \\
$[0.9375,\,0.9400]$ & 2486 & $[0.4759,\,0.4793]$ & $[0.99772,\,0.99781]$ \\
$[0.9400,\,0.9425]$ & 2480 & $[0.4766,\,0.4797]$ & $[0.99780,\,0.99790]$ \\
$[0.9425,\,0.9450]$ & 2474 & $[0.4773,\,0.4803]$ & $[0.99789,\,0.99799]$ \\
$[0.9450,\,0.9475]$ & 2470 & $[0.4780,\,0.4812]$ & $[0.99798,\,0.99807]$ \\
$[0.9475,\,0.9500]$ & 2470 & $[0.4783,\,0.4822]$ & $[0.99806,\,0.99816]$ \\
$[0.9500,\,0.9525]$ & 2468 & $[0.4790,\,0.4828]$ & $[0.99815,\,0.99825]$ \\
$[0.9525,\,0.9550]$ & 2468 & $[0.4799,\,0.4837]$ & $[0.99824,\,0.99834]$ \\
$[0.9550,\,0.9575]$ & 2468 & $[0.4803,\,0.4840]$ & $[0.99833,\,0.99843]$ \\
$[0.9575,\,0.9600]$ & 2468 & $[0.4815,\,0.4853]$ & $[0.99842,\,0.99852]$ \\
$[0.9600,\,0.9625]$ & 2470 & $[0.4822,\,0.4859]$ & $[0.99851,\,0.99861]$ \\
$[0.9625,\,0.9650]$ & 2472 & $[0.4831,\,0.4868]$ & $[0.99860,\,0.99870]$ \\
$[0.9650,\,0.9675]$ & 2474 & $[0.4837,\,0.4874]$ & $[0.99869,\,0.99879]$ \\
$[0.9675,\,0.9700]$ & 2480 & $[0.4847,\,0.4884]$ & $[0.99878,\,0.99888]$ \\
$[0.9700,\,0.9725]$ & 2482 & $[0.4855,\,0.4896]$ & $[0.99887,\,0.99897]$ \\
$[0.9725,\,0.9750]$ & 2492 & $[0.4867,\,0.4909]$ & $[0.99896,\,0.99906]$ \\
$[0.9750,\,0.9775]$ & 2498 & $[0.4876,\,0.4918]$ & $[0.99906,\,0.99916]$ \\
$[0.9775,\,0.9800]$ & 2514 & $[0.4878,\,0.4931]$ & $[0.99915,\,0.99925]$ \\
$[0.9800,\,0.9825]$ & 2538 & $[0.4890,\,0.4943]$ & $[0.99924,\,0.99934]$ \\
$[0.9825,\,0.9850]$ & 2568 & $[0.4906,\,0.4952]$ & $[0.99934,\,0.99943]$ \\
$[0.9850,\,0.9875]$ & 2572 & $[0.4919,\,0.4965]$ & $[0.99943,\,0.99953]$ \\
$[0.9875,\,0.9900]$ & 2572 & $[0.4934,\,0.4985]$ & $[0.99952,\,0.99962]$ \\
\end{longtable}
}

{\small
\begin{longtable}{cccccc}
\caption{Results on $\tau\in[0.05,0.3]$ (no critical point in any box).}\label{tab:A2}\\
\toprule
$\pi_0$ box & $\tau$ boxes & $\pi_0$ box & $\tau$ boxes & $\pi_0$ box & $\tau$ boxes\\
\midrule
\endfirsthead
\toprule
$\pi_0$ box & $\tau$ boxes & $\pi_0$ box & $\tau$ boxes & $\pi_0$ box & $\tau$ boxes\\
\midrule
\endhead
\bottomrule
\endfoot
$[0.10,\,0.11]$ & 1390 & $[0.11,\,0.12]$ & 1392 & $[0.12,\,0.13]$ & 1392 \\
$[0.13,\,0.14]$ & 1394 & $[0.14,\,0.15]$ & 1394 & $[0.15,\,0.16]$ & 1396 \\
$[0.16,\,0.17]$ & 1396 & $[0.17,\,0.18]$ & 1398 & $[0.18,\,0.19]$ & 1398 \\
$[0.19,\,0.20]$ & 1400 & $[0.20,\,0.21]$ & 1400 & $[0.21,\,0.22]$ & 1402 \\
$[0.22,\,0.23]$ & 1402 & $[0.23,\,0.24]$ & 1404 & $[0.24,\,0.25]$ & 1404 \\
$[0.25,\,0.26]$ & 1406 & $[0.26,\,0.27]$ & 1406 & $[0.27,\,0.28]$ & 1408 \\
$[0.28,\,0.29]$ & 1408 & $[0.29,\,0.30]$ & 1410 & $[0.30,\,0.31]$ & 1410 \\
$[0.31,\,0.32]$ & 1410 & $[0.32,\,0.33]$ & 1412 & $[0.33,\,0.34]$ & 1412 \\
$[0.34,\,0.35]$ & 1414 & $[0.35,\,0.36]$ & 1414 & $[0.36,\,0.37]$ & 1416 \\
$[0.37,\,0.38]$ & 1416 & $[0.38,\,0.39]$ & 1416 & $[0.39,\,0.40]$ & 1418 \\
$[0.40,\,0.41]$ & 1418 & $[0.41,\,0.42]$ & 1420 & $[0.42,\,0.43]$ & 1420 \\
$[0.43,\,0.44]$ & 1420 & $[0.44,\,0.45]$ & 1422 & $[0.45,\,0.46]$ & 1422 \\
$[0.46,\,0.47]$ & 1422 & $[0.47,\,0.48]$ & 1424 & $[0.48,\,0.49]$ & 1424 \\
$[0.49,\,0.50]$ & 1426 & $[0.50,\,0.51]$ & 1426 & $[0.51,\,0.52]$ & 1426 \\
$[0.52,\,0.53]$ & 1428 & $[0.53,\,0.54]$ & 1428 & $[0.54,\,0.55]$ & 1428 \\
$[0.55,\,0.56]$ & 1430 & $[0.56,\,0.57]$ & 1430 & $[0.57,\,0.58]$ & 1430 \\
$[0.58,\,0.59]$ & 1432 & $[0.59,\,0.60]$ & 1432 & $[0.60,\,0.61]$ & 1432 \\
$[0.61,\,0.62]$ & 1434 & $[0.62,\,0.63]$ & 1434 & $[0.63,\,0.64]$ & 1434 \\
$[0.64,\,0.65]$ & 1436 & $[0.65,\,0.66]$ & 1436 & $[0.66,\,0.67]$ & 1436 \\
$[0.67,\,0.68]$ & 1438 & $[0.68,\,0.69]$ & 1438 & $[0.69,\,0.70]$ & 1438 \\
$[0.70,\,0.71]$ & 1438 & $[0.71,\,0.72]$ & 1440 & $[0.72,\,0.73]$ & 1440 \\
$[0.73,\,0.74]$ & 1440 & $[0.74,\,0.75]$ & 1442 & $[0.75,\,0.76]$ & 1442 \\
$[0.76,\,0.77]$ & 1442 & $[0.77,\,0.78]$ & 1444 & $[0.78,\,0.79]$ & 1444 \\
$[0.79,\,0.80]$ & 1444 & $[0.80,\,0.81]$ & 1444 & $[0.81,\,0.82]$ & 1350 \\
$[0.82,\,0.83]$ & 1308 & $[0.83,\,0.84]$ & 1286 & $[0.84,\,0.85]$ & 1268 \\
$[0.85,\,0.86]$ & 1256 & $[0.86,\,0.87]$ & 1242 & $[0.87,\,0.88]$ & 1232 \\
$[0.88,\,0.89]$ & 1222 & $[0.89,\,0.90]$ & 1214 & $[0.90,\,0.91]$ & 1206 \\
$[0.91,\,0.92]$ & 1198 & $[0.92,\,0.93]$ & 1192 & $[0.93,\,0.94]$ & 1186 \\
$[0.94,\,0.95]$ & 1180 & $[0.95,\,0.96]$ & 1174 & $[0.96,\,0.97]$ & 1170 \\
$[0.97,\,0.98]$ & 1164 & $[0.98,\,0.99]$ & 1160 &  &  \\
\end{longtable}
}


\begin{thebibliography}{99}
\bibitem{Taguchi2024} Taguchi Y-h. Unsupervised feature extraction applied to bioinformatics: a PCA based and TD based approach. 2nd ed. Cham: Springer; 2024.
\bibitem{DeLathauwer2000} De Lathauwer L, De Moor B, Vandewalle J. A multilinear singular value decomposition. SIAM J Matrix Anal Appl. 2000;21:1253--78.
\bibitem{Kolda2009} Kolda TG, Bader BW. Tensor decompositions and applications. SIAM Rev. 2009;51:455--500.
\bibitem{Taguchi2022} Taguchi Y-h, Turki T. Adapted tensor decomposition and PCA based unsupervised feature extraction select more biologically reasonable differentially expressed genes than conventional methods. Sci Rep. 2022;12:17438.
\bibitem{Efron2004} Efron B. Large-scale simultaneous hypothesis testing: the choice of a null hypothesis. J Am Stat Assoc. 2004;99:96--104.
\bibitem{JinCai2007} Jin J, Cai TT. Estimating the null and the proportion of nonnull effects in large-scale multiple comparisons. J Am Stat Assoc. 2007;102:495--506.
\bibitem{Storey2002} Storey JD. A direct approach to false discovery rates. J R Stat Soc Series B Stat Methodol. 2002;64:479--98.
\bibitem{BH1995} Benjamini Y, Hochberg Y. Controlling the false discovery rate: a practical and powerful approach to multiple testing. J R Stat Soc Series B Stat Methodol. 1995;57:289--300.
\bibitem{Johansson2017} Johansson F. Arb: efficient arbitrary-precision midpoint-radius interval arithmetic. IEEE Trans Comput. 2017;66:1281--92.
\bibitem{DiaconisFreedman1987} Diaconis P, Freedman D. A dozen de Finetti-style results in search of a theory. Ann Inst Henri Poincar\'e Probab Stat. 1987;23:397--423.
\bibitem{BenaychNadakuditi2012} Benaych-Georges F, Nadakuditi RR. The singular values and vectors of low rank perturbations of large rectangular random matrices. J Multivar Anal. 2012;111:120--35.
\bibitem{Shevtsova2010} Shevtsova IG. An improvement of convergence rate estimates in the Lyapunov theorem. Dokl Math. 2010;82:862--4.
\bibitem{YinBaiKrishnaiah1988} Yin YQ, Bai ZD, Krishnaiah PR. On the limit of the largest eigenvalue of the large dimensional sample covariance matrix. Probab Theory Relat Fields. 1988;78:509--21.
\bibitem{Wedin1972} Wedin P-{\AA}. Perturbation bounds in connection with singular value decomposition. BIT. 1972;12:99--111.
\bibitem{GenoveseWasserman2004} Genovese C, Wasserman L. A stochastic process approach to false discovery control. Ann Stat. 2004;32:1035--61.
\bibitem{Tucker2011} Tucker W. Validated numerics: a short introduction to rigorous computations. Princeton: Princeton University Press; 2011.
\bibitem{GenoveseWasserman2002} Genovese C, Wasserman L. Operating characteristics and extensions of the false discovery rate procedure. J R Stat Soc Series B Stat Methodol. 2002;64:499--517.
\bibitem{vanderVaart1998} van der Vaart AW. Asymptotic statistics. Cambridge: Cambridge University Press; 1998.
\end{thebibliography}
\end{document}